\documentclass[a4paper,11pt]{article}
\pdfoutput=1 % if your are submitting a pdflatex (i.e. if you have
\usepackage{jcappub,setup} % for details on the use of the package, please
\usepackage[T1]{fontenc} % if needed

\def\frac#1#2{{{#1}\over {#2}}}
\def\gsim{\gtrsim}
\def\lsim{\lesssim}    
\newcommand{\be}{\begin{equation}}
\newcommand{\ee}{\end{equation}}
\newcommand{\bea}{\begin{eqnarray}}
\newcommand{\eea}{\end{eqnarray}}
\newcommand{\bi}{\begin{itemize}}
\newcommand{\ei}{\end{itemize}}
\newcommand{\ben}{\begin{enumerate}}
\newcommand{\een}{\end{enumerate}}
\newcommand{\la}{\left\langle}
\newcommand{\ra}{\right\rangle}

\newcommand{\lp}{\left(}
\newcommand{\rp}{\right)}

\usepackage{listings,color}

\numberwithin{equation}{section}
\numberwithin{figure}{section}
\numberwithin{table}{section}

\definecolor{mygreen}{rgb}{0,0.6,0}
\definecolor{mygray}{rgb}{0.5,0.5,0.5}
\definecolor{mymauve}{rgb}{0.58,0,0.82}

\addtocontents{toc}{\protect\setcounter{tocdepth}{2}}

\title{\boldmath Complete predictions for high-energy neutrino propagation in matter}

\author[a]{Alfonso Garcia,}
\author[a]{Rhorry Gauld,}
\author[a]{Aart Heijboer,}
\author[a,b]{Juan Rojo}
\affiliation[a]{Nikhef, Science Park 105, 1098 XG Amsterdam, The Netherlands}
\affiliation[b]{Department of Physics and Astronomy, VU, 1081 HV Amsterdam, The Netherlands}

\emailAdd{alfonsog@nikhef.nl}
\emailAdd{aart.heijboer@nikhef.nl}
\emailAdd{r.gauld@nikhef.nl}
\emailAdd{j.rojo@vu.nl}

\abstract{
We present predictions for the interactions of energetic neutrinos with matter as they propagate through Earth towards large-volume detectors. Our results are based on state-of-the-art calculations of the high-energy neutrino-matter interaction cross-sections, which we have implemented in the {\tt HEDIS} module of {\tt GENIE}. In addition to the dominant interaction process, deep inelastic scattering off quarks and gluons, we include the relevant subdominant channels: (in)elastic scattering off the photon field of nucleons, coherent scattering off the photon field of nuclei, as well as the scattering on atomic electrons via the Glashow resonance. Our predictions for the neutrino attenuation rates are provided by a new software package, {\tt NuPropEarth}. We quantify the dependence of our results on the cross-section model, including nuclear corrections, the incidence angle, and the spectral index, and compare them with other publicly available tools.
}


\begin{document}
\maketitle
\flushbottom

\section{Introduction}
\label{sec:intro}

The detection of high-energy neutrinos from astrophysical sources has the potential to greatly advance our knowledge of both astronomy and particle physics. 
For several ongoing and planned experiments, including ANTARES~\cite{Collaboration:2011nsa}, IceCube~\cite{Halzen:2010yj,Aartsen:2014gkd}, KM3NeT~\cite{Adrian-Martinez:2016fdl}, ANITA~\cite{Gorham:2010kv}, BAIKAL~\cite{Aynutdinov:2005dq} and IceCube-Gen2~\cite{Aartsen:2014njl}, the detection of such energetic neutrinos is one of the main goals.
At higher energies, Earth-skimming neutrinos are the target for a slew of running and proposed experiments, like the Askaryan Radio Array (ARA)~\cite{ALLISON2012457}, ARIANNA~\cite{Barwick:2014pca}, the Pierre Auger Observatory~\cite{aab:2019auo}, the Antarctic Impulsive Transient Antenna (ANITA)~\cite{Gorham:2019guw}, and the Giant Radio Array for Neutrino Detection (GRAND)~\cite{Alvarez-Muniz:2018bhp}.
The interpretation of the measured event rates from these experiments depends on precise knowledge of the interaction cross section of high-energy neutrinos in two ways.
First of all, these cross sections determine the probability of a neutrino interacting within the detector's sensitive volume.
Secondly, interactions with Earth matter modify the incoming neutrinos as they travel towards the detector, leading to either absorption or to a change in the neutrino energy and direction.

High-energy neutrinos can interact with Earth matter via different processes.
The total cross section is dominated by deep inelastic scattering (DIS) off the quarks and gluons from matter nucleons.
Neutrino DIS can be either mediated by a \textit{W}-boson in charged-current (CC) scattering,
or by a \textit{Z}-boson in neutral-current (NC) scattering.
The former leads to the production of a charged lepton and in general to the attenuation of the neutrino flux, while the latter results in a degradation of the neutrino energies.
Further, tau leptons produced in the CC scattering of tau neutrinos will decay within the Earth volume,
feeding the tau neutrino flux at lower energies by means of a regeneration process.

Beyond the dominant NC and CC deep inelastic scattering mechanisms, other subdominant channels contribute to neutrino-matter interactions at high energies. These include the (in)elastic scattering off the photon field of nucleons, coherent scattering off the photon field of nuclei, as well as the scattering on atomic electrons via the Glashow resonance.
The latter process is restricted to electron anti-neutrinos which undergo CC interactions with atomic electrons
via the resonant exchange of a \textit{W}-boson, which occurs for $E_{\nu}\simeq 6.3~\PeV$.
These channels can modify the neutrino-matter interaction cross section by up several percent as compared to the leading DIS process, and it is therefore important to include them to precisely model the propagation of neutrinos through Earth.
 
The modification of the astrophysical neutrino flux due to these various interaction processes depends, among other factors, on the neutrino energy, the flux spectral index, the amount of matter traveled (equivalently on the zenith angle), the model for the Earth structure, and the cross-sections modeling.
Furthermore, as the neutrino interaction cross sections grow with $E_{\nu}$ (power-like for $E_{\nu}\lsim m_{\PW}$, logarithmically above), these matter-induced attenuation effects are particularly effective in suppressing the high-energy component of the flux.
This sensitivity of the expected event rates on the neutrino-matter interactions was exploited by IceCube to present a first measurement of the neutrino-nucleon interaction cross section up to $E_{\nu}=980~\TeV$~\cite{Aartsen:2017kpd} (see also~\cite{Bustamante:2017xuy}), recently updated to energies of $10~\PeV$ by using 7.5 years of the high-energy starting events (HESE) sample~\cite{Yuan:2019wil}.
This effect has been also used to study the Earth’s internal structure~\cite{Donini:2018tsg}.

Theoretical calculations of the neutrino-matter interaction cross section at high energy rely on inputs such as
higher-order perturbative QCD and electroweak calculations, heavy quark mass effects, and the quark and gluon substructure of nucleons and nuclei encoded in the parton distributions (PDFs)~\cite{Gao:2017yyd,Rojo:2019uip,Ethier:2020way}.
Several groups have presented predictions for the neutrino-nucleon DIS cross sections at high-energies, both based on the collinear DGLAP framework~\cite{CooperSarkar:2007cv,Gluck:2010rw,Connolly:2011,CooperSarkar:2011pa,Bertone:2018dse,Gandhi:1998ri} and in alternative approaches~\cite{Albacete:2015zra,Arguelles:2015wba,Goncalves:2010ay} such as accounting for non-linear corrections to the QCD evolution equations.
For energies in the range $E_\nu \gsim 10^7~\GeV (10~\PeV)$, these cross sections become sensitive to the behaviour of the nucleon and nuclear PDFs at  small values of the momentum fraction $x \lsim 10^{-5}$, for which limited experimental constraints are available.
None of these studies account for the effects of subdominant interaction processes in the total cross section, which
have been considered in~\cite{Seckel:1997kk,Alikhanov:2015kla,Gauld:2019pgt,Zhou:2019vxt,Beacom:2019pzs}.

%%%%%%

In this work we present predictions for the interactions of high-energy neutrinos with matter as they propagate through Earth. Our results are based on state-of-the-art calculations of the high-energy neutrino-matter interaction cross-sections, which we have implemented in an updated version of the {\tt HEDIS}~\cite{Garcia:2019hze} module of the {\tt GENIE}~\cite{Andreopoulos:2009rq} neutrino event generator.
The predictions for the neutrino DIS cross section presented in this work are based on the BGR18 calculation~\cite{Bertone:2018dse}, which combines higher-order QCD calculations with the NNPDF nucleon PDFs
constrained in the small-$x$ region by the LHCb $D$-meson production data~\cite{Gauld:2016kpd,Gauld:2015yia}.
The description of the subleading interactions is based on~\cite{Gauld:2019pgt}, and we adopt the formalism for coherent scattering of~\cite{Ballett:2018uuc} which was presented for \textit{W}-boson production in~\cite{Beacom:2019pzs,Zhou:2019vxt}.

This calculation of the interaction cross-sections provided by {\tt HEDIS} has been implemented in a novel software package for the modeling of high-energy neutrino propagation in matter, {\tt NuPropEarth}.
This framework is based on a Monte Carlo simulation, which allows keeping track of the trajectory that each of the incoming neutrinos follow as they travel through Earth.
By means of {\tt NuPropEarth} we are able to study the sensitivity of the resulting neutrino attenuation rates with respect to the incoming neutrino flavour, the cross-section model, the incidence angle, the spectral index $\gamma$ of the incoming flux, and the subleading interaction mechanisms.
We also assess the impact of nuclear corrections in the 
attenuation rates using the nNNPDF1.0 determination~\cite{AbdulKhalek:2019mzd}, finding that these represent one of the dominant sources of theoretical uncertainty.
We  compare our results for the neutrino attenuation rates to those from other publicly available tools such as  {\tt NuFate}~\cite{Vincent:2017svp}, {\tt NuTauSim}~\cite{Alvarez-Muniz:2017mpk} and {\tt TauRunner}~\cite{Safa:2019ege}, and identify the underlying origin of their differences when present.
Instructions for the installation and running of {\tt NuPropEarth},
which can be obtained from its public {\tt GitHub} repository, are also provided.

The outline of this paper is the following.
To begin with, in Sect.~\ref{sec:uheneuttheory} we review the theoretical ingredients that determine the interactions
between high energy neutrinos and Earth matter.
Then in Sect.~\ref{sec:simulation} we present the details of our simulation of the propagation of neutrinos through the Earth based on the {\tt NuPropEarth} framework.
In Sect.~\ref{sec:results_p} we discuss the results of our calculation for the neutrino flux attenuation and compare with other approaches presented in the literature.
In Sect.~\ref{sec:results_nucl} we revisit these results now accounting for the uncertainties associated to the modelling of the nuclear structure.
Finally, in Sect.~\ref{sec:summary} we conclude and outline possible future developments for our work.
The dependence of the attenuation rates on the spectral index is collected in the Appendix~\ref{app:spectral},
we provide recommendations for the calculation of neutrino-nucleon DIS predictions in Appendix~\ref{app:DIS},
and in Appendix~\ref{sec:installation} we describe how {\tt  NuPropEarth} should be installed and executed.

%%%%%%%%%%%%%%%%%%%%%%%%%%%%%%%%%%%%%%%%%%%%%%%%%%%%%%%%%%%%%%%%%%%%%%%%%%%%%%%%%%
%%%%%%%%%%%%%%%%%%%%%%%%%%%%%%%%%%%%%%%%%%%%%%%%%%%%%%%%%%%%%%%%%%%%%%%%%%%%%%%%%%

\section{Neutrino-matter interactions at high energies and {\tt HEDIS}}
\label{sec:uheneuttheory}

The main goal of this work is to provide a flexible framework suitable for describing the attenuation processes that high-energy neutrinos undergo as they travel through the Earth, accounting for all relevant neutrino-matter interaction mechanisms.
The core of this framework are the theoretical predictions for the differential cross sections describing how high-energy neutrinos can scatter with the possible matter targets encountered as they traverses the Earth.

In this section, we review the theoretical formalism which is used to describe such high-energy neutrino ($E_{\nu}\gtrsim 1~\TeV$) scattering.
Furthermore, we provide here specific details on how each of the relevant mechanisms have been implemented within {\tt HEDIS}~\cite{Garcia:2019hze}.
As part of this implementation, we also benchmark the neutrino-nucleon Deep Inelastic Scattering (DIS) calculation against the previous results provided in BGR18~\cite{Bertone:2018dse} and CMS11~\cite{CooperSarkar:2011pa}.
As will be discussed, this is the dominant scattering process for neutrino scattering at high-energies, and these models will be used as benchmarks in the following sections when we study specific features of the propagation of neutrinos in Earth.
As these processes have all been implemented within a common framework, we take the opportunity to provide a prediction for the inclusive cross section taking into account each of these scattering regimes (with modern inputs in all cases). This prediction represents an improvement over those made in all previous studies.

%%%%%%%%%%%%%%%%%%%%%%%%%%%%%%%%%%%%%%%%%%%%%%%%%%
\begin{figure}[tbp]
\centering 
\includegraphics[width=.4\textwidth]{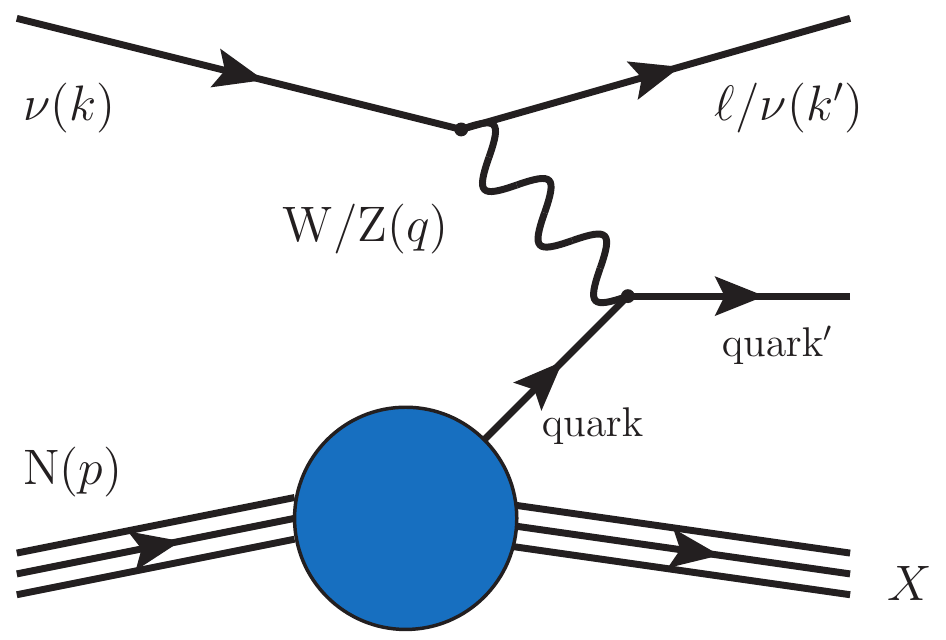}\hspace{2cm}
\includegraphics[width=.4\textwidth]{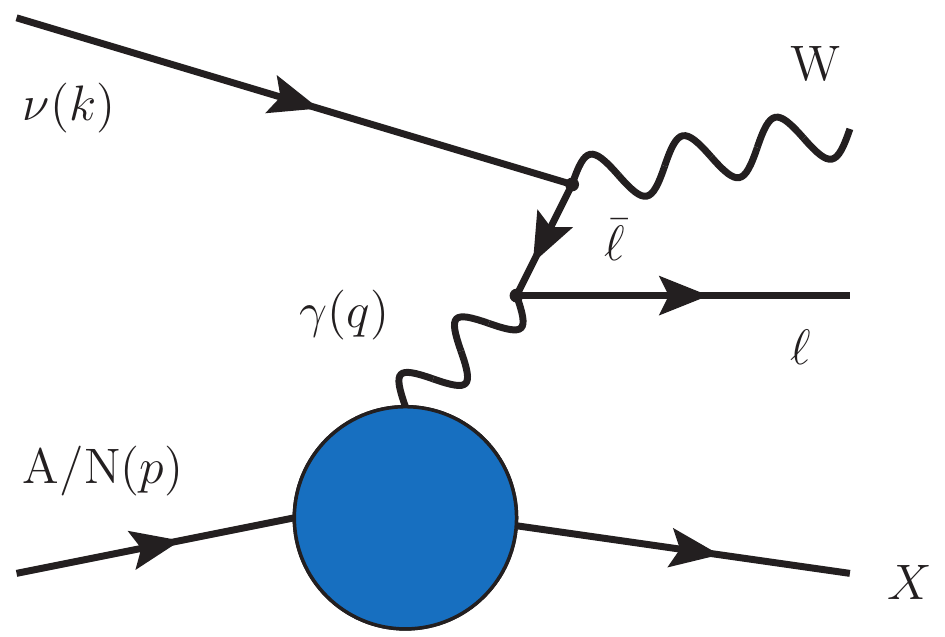}
\caption{Left: neutrinos interact with the nucleons of Earth matter either via
  charged current scattering, mediated by a \textit{W}-boson, or via
  neutral current scattering, mediated by a \textit{Z}-boson.
  The corresponding cross-section calculation requires as input knowledge on
  the quark and gluon PDFs in the nucleon.
  Right: neutrinos may interact coherently with the photon field of a nucleus to produce an on-shell \textit{W}-boson. A similar process also exists when the neutrino interacts with the photon field of individual nucleons.
}
\label{fig:xsec-diagram}
\end{figure}
%%%%%%%%%%%%%%%%%%%%%%%%%%%%%%%%%%%%%%%%%%%%%%%%%%

\subsection{Theoretical formalism}
\label{sec:formalism}

To structure the discussion of the various neutrino scattering mechanisms relevant to the attenuation calculation, it will be useful to recognise that although we are interested in the scattering rate of high-energy neutrinos, the total momentum exchange ($Q$) of the neutrino to the possible targets (atoms) in the scattering process may range from high to very small values ($Q \leq \GeV$).

The distinction between $E_\nu$ and $Q$ emphasizes that, even for neutrinos with a common energy $E_{\nu}$, depending on the actual momentum transfer $Q$ the scattering process will probe different structures of the target atom, which in turn determines which should be the appropriate theoretical description.
Taking into account this consideration, we can summarise the relevant scattering mechanisms for the calculation of neutrino attenuation due to matter effects as follows:
\begin{itemize}
\item Deep inelastic scattering (DIS) off nucleons.
In this process, the neutrino scatters upon the internal content of bound nucleons.
This channel is dominated by the subprocess where the neutrino exchanges an electroweak gauge-boson, either a \textit{W} or a \textit{Z} in CC and NC scattering respectively, with the partonic content of the bound nucleon.
The momentum transfer $Q$ associated to this channel corresponds to a continuum of energies $Q\gtrsim0$, although the numerically most important contribution arises from the region where the exchanged momentum is large, $Q\approx m_{\PW}$.

In addition to scattering off quarks and gluons, DIS also receives sub-leading contributions from processes where the neutrino (indirectly) interacts with the photon field of the nucleon.
Such interactions are relevant when it becomes kinematically possible to produce a \textit{W}-boson, leading to a resonant enhancement of the cross section.

\item Coherent scattering on a nucleus.
  At very small values of the momentum transfer, $Q\leq 1~\GeV$, the neutrino can resolve the photon field of the whole nucleus~\cite{Freedman:1973yd}.
  The contribution from this process becomes relevant for neutrino energies of $E_{\nu} \gtrsim~30~\TeV$, for
  which despite the small value of $Q$ it becomes kinematically possible to produce a \textit{W}-boson.  The cross section for this process is proportional to the atomic number of the nucleus squared ($\propto Z^2$) and therefore becomes increasingly more important for heavy nuclei, where it can reach up to (5-10)\% of the DIS cross section.
\item Elastic/diffractive scattering on nucleons.
  At momentum transfers of $Q\approx 1~\GeV$, the neutrino can resolve the photon field of an individual bound nucleon. Typically, this contribution is numerically unimportant (below 1\% of the total cross section).
\item Scattering upon atomic electrons.
  Electron anti-neutrinos can interact resonantly with atomic electrons to produce an on-shell \textit{W}-boson. This process, known as the Glashow resonance~\cite{Glashow:1960zz}, becomes important when the centre-of-mass energy of the scattering satisfies $\sqrt{s}=\sqrt{2 m_e E_{\bar{\nu}_e}} \simeq m_W$, which occurs for neutrino energies of $E_{\nu}\simeq 6.3~\PeV$.
\end{itemize}
A schematic representation of the CC and NC DIS processes is shown in Fig.~\ref{fig:xsec-diagram} (left), while the interaction of a neutrino with the photon field of either a nucleus or nucleon is depicted in Fig.~\ref{fig:xsec-diagram} (right). A technical description of these processes and their implementation is provided below.

\subsubsection{DIS: charged- and neutral-current scattering}
\label{sec:disQCD}

The scattering between high-energy neutrinos and Earth matter is dominated by CC and NC DIS interactions.
In these processes, the incoming neutrino probes the internal structure of bound nucleons through the exchange of a either a \textit{W}-boson or a \textit{Z}-boson respectively, as depicted in Fig.~\ref{fig:xsec-diagram}.
In the context of the neutrino flux attenuation calculation, CC scattering leads to both neutrino absorption and regeneration (the latter through $\tau$-lepton decay), while NC scattering results in a softening of the
neutrino energy distribution.

The modelling of neutrino attenuation due to matter effects requires a differential description of the lepton kinematics following the scattering processes, which is provided by the DIS structure functions $F_i^{\nu N}(x,Q^2)$. 
Following the notation of~\cite{Bertone:2018dse}, we can express the double-differential cross section for the CC process $\nu(k) + N(p) \to \ell(k^{\prime})+X(W)$ as:
\beq \label{eq:CCxsec}
\frac{{\rm d}^2 \sigma^{\rm CC}_{\nu N}}{{\rm d} x\,{\rm d}Q^2} = \frac{G_{F}^2 m_{\PW}^{4}}{4\pi x (Q^2+m_{\PW}^2)^2} 
	\Bigg(Y_+ F_{2,{\rm CC}}^{\nu N}(x,Q^2)   + Y_- x F_{3,{\rm CC}}^{\nu N}(x,Q^2) - y^2 F_{L,{\rm CC}}^{\nu N}(x,Q^2)   \Bigg) \,,
\eeq
where the momentum transfer for this process defined as $Q^2 = -q^2$ with $q=k'-k$; the inelasticity variable is $y = (q\cdot p)/(k\cdot p)$;  $Y_{\pm} = 1\pm(1-y)^2$; and $x = Q^2/(2\,q\cdot p)$.
A similar expression also holds for NC scattering (see Appendix A of Ref.~\cite{Bertone:2018dse}).
The corresponding formula for the case of an incoming anti-neutrino is obtained by reversing the sign of the $xF_3$ term in Eq.~(\ref{eq:CCxsec}).

For a given neutrino energy $E_{\nu}$, the neutrino-nucleon scattering process quantified by Eq.~(\ref{eq:CCxsec}) can occur for a continuum of values of the momentum transfer, $Q \gtrsim 0$, and therefore in principle knowledge of the structure functions across the entire kinematic regime is required.
In the non-perturbative regime, defined by $Q \lesssim 1~\GeV$, the structure functions can be obtained from phenomenological fit to data, or may be constructed from models which have been tuned to data such as in~\cite{Bodek:2004pc}.
However, as highlighted in~\cite{Bertone:2018dse}, the contribution of this low-$Q$ region to the total cross section is negligible for neutrino energies $E_{\nu} \gtrsim 1~\TeV$ and so we do not consider it further in this work.
See also the discussion in Appendix~\ref{app:DIS}.

The relevant kinematical region for the neutrino attenuation calculation is the deep-inelastic regime, defined by $Q \gg 1~\GeV$, where the DIS structure functions $F_{i}^{\nu N}(x,Q^2)$ admit a factorised expression in terms of perturbative coefficient functions and non-perturbative parton distribution functions (PDFs),
%th
\begin{align} \label{eq:SF}
  F^{\nu N}_i (x, Q^2) = \sum_{a = g, q} \int_x^1 \frac{\rd z}{z} C^{\nu}_{i,a} \left( \frac{x}{z}, Q^2 \right) f_a^{(N)}\left(z,Q^2\right) \,,
\end{align}
corresponding to a convolution of the process-independent PDFs ($f_a^{(N)}$) of the (potentially bound) nucleon $N$ with process-dependent coefficient functions ($C^{\nu}_{i,a}$). The latter can be computed in perturbation theory as a power expansion in the strong coupling $\alpha_s$.
The PDFs are determined by non-perturbative QCD dynamics which need to be extracted from experimental data by means of global analyses,\footnote{See~\cite{Lin:2017snn} for recent progress in
  first-principle calculations of PDFs via lattice QCD.} while the evolution of these PDFs is instead perturbative and can be determined from the solutions of the DGLAP evolution equations (which are $x$-space convolutions of splitting functions with PDFs).

Therefore, in order to evaluate the CC and NC DIS cross sections relevant for high-energy neutrino attenuation one needs to compute  the structure functions $F_i(x,Q^2)$ using Eq.~(\ref{eq:SF}).
In this work, these DIS structure functions are computed using the {\tt APFEL} program~\cite{Bertone:2013vaa}.
{\tt APFEL} contains expressions for the massless CC and NC coefficient functions up to $\mathcal{O}(\alpha_s^2)$~\cite{vanNeerven:1991nn,Zijlstra:1991qc,Zijlstra:1992qd} and the massive CC and NC coefficient functions up to $\mathcal{O}(\alpha_s^2)$~\cite{Laenen:1992zk} and $\mathcal{O}(\alpha_s)$~\cite{Gottschalk:1980rv,Gluck:1997sj,Blumlein:2011zu} respectively (higher-order corrections for the CC process have been completed~\cite{Berger:2016inr,Gao:2017kkx}, but are not in a suitable format for our current purposes).
Within {\tt APFEL}, options are also available to combine these computations to construct structure function predictions in particular mass schemes, such as the FONLL general-mass variable flavour number scheme~\cite{Forte:2010ta}.
Note that the possibility to account for the impact of small-$x$ resummation of coefficient functions and PDF evolution~\cite{Bonvini:2016wki} is also possible, which is provided though an interface with the {\tt HELL} program~\cite{Bonvini:2017ogt,Bonvini:2018iwt}. 
Further, the impact of nuclear corrections in the DIS structure functions can be taken into account by using a set of PDFs for a bound nucleon as an input to Eq.~\eqref{eq:SF}. These input PDFs can either be evolved by {\tt APFEL}, or obtained from the pre-tabulated grids with the {\tt LHAPDF6} interpolator~\cite{Buckley:2014ana}.

The DIS predictions which enter the neutrino attenuation calculation therefore depend on the choice of the input PDFs, the scheme choice for including/ignoring heavy quark mass effects, as well as the perturbative order of the computation.
Throughout this paper, we will adopt the BGR18 calculation of DIS structure functions as baseline, and present also the results corresponding to the CMS11 calculation.
We summarise below the main features of these two calculations, while the validation of their implementation within {\tt HEDIS} is given towards the end of this section.

\paragraph{CMS11.} This calculation was presented in~\cite{CooperSarkar:2011pa} (updating earlier work~\cite{CooperSarkar:2007cv}) and has been an important benchmark for the (ultra)-high-energy neutrino-nucleon cross section.
As inputs, this calculation uses the NLO HERA1.5 PDF set~\cite{CooperSarkar:2010wm} which is fitted to the combined HERA data~\cite{Aaron:2009aa}, and uses expressions for the massless coefficient functions at NLO as made available in {\tt DISPred}~\cite{Ferrando:2010dx} and {\tt QCDNUM}~\cite{Botje:2011sn}.

\paragraph{BGR18.}\label{BGR18} This calculation was presented in~\cite{Bertone:2018dse}, and includes predictions at (N)NLO including the impact of small-$x$ resummation, as well as at NLO accuracy. The input PDF sets are obtained from the NNPDF3.1sx global analyses of collider data~\cite{Ball:2017otu}. 
In addition to the data considered in that analysis, the impact of $D$-meson production in $\Pp\Pp$ collisions at 5,7, and 13~\TeV~\cite{Aaij:2013mga,Aaij:2015bpa,Aaij:2016jht} is also accounted for by performing a reweighting of these PDF sets---this data is relevant as it constrains the PDFs at small-$x$ values (beyond the kinematic range of HERA data) which enters the neutrino-nucleon cross section for $E_{\nu} \gsim 1~\EeV$~\cite{Gauld:2016kpd}.
These calculations use the FONLL general-mass variable flavour number scheme~\cite{Forte:2010ta} to account for the impact of heavy quark mass effects on the cross section. %The inclusion of mass corrections is mostly relevant when considering top-quark production in CC scattering, where the massless approximation can lead to a large overestimation of the cross section.
The impact of nuclear corrections in~\cite{Bertone:2018dse} were accounted for by computing a nuclear modification using the EPPS16~\cite{Eskola:2016oht,Dulat:2015mca} nPDFs, and applying this to the free-nucleon predictions as described above.

For the results shown in this work, we always use the BGR18 calculation where all inputs are computed at NLO. This leads to the most consistent prediction as heavy quark mass effects, nuclear corrections, and the description of the LHCb $D$-meson data can all be computed at this order.
The impact of small-$x$ resummation on the total cross-section is less than a few percent provided $E_{\nu} \lesssim 10^{10}~\GeV$~\cite{Bertone:2018dse}. This is a small effect as compared to the uncertainties associated to nuclear corrections at this energy, and so it can be neglected.
The role of nuclear corrections on the attenuation calculation will be discussed in detail in Sect.~\ref{sec:results_nucl}.

\subsubsection{DIS: resonant interactions with the nucleon's photon field}
\label{sec:DISresonant}

In addition to the dominant deep-inelastic scattering processes described above, the incoming neutrino may instead interact indirectly with the photon field of the nucleon.
In this channel the incoming neutrino may interact with a lepton which has been generated by the photon field of the nucleon.
This CC scattering process is negligibly small under most circumstances, the exception being when the neutrino has enough energy ($\sqrt{2 m_{\rm N} E_{\nu}} \gtrsim m_{\PW}$) to produce an on-shell \textit{W}-boson which leads to a resonant enhancement of the cross section. For this reason, we refer to it as `DIS resonant'. See Fig.~\ref{fig:xsec-diagram} (right) for a schematic depiction.
This channel can amount up to a 3\% correction of the total DIS cross section~\cite{Seckel:1997kk,Alikhanov:2015kla,Gauld:2019pgt,Zhou:2019vxt,Beacom:2019pzs}, contributes to both neutrino absorption and regeneration effects (through leptonic \textit{W}-boson decays), and is therefore relevant for the description of the neutrino attenuation process due to Earth matter.

The contribution from this channel can be taken into account by convoluting a partonic cross section ($\widehat{\sigma}$) for the process $\nu\gamma \to \ell\PW$ with the inelastic photon PDF of the nucleon according to
\beq \label{eq:DISgamma}
\sigma_{\nu {\rm N}}(E_{\nu}) = \int {\rm d}x \, \gamma^{\rm N}_{\rm inel}\left(x,\mu_F^2\right)\widehat{\sigma}_{\nu \gamma}(x,\mu_F^2,E_{\nu})\,,
\eeq
where $\gamma^{\rm N}_{\rm inel}(x,\mu_F^2)$ is the inelastic photon PDF of nucleon ${\rm N}$.
The partonic cross section for the neutrino-photon scattering process is given by
\beq \label{eq:DISgammaSig}
\widehat{\sigma}_{\nu \gamma}(x,\mu_F^2,E_{\nu}) = \frac{1}{2 s_{\nu\gamma}} \int \overline{\sum} |\mathcal{M}_{\nu \gamma\to f}|^2\,{\rm d} \Phi_f\,,
\eeq
where $\Phi_f$ is the phase space for the the final state $f$ ($f = \PW\ell$ in this case), $\mathcal{M}$ is the matrix element for the partonic process, the squared partonic centre-of-mass energy is $s_{\nu\gamma} = 2xm_NE_{\nu}$, and a sum (average) of final (initial) state spins/polarisations is assumed.
Practically, we have implemented analytic expressions for the spin-averaged squared matrix-element and performed the phase space integration numerically, which makes it straightforward to compute differential observables of interest (by placing a constraint on the phase space integration).

This computation can also be performed in the zero-mass limit for the leptons by performing mass factorisation as noted in~\cite{Gauld:2019pgt}.
In this approach, the leptons are included as parton distribution functions of the nucleon and the convolution in Eq.~\eqref{eq:DISgamma} is between the lepton PDFs and a partonic cross section for the subprocess $\nu + \bar{\ell} \to \PW$.
The benefits of this approach are that the large collinear logarithms which appear at $\mathcal{O}(\alpha)$ for the massive computation are resummed to all orders by the joint QCD$\otimes$QED evolution of the PDFs~\cite{Bertone:2015lqa}, and that the calculation is easier to extend to higher-orders or to include off-shell effects for the \textit{W}-boson decays.
Within this approach, all relevant $\mathcal{O}(\alpha)$ corrections for the off-shell computation were actually already computed in~\cite{Gauld:2019pgt}.

In either case, the inelastic photon PDF of bound neutrons and protons is required. For the proton, we use the NNPDF3.1 PDF fit with a photon PDF~\cite{Bertone:2017bme} obtained using the luxQED formalism of Ref.~\cite{Manohar:2016nzj,Manohar:2017eqh}. 
For the neutron, we instead use the boundary condition of the QCD-only NNPDF3.1 PDF fit of the proton~\cite{Ball:2017nwa}, and use isospin symmetry to extract the QCD PDFs of the neutron at $Q_0 = 1.65~\GeV$. 
These PDF sets are then evolved according to the joint QCD$\otimes$QED solution to the DGLAP equation implemented within {\tt APFEL}~\cite{Bertone:2013vaa}. 
When the massless calculation is performed, this evolution includes the lepton PDFs. In this case, we also use a fixed-order ansatz to approximate the $x$-dependence of the lepton PDFs of the proton at the initial evolution scale $Q_0 = 1.65~\GeV$.
We note that there has recently been a direct extraction of lepton PDFs~\cite{Buonocore:2020nai} which may be used as an alternative to the procedure discussed above.
It should also be noted that here the impact of nuclear effects on the photon PDF are not accounted for. This is because, currently, available analyses of nPDFs are limited to only QCD partons.

A detailed study of both massive and massless approaches, as well as the impact of off-shell effects, to describing this process has been performed. In general it was found that these approaches lead to similar results. The massless LO calculation approximates well the full NLO calculation~\cite{Gauld:2019pgt} for $E_{\nu} \lesssim 1~\PeV$, whereas at higher energies it was important to include the $\nu\gamma$-induced channel (part of the NLO calculation). In addition, it was found that the massive computation overestimates (as compared to the NLO accurate massless computation with off-shell effects) the cross-section in the electron channel. At the level of the total cross-section, these differences amount to corrections of $\approx1\%$.
In this work we have chosen to use the LO massless computation our baseline results.

\subsubsection{Coherent neutrino scattering}

At small values of the momentum transfer, $Q\lsim1~\GeV$, the neutrino may interact
coherently with the photon field of the entire target nucleus.
For sufficiently high-energy neutrinos ($E_{\nu}\gtrsim 3\times10^{4}~\GeV$),
the production of a \textit{W}-boson via coherent scattering becomes kinematically possible.
As noted before, the cross section for this process is proportional to the atomic number of nucleus squared ($Z^2$) which makes it particularly relevant for the attenuation process in the Earth where $\langle Z \rangle \approx16$.
For such a nucleus, coherent neutrino scattering can impact the total cross section by up to 10\%, which in turn implies that it also affects the total rate of absorption and regeneration of neutrinos which traverse the Earth.
 
For the description of this process, we use the formalism presented for neutrino trident production in~\cite{Ballett:2018uuc} (based on earlier work~\cite{Czyz:1964zz,Lovseth:1971vv}), as well as for the case of \textit{W}-boson production in~\cite{Beacom:2019pzs,Zhou:2019vxt}. Following Ref.~\cite{Ballett:2018uuc}, the differential cross section for the coherent scattering process ($C$) may be written as 
\beq \label{eq:coh_had}
\frac{{\rm d}^2\sigma_{\nu C}}{{\rm d}Q^2\,{\rm d}\hat{s}} = \frac{1}{32\pi^2}\frac{1}{\hat{s}Q^2} 
	 \left[h_C^T(Q^2,\hat{s}) \widehat{\sigma}^T_{\nu\gamma}(Q^2,\hat{s})
	      +h_C^L(Q^2,\hat{s}) \widehat{\sigma}^L_{\nu\gamma}(Q^2,\hat{s})
	 \right]\,,
\eeq
where $h_C^{T/L}(Q^2,\hat{s})$ are transverse/longitudinal hadronic flux functions, $\widehat{\sigma}^{T/L}_{\nu\gamma}(Q^2,\hat{s})$ are the corresponding partonic cross sections for the leptonic subprocess, $Q^2$ is the negative virtuality of the off-shell photon momentum (with four momentum $q$), and $\hat{s} = 2\,k \cdot q$ (two times the dot product of the incoming neutrino and photon four-momenta). Notice that this process is governed by different kinematics as compared to the DIS process. %, and the definition of the momentum exchange is different.
The exact form of the functions $h_C^{T/L}(Q^2,\hat{s})$ can be found in Eq.~(2.15a,\,2.15b) of Ref.~\cite{Ballett:2018uuc}. These flux functions are proportional to $Z^2$ and the squared modulus of the electromagnetic nuclear form factor, and for the latter we also use an analytic expression for the symmetrised Fermi function~\cite{Sprung_1997,Acciarri:2016sli}.
The computation of the leptonic partonic cross section in terms of transverse and longitudinal components is performed according to
\begin{align} \label{eq:coh_lep}\nonumber
\widehat{\sigma}^{T}_{\nu\gamma}(Q^2,\hat{s}) &= \frac{1}{2\hat{s}} \int \frac{1}{2} \sum \left( -g^{\mu\nu} +\frac{4Q^2}{\hat{s}^2} k^{\nu}k^{\nu}\right)   |\mathcal{M}|^2_{\mu\nu}\, {\rm d}\Phi_f \,,\\
\widehat{\sigma}^{L}_{\nu\gamma}(Q^2,\hat{s}) &= \frac{1}{\hat{s}} \int \sum \frac{4Q^2}{\hat{s}^2} k^{\nu}k^{\nu} |\mathcal{M}|^2_{\mu\nu}\, {\rm d}\Phi_f \,,
\end{align}
where $|\mathcal{M}|^2_{\mu\nu}$ is the squared matrix-element for the leptonic process $\nu(k) + \gamma(q) \to f$ with the spin-index of the off-shell photon left open.
In this work we are interested in the contribution to \textit{W}-boson production, where $f = \PW \ell$. The relevant amplitude is given in the Appendix of Ref.~\cite{Zhou:2019vxt}, and we have implemented analytic expressions for the contracted squared matrix-elements appearing in Eq.~\eqref{eq:coh_lep}.
As before, the phase space integration is performed numerically to allow for maximum flexibility.
In practice, the phase space is parameterised in the centre-of-mass frame of the incoming neutrino and off-shell photon, and a transformation to the lab frame is performed with a series of boosts and rotations (see Appendix~A of Ref.~\cite{Beacom:2019pzs}). In this way, lab-frame observables are easily accessible without the need to rely
on any approximation.

As a cross-check, we have also implemented an alternative computation of this process based on the Equivalent Photon Approximation (EPA)~\cite{Fermi:1924tc,vonWeizsacker:1934nji,Williams:1934ad} for \textit{W}-boson production as in~\cite{Alikhanov:2015kla}.
We find (in agreement with~\cite{Zhou:2019vxt}) that the EPA leads to an over-estimation of the inclusive cross section in the electron channel (breakdown of the EPA), while for muon and tau channels we find agreement between the two methods. This behaviour is expected as the impact of the off-shell effects is most relevant when $Q\gtrsim m_{\ell}$, which is a kinematic region frequently encountered for the electron channel, but not for the muon and tau channels as the electro-magnetic nuclear form factor falls off steeply for $Q\gtrsim0.1~\GeV$.

\subsubsection{Elastic and diffractive scattering}

At momentum transfer values of  $Q \sim \GeV$, the neutrino may instead resolve the photon field of individual nucleons. In this case, similarly as in coherent neutrino scattering, it is possible to produce an on-shell \textit{W}-boson, which can occur for $E_{\nu}\gtrsim 3\times10^{3}~\GeV$.
As previously discussed, when describing the DIS resonant process, we use a description of the photon PDF of the proton which has been obtained using the luxQED formalism. This PDF includes an elastic component which is extracted from knowledge of the electric and magnetic Sachs form factors of the proton, which have in turn been fitted from low momentum-exchange $e^-\Pp$ scattering data~\cite{Bernauer:2013tpr}\footnote{Note that these fits of the Sachs form factors deviate from those obtained from the simple dipole form.}. This elastic component therefore contributes to the calculation of the DIS resonant process on proton targets as discussed above. This component is therefore included in the contributions we label as `DIS resonant'.

Alternatively, this contribution can be directly computed using the same formalism as presented for coherent scattering discussed above. The differential cross section as presented in Eq.~\eqref{eq:coh_had} is also applied in this case.
The hadronic flux-functions must then be replaced with those for individual nucleons (neutrons and protons). The relevant functions are provided in Eq.~(2.19a,\,2.19b) of Ref.~\cite{Ballett:2018uuc}, which can be written in terms of the electric and magnetic Sachs form factors (see Appendix A of the same work).
An implementation of this process has also been made available in {\tt HEDIS}, where an analytic dipole form of the Sachs form factors has been used, with the option to also include Pauli blocking effects~\cite{Brown:1971qr}.

\subsubsection{Scattering upon atomic electrons}
\label{sec:glashow}

In addition to the scattering processes off nucleons and nuclei from matter targets discussed above, it is also possible that the incoming neutrinos interact with atomic electrons.
This process is negligible under most circumstances, with the exception of high-energy electron anti-neutrinos.
For that case, neutrino scattering upon atomic electrons receives a large resonant enhancement when the centre-of-mass energy is $\sqrt{s} = \sqrt{2 m_e E_{\bar{\nu}}} \approx m_{\PW}$, corresponding to the production of an on-shell \textit{W}-boson and known as the Glashow resonance.
Close to this resonance region, corresponding to $E_{\bar{\nu}} \in [3,10]~\PeV$, neutrino
scattering upon atomic electrons dominates over all other scattering processes.
It is therefore important to provide a detailed description of this channel in calculations of neutrino attenuation due to matter effects.
A precision computation for this process was presented in~\cite{Gauld:2019pgt}, where all NLO corrections, both in QCD and in electroweak theory, were evaluated.
This calculation also accounted for the impact of higher-order initial-state-radiation (ISR) effects.
In addition to these perturbative corrections, this process may also be subject to atomic broadening effects which have been discussed in~\cite{Loewy:2014zva}.

In this work, instead of adopting the complete calculation of Glashow resonant scattering for the attenuation process, we have implemented an approximation using a structure function based approach to account for the impact of the emission of initial-state photons.
In this approach, the cross section takes the form
\begin{align}
\sigma(E_{\bar{\nu}}) = \int {\rm d}x \, \Gamma_{ee}(x,\muf^2) \, \widehat{\sigma}_{e\bar{\nu}_e}(x,\muf^2,E_{\bar{\nu}}) \,,
\end{align}
where the partonic cross section is the same as that in Eq.~\eqref{eq:DISgammaSig}, and we use the exponentiated form of the structure function given by
\begin{align} \label{eq:eSF}
\Gamma_{ee}(x,\muf^2) = \frac{\rm exp\left( -\beta_l \gamma_E + \frac{3}{4}\beta_l\right)}{\Gamma\left(1+\beta_l\right)} \beta_l(1-x)^{\beta_l-1} - \frac{\beta_l}{2} (1+x) \,, \quad \beta_l = \frac{\alpha}{\pi} \left( \ln\left[\frac{\muf^2}{m_e^2}\right]-1 \right)\,,
\end{align}
with $\gamma_E$ being the Euler-Mascheroni constant, and where the function $\Gamma$ appearing in the denominator of Eq.~\eqref{eq:eSF} (which has no subscripts) is the usual gamma function. This approach accounts for the leading logarithmic correction generated by photon emission from the atomic electron, with an all-order treatment of the soft contribution~\cite{Yennie:1961ad,Gribov:1972ri,Kuraev:1985hb,Nicrosini:1986sm,Berends:1987ab,Blumlein:2019pqb,Bertone:2019hks}.
The approximation of Eq.~\eqref{eq:eSF} is found to well approximate the \NLO{\rm+LL} result of~\cite{Gauld:2019pgt} (computed in the $G_{F}$-scheme) within the vicinity of the resonance.
More sophisticated approaches are possible (see~\cite{Denner:2019vbn} for a summary) and necessary for precision collider experiments, but the approach taken here is sufficient to achieve the percent-level accuracy relevant for the attenuation calculation.
As for the other subprocesses describe above, here we implement an expression for the partonic cross section according to Eq.~\eqref{eq:DISgammaSig} and perform the phase-space integration numerically.

\subsection{Validation and cross-section predictions with {\tt HEDIS}}
\label{sec:hedis}

In the context of calculations of high-energy neutrino propagation through Earth matter, the main advantage of adopting the {\tt HEDIS} module~\cite{Garcia:2019hze} is that it makes possible using state-of-the-art calculations of the neutrino-matter cross sections, in particular for the processes described in the previous section, into a full-fledged neutrino event generator such as {\tt GENIE}.
Originally, the  {\tt HEDIS} program accounted only for the DIS structure function contributions (see Sect.~\ref{sec:disQCD}) to the neutrino-matter interaction
cross sections~\cite{Garcia:2019hze}.
As compared to those first studies, in this work we have extended {\tt HEDIS} with the addition of three sub-dominant contributions to the total interaction cross section: the Glashow resonance process (scattering on atomic electrons), the scattering of neutrinos with the photon-field of nucleons, as well as the coherent scattering process where the neutrino interacts with the photon field of the entire nucleus.

We note that other popular neutrino event generator codes such
as {\tt GENHEN}~\cite{genhen} (based on {\tt LEPTO}~\cite{Ingelman:1996mq}), {\tt ANIS}~\cite{Gazizov:2004va} are restricted to leading-order calculations, use obsolete PDF sets (which
do not account for the latest constraints from HERA and LHC data), and are often restricted to
precomputed look-up cross-section tables.
Relying on obsolete PDF sets is specially problematic if such generators would be used to
model high-energy neutrino interactions, since as mentioned above the DIS neutrino-nucleon cross sections become sensitive to small-$x$ PDFs whose behaviour is very different in modern PDF sets constrained by the latest experimental measurements from HERA and the LHC.

\paragraph{DIS CC and NC scattering.} \label{sec:DISvalidation}
As explained in Sect.~\ref{sec:formalism}, the dominant contribution for high-energy neutrino matter interactions is the DIS neutrino-nucleon scattering. As reported in~\cite{Garcia:2019hze}, both
the BGR18 and CMS11 calculations have been implemented in {\tt HEDIS}.
These cross-section calculations are based on the evaluation of the double-differential DIS neutrino cross section Eq.~(\ref{eq:CCxsec}), using {\tt APFEL}~\cite{Bertone:2013vaa} to compute structure functions and {\tt LHAPDF6}~\cite{Buckley:2014ana} to access the PDFs.

\begin{figure}[tbp]
\centering 
\includegraphics[width=1.\textwidth]{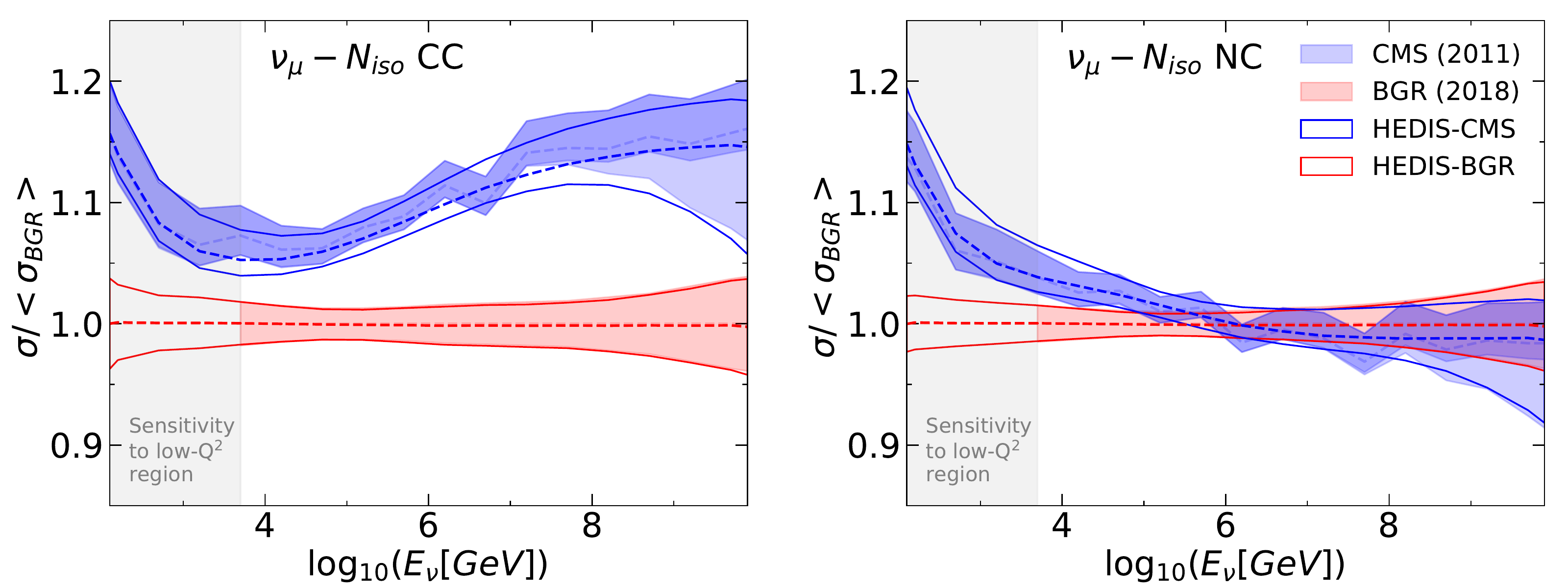}
\caption{The neutrino-nucleon interaction cross section as a function of the neutrino energy $E_{\nu}$ for charged-current (left) and neutral-current (right plot)
scattering.
We compare the results of the original (NLO) BGR18 and CMS11 calculations among them as well as with their corresponding implementation in {\tt HEDIS}. 
Results are shown normalised to the central value of
the BGR18 calculation, and the bands correspond
to the PDF uncertainties in each case.
The relative rise ($\approx10\%$) of the CMS11 calculation in the low-$E_{\nu}$ region is due to the inclusion of low-momentum contributions ($1.0 < Q^2 < 1.64^2~\GeV^2$) which are absent in BGR18 (see Appendix~\ref{app:DIS}).
}
\label{fig:xsec_nucleon}
\end{figure}

In Fig.~\ref{fig:xsec_nucleon} we display the neutrino-nucleon DIS cross section as a function of the neutrino energy $E_{\nu}$ for CC (left) and NC (right) scattering. We compare the results of the original BGR18 and CMS11 calculations with the predictions computed with {\tt HEDIS} using the same theoretical set-up in each case. As noted in Sect.~\ref{BGR18}, in this work we use the BGR18 model with all inputs computed at NLO, and compare to the corresponding results computed in~\cite{Bertone:2018dse}.
Results are shown normalised to the central value of the BGR18 calculation, and the bands correspond to the PDF uncertainties in each case. In the CMS11 case the darker and lighter bands correspond to two different prescriptions of estimating the PDF uncertainties.\footnote{In the darker band, one of the HERAPDF1.5 error eigenvectors has been excluded from the calculation.}
These results are for an isoscalar target without nuclear effects.
One can observe that the {\tt HEDIS} implementation reproduces both the central values and the uncertainties of the original BGR18 and CMS11 calculations.
Notice in this plot that the BGR18 calculation has been extended to $E_{\nu}$ values which are below the region of recommended use (as indicated by the filled band on plots). In this region, the uncertainty of the BGR18 calculation is that of the DIS prediction obtained with the restriction $Q \geq 1.64~\GeV$, and does not include the (non-perturbative) low-momentum exchange contribution, or those due to quasi-elastic scattering or resonant processes. Further comments on the use of the calculation in this region are given in Appendix~\ref{app:DIS}.

This comparison completes the validation of the implementation of these models within {\tt HEDIS}, which are used as a benchmark for the study of the attenuation rate throughout the paper.
A quantitative comparison between the BGR18 and CMS11 models is provided in Appendix~\ref{app:DIS}. In addition to this comparison, this appendix also includes a description of the DIS model which is currently the default within {\tt HEDIS}, and is our recommended model for users.

%%%%%%%%%%%%%%%%%%%%%%%%%%%%%%%%%%%%%%%%%%%%%%%%%%
%%%%%%%%%%%%%%%%%%%%%%%%%%%%%%%%%%%%%%%%%%%%%%%%%%

Finally, we note that the validation of these cross sections is performed under the assumption of massless leptons. Nevertheless, in this version of the software, an option to include the impact of the final-state charged lepton mass when defining the kinematic limits of the scattering process has been included.
This correction is relevant for the inclusive $\nu_\tau$ CC cross section at low neutrino energies, where the this constraint leads to a suppression of the cross section by a factor 0.87 (0.97) at 100 $\GeV$ ($1~\TeV$). This approximation has been shown to capture the main effects of the exact mass calculation~\cite{Jeong:2010nt}.

\begin{figure}[tbp]
\centering 
\includegraphics[width=1.\textwidth]{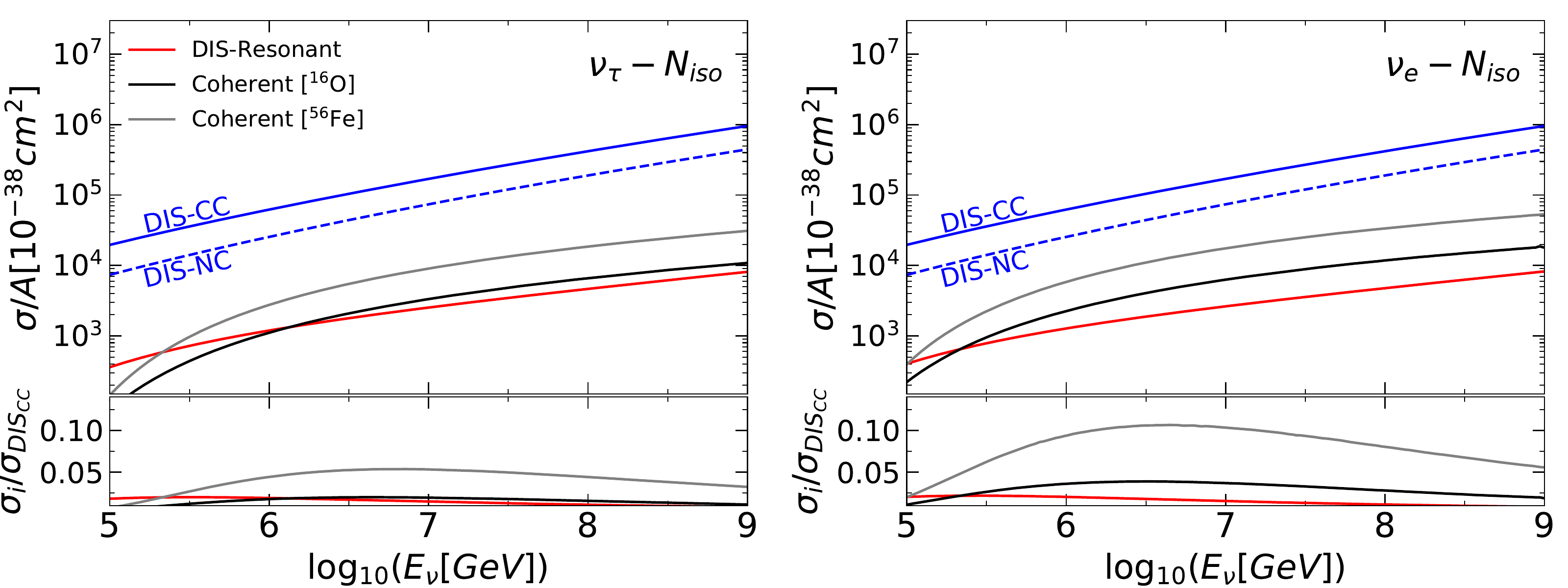}
\caption{Upper panels: the neutrino-matter interaction cross sections per nucleon
as a function of the neutrino energy $E_{\nu}$ for
$\nu_\tau$ (left) and $\nu_e$ (right plot)  computed with \texttt{HEDIS}.
The results for the DIS CC and NC channels were obtained using the BGR18 model,
used also for Fig.~\ref{fig:xsec_nucleon}.
The coherent channels are shown for two different nuclei (O and Fe)
in order to illustrate the $Z^2/A$ scaling effect.
For the other channels, an isoscalar target is assumed.
Bottoms panels: the ratio of the different channels to the  
DIS charged-current cross sections.
}
\label{fig:xsec_subleading}
\end{figure}

\paragraph{Coherent and resonant scattering.}
In addition to DIS neutrino-nucleon scattering, the updated version of {\tt HEDIS} presented
here includes also the contributions from the sub-dominant processes described in
Sects.~\ref{sec:DISresonant}-\ref{sec:glashow}.
%
%As explained there, differential cross sections are used to evaluate the interaction probabilities
%associated to each of those contributions.
%
Fig.~\ref{fig:xsec_subleading} displays the neutrino-matter cross sections as a function of the neutrino energy $E_{\nu}$ for $\nu_\tau$ (left) and $\nu_e$ (right) scattering computed using {\tt HEDIS}.
One can observe that the contribution from the sub-dominant processes can be up to 10$\%$ of the DIS CC neutrino-nucleon cross section for some energies, mainly due to the enhancement of the coherent contribution as a result of the $Z^2/A$ scaling effect.

%%%%%%%%%%%%%%%%%%%%%%%%%%%%%%%%%%%%%%%%%%%%%%%%%%
%%%%%%%%%%%%%%%%%%%%%%%%%%%%%%%%%%%%%%%%%%%%%%%%%%

\begin{figure}[tbp]
\centering 
\includegraphics[width=.6\textwidth]{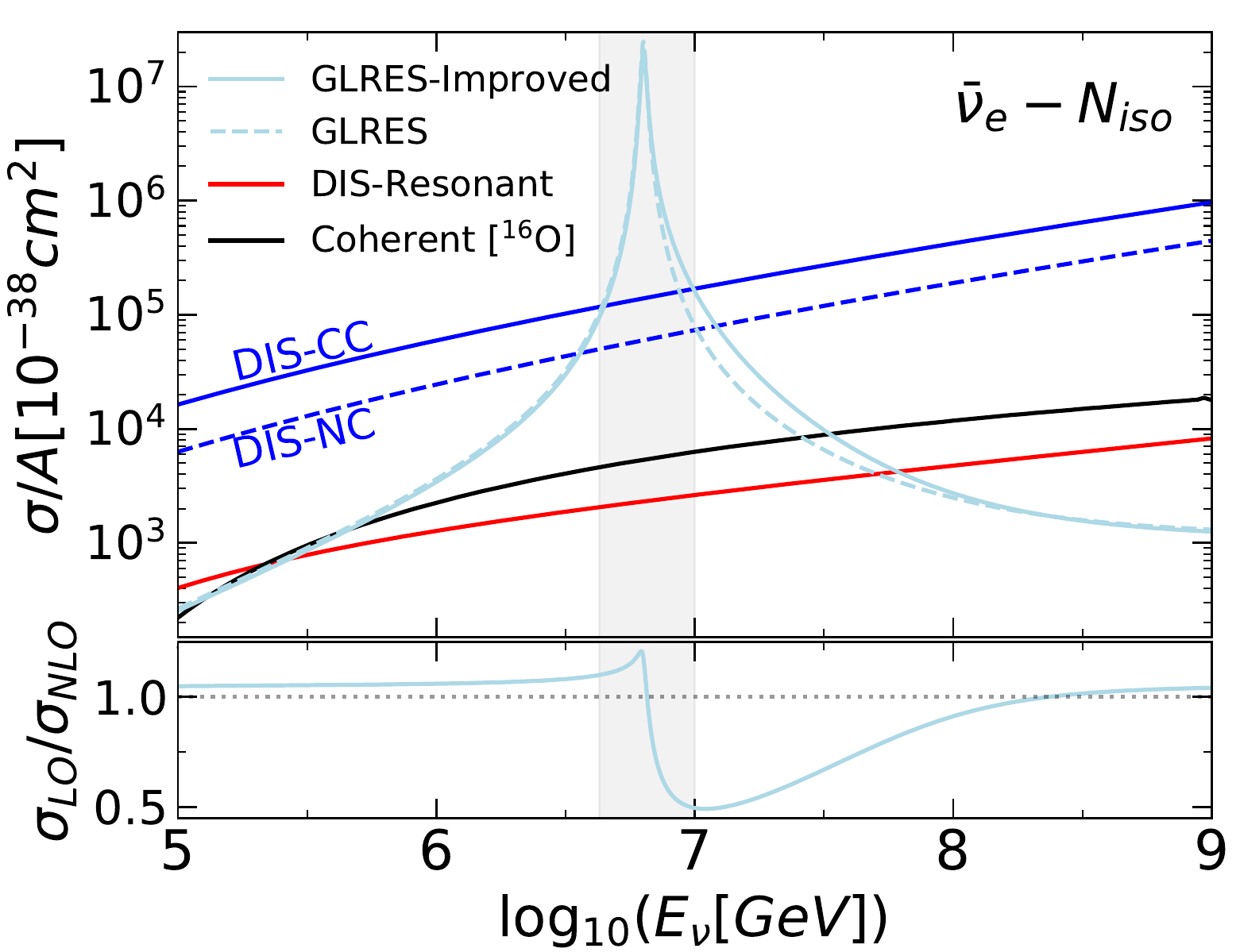}
\caption{Upper panel: same as Fig.~\ref{fig:xsec_subleading} now
  for electron anti-neutrino scattering.
  In this case the Glashow resonance channel is also shown for the cases in which it is computed at the Born level
(“GLRES”) and when including higher-order electroweak corrections (“GLRES-Improved”).
  The shaded area indicates the neutrino energy region were
  the Glashow resonance dominates over the DIS cross section.
  Bottom panel: the ratio between the Glashow resonance cross sections with and without
 higher-order radiative corrections.
}
\label{fig:xsec_glashow}
\end{figure}

\paragraph{Glashow resonance.}
Finally, the scattering off atomic electrons described in Sect.~\ref{sec:glashow} is now also included in the {\tt HEDIS} module.
Fig.~\ref{fig:xsec_glashow} displays the computed neutrino-matter cross sections for electron 
anti-neutrinos, including the Glashow resonance, which dominates at neutrino energies around $E_{\nu}=6.3~\PeV$, as is shown in Fig.~\ref{fig:xsec_glashow}.
%
%As explained above, one notices from  that resonant scattering upon atomic electrons drastically enhances the interaction probability, becoming the dominant contribution in the energy range.
%
In {\tt HEDIS} this important process is computed at LO accuracy including the impact of leading higher-order corrections. The bottom panel of Fig.~\ref{fig:xsec_glashow} shows the ratio between the Glashow resonance cross sections with and without higher-order radiative corrections, which represent up to a 50$\%$ effect.

%%%%%%%%%%%%%%%%%%%%%%%%%%%%%%%%%%%%%%%%%%%%%%%%%%
%%%%%%%%%%%%%%%%%%%%%%%%%%%%%%%%%%%%%%%%%%%%%%%%%%
\paragraph{Simulation of final-state particles.}
Although the final-state particles play a secondary role to study in the neutrino propagation through the Earth, it is worth mentioning how {\tt HEDIS} simulates them.
For DIS neutrino-nucleon interactions, the kinematics of the outgoing lepton are derived from the differential cross section and the hadronisation is performed using {\tt PYTHIA6} as described in~\cite{Garcia:2019hze}. Similarly, for the scattering upon atomic electrons (and the diffractive scattering) the differential information allow us to extract all the kinematics of the final state particles (except when the \textit{W}-boson decays into hadrons, in which {\tt PYTHIA6} is used).
Finally, in the coherent-scattering channel the \textit{W}-boson is generated on-shell according to the differential cross section and decayed afterwards using {\tt PYTHIA6}.
%%%%%%%%%%%%%%%%%%%%%%%%%%%%%%%%%%%%%%%%%%%%%%%%%%%%%%%%%5

\section{Simulation framework: \texttt{NuPropEarth}}
\label{sec:simulation}

In this work the propagation of neutrinos through Earth matter is simulated by means of the newly developed \texttt{NuPropEarth} simulation framework.
\texttt{NuPropEarth} has the structure of a general-purpose Monte Carlo event generator, and therefore allows following the path and interactions of individual neutrinos as they travel through Earth on an event-by-event basis.
In this section, we present the main details of the \texttt{NuPropEarth} simulation framework, before moving in the next section to a discussion of the results for the neutrino attenuation obtained with it.

\begin{figure}[t]
\centering 
\includegraphics[width=1.0\textwidth]{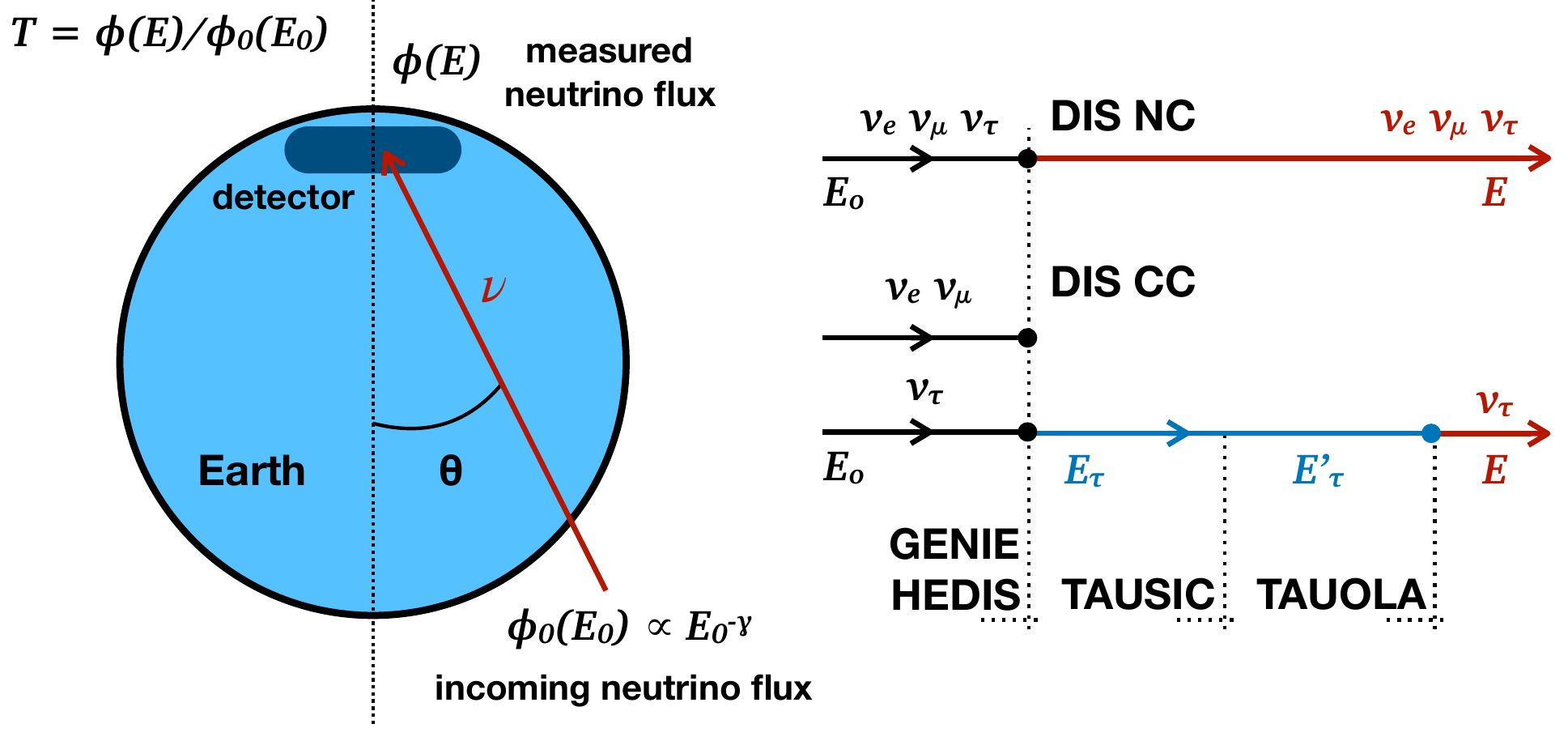}
\caption{Left plot: schematic representation of the path through the Earth
followed by an incoming high-energy neutrino before reaching
the detector.
The \texttt{NuPropEarth} framework evaluates the transmission coefficient $T(E_\nu)=\phi(E_\nu)/\phi_0(E_\nu)$ as a function of neutrino energy $E_\nu$, the nadir angle $\theta$, the spectral index $\gamma$ of the incoming neutrino flux $\phi_0\propto E^{-\gamma}$, the Earth model, and the neutrino-matter interaction cross sections described in Sect.~\ref{sec:uheneuttheory}.
Right plot: diagram of the simulation chain used 
in \texttt{NuPropEarth} indicating which 
interactions take place at each stage
and the associated tools. To simplify the diagram, only the leading contribution, neutrino-nucleon DIS, is included but the sub-leading processes are also propagated.
The incoming and outgoing neutrinos are marked in 
black and red, respectively.
}
\label{fig:scheme}
\end{figure}

Let us start by providing an overview of the calculations that are aiming to simulate.
The left plot of Fig.~\ref{fig:scheme} displays a schematic representation of the path through the Earth followed by an incoming high-energy neutrino before reaching the detector.
The incoming flux scales with the neutrino energy $E_\nu$ as $\phi_0(E_\nu)\propto E_\nu^{-\gamma}$, where $\gamma$ is a parameter known as the spectral index and whose value depends on the astrophysical process that generate these high-energy neutrinos.
Different assumptions in the data analysis can lead to different values for $\gamma$.
An analysis of the 4-year IceCube high-energy starting events determined that $\gamma = 2.83 \pm 0.50$~\cite{Vincent:2016nut}, which is compatible with updated work given in~\cite{Schneider:2019ayi}.

%%%%%%%%%%%%%%%%%%%%%%%%%%%%%%%%%%%%%%%%%%%%%%%%%
%%%%%%%%%%%%%%%%%%%%%%%%%%%%%%%%%%%%%%%%%%%%%%%%%%

The main goal of the \texttt{NuPropEarth} framework then is to compute the value of the transmission $T(E_\nu)$ and attenuation ${\rm Att}(E_\nu)$ coefficients, defined as ratios between the incoming neutrino flux $\phi_0(E_\nu)$ and the flux arriving at the detector volume $\phi(E_\nu)$,
\be
\label{eq:transmissioncoeff}
T(E_\nu)=\frac{\phi(E_\nu)}{\phi_0(E_\nu)} \, , \qquad {\rm Att}(E_\nu) = 1-T(E_\nu) = \frac{\phi_0(E_\nu)-\phi(E_\nu)}{\phi_0(E_\nu)} \, .
\ee
The values of $T(E_\nu)$ will depend in general on the nadir angle $\theta$, the spectral index $\gamma$, the Earth model, the neutrino-matter interaction cross sections described in Sect.~\ref{sec:uheneuttheory}, as well as on other input parameters required for the calculation.
Note that the dependence on the nadir angle $\theta$ measures how much matter will such neutrinos encounter.
For $\cos\theta\simeq 0$ (Earth-skimming neutrinos) very little matter is encountered and thus the resulting attenuation is small ($T\to 1$), while for $\cos\theta\simeq 1$ the largest possible amount of Earth matter is traversed, resulting in a larger attenuation ($T\ll 1$).
Therefore, for a given neutrino energy $E_\nu$ one expects neutrinos with larger $\cos\theta$ to be attenuated more significantly.

The incoming neutrino flux as a function of $E_\nu$ and $\theta$ is the starting point of the \texttt{NuPropEarth} simulation.
These incoming neutrinos are injected in a medium that models the Earth matter and then propagated until they reach the location of the detector.
The output of \texttt{NuPropEarth} is the number and kinematics (energy, direction) of the neutrinos that reach the detector, that can be subsequently used to {\it e.g.} model the detector response and evaluate the expected event rates.
Here we assume that the neutrino flux is isotropic, namely that the number of incoming neutrinos does not depend on the value of the nadir angle $\theta$.
With this framework, we can thus evaluate the transmission and attenuation coefficients for the different possible trajectories that the incident neutrinos will follow on their way to the detector, and study its dependence on the various inputs of the calculation, such as the modelling of the neutrino-matter interaction cross sections.

The main features of the \texttt{NuPropEarth} simulation chain are summarized in the right diagram of Fig.~\ref{fig:scheme}.
There we indicate which interactions take place at each stage of the chain, and the associated tools used in their modelling.
The incoming and outgoing neutrinos are marked in black and red, respectively.
This scheme illustrates the main types of physical processes whereby the energy of the incoming neutrinos can be modified for the dominant contribution, neutrino-nucleon DIS.
First of all, neutrinos of the three flavours can lose energy via neutral current interactions.
Second, electron and muon neutrinos can interact via charged current scattering and disappear from the simulation chain, since they transform into the corresponding charged leptons.
These first two processes are simulated with {\tt GENIE} supplemented by the {\tt HEDIS} module, as reviewed in Sect.~\ref{sec:hedis}. 
Finally, tau neutrinos can also interact via CC scattering, but in this case the produced tau lepton will decay within the Earth volume resulting in another tau neutrino, with essentially the same propagation direction but smaller energy as compared to the original one (the $\nu_\tau$ regeneration process).
The energy losses due to electromagnetic interactions of tau leptons are modeled with {\tt TAUSIC}~\cite{ANTONIOLI1997357} while its decays are computed with {\tt TAUOLA}~\cite{DAVIDSON2012821}, as discussed in more detail below.

One of the central ingredients in the calculation of high-energy neutrino attenuation consists on the modelling of the Earth matter.
The Preliminary Reference Earth Model (PREM)~\cite{DZIEWONSKI1981297} parametrises the density profile of the Earth which is assumed to be spherical.
The atmosphere, which has a negligible contribution for the present study, is excluded.
According to this model, the average atomic mass number $A$ of the Earth's matter turns out to be around $A\simeq 31$.
In general, the elements that compose the Earth are not isoscalar ($A \ne 2Z$) and thus one has to account
for the effects of non-isoscalarity, which are especially relevant for neutrino energies below $1~\TeV$.
We account for non-isoscalar effects by using the target number density in each of the Earth's layer and then correcting for excess of neutrons or protons for each of the corresponding nuclei.

The two quantities that determine the probability of a neutrino interaction are the path length (the more matter traversed, the more likely the interaction) and the total cross section.
Within {\tt NuPropEarth}, both the dominant CC and NC neutrino DIS interactions as well as the subdominant channels described in Sect.~\ref{sec:formalism} are simulated by means of the {\tt HEDIS} module of {\tt GENIE} as a function of the neutrino energy $E_\nu$ and incident angle $\theta$.
More precisely, we evaluate for each neutrino the survival probability for a given path length and energy, and then when this path length is comparable to the distance to the detector, we simulate the corresponding  neutrino interaction, including the kinematics of the final state particles.

In the case of a NC interaction, the incoming neutrino will experience a energy degradation but it will keep propagating in essentially the same direction.
The energy and scattered angle of the outgoing neutrino following a NC scattering are computed using {\tt HEDIS} as described in Sect.~\ref{sec:uheneuttheory}.
At high energies, the neutrino scattering angle is small, so that when neutrinos reach the detector they still point directly to their original source.

A different approach must be adopted in the case of CC interactions.
For interactions involving either a $\nu_\mu$ or a $\nu_e$ the simulation chain is stopped because the outgoing charged lepton is a long-lived particle and therefore does not lead to any other neutrino.
Instead, for $\nu_\tau$ scattering the outgoing $\tau$ lepton will decay  within the Earth model creating a secondary $\nu_\tau$.
This effect, the so-called tau neutrino regeneration, is included accounting for the energy loss experienced by the $\tau$ lepton (via ionisation, bremsstrahlung, pair production and photo-nuclear interactions) before it decays.
These energy losses are computed using the \texttt{TAUSIC} program, and start being relevant at tau lepton energies above $10^7~\GeV$.
In particular, at high energies one can consider the $\tau$ produced in the DIS as fully polarised~\cite{KUZMIN}.
This is taken into account to extract the kinematics of the associated decayed products by means of the \texttt{TAUOLA} software.
In addition to tau neutrinos, we also account for the secondary $\nu_\mu$ and $\nu_e$ neutrinos produced from the $\tau$ decay and that contribute to the corresponding fluxes.
Note that for some values of the spectral index $\gamma$, the tau regeneration process might lead to an enhancement (rather than a suppression) of the neutrino flux, that is, $T(E)\ge 1$, explained by the fact the high-energy $\tau$ neutrinos are effectively converted into lower energy neutrinos of the three species.

We note that the same considerations apply both for the dominant DIS component of the total interaction cross section as well to the subleading contributions listed in Sect.~\ref{sec:formalism}.
Hence, whenever a neutrino or tau-lepton is produced as a final state particle, it is then propagated further using the same simulation approach in the previous paragraphs.
In this respect, given the structure of the {\tt NuPropEarth} simulation framework, all interaction processes can be treated in exactly the same way, and their output is always the flavour, energy, and direction of the produced neutrino or charged lepton which are then used for the next step of the simulation chain.

\section{Results}
\label{sec:results_p}

In this section we present the main results of this work, namely the calculation of the transmission and attenuation coefficients defined Eq.~(\ref{eq:transmissioncoeff}) for the high-energy neutrino flux by means of the Monte Carlo simulation chain provided by the {\tt NuPropEarth} framework.
Unless otherwise stated, here we will assume that the energy scaling of the incoming neutrino flux is $\phi_0(E_0) \propto E_0^{-\gamma}$ with an spectral index of $\gamma=2$ and a cut-off at $E_0=10^{10}~\GeV$.
Then in Appendix~\ref{app:spectral} we will present results for other values of the flux spectral index $\gamma$.

First of all, we will present our baseline results for the transmission coefficient $T(E_\nu)$, based on the BGR18 setup of the DIS neutrino-nucleon cross section calculation, and with the subleading interaction channels neglected.
Nuclear corrections are ignored in this section, and they will be quantified separately in Sect.~\ref{sec:results_nucl}.
Next, we assess the impact of different cross-section models and their associated uncertainties, by comparing results based on the BGR18 and CMS11 calculations.
After that, we compare our results with the predictions obtained from other publicly available frameworks to model high-energy neutrino propagation in matter, in particular with the {\tt NuFate}, {\tt NuTauSim} and {\tt TauRunner} codes.
Finally, we quantify how the transmission coefficients vary as a function of $E_\nu$ when the DIS cross sections
are supplemented by the subleading interaction processes described in Sect.~\ref{sec:uheneuttheory}.

\subsection{Baseline results}

To begin, we present the predictions for the transmission coefficient $T(E_\nu)$ obtained with {\tt NuPropEarth} using
the baseline theory settings, namely those from the BGR18 calculation for the neutrino-nucleon DIS cross-section model.
As mentioned above, these results do not include the contribution from the subleading interaction mechanisms or the impact of nuclear corrections which will instead be discussed in in Sect.~\ref{sec:results_subleading} and Sect.~\ref{sec:results_nucl} respectively.

Our calculation of $T(E_\nu)$ is provided with an estimate of the dominant theory uncertainty associated to the cross-section interaction model, namely the proton PDF uncertainties.
This uncertainty propagation is carried out without introducing any approximation, by redoing the full simulation chain for each of the members in the PDF set, and then evaluating the corresponding PDF uncertainty using the prescription associated for each PDF set.
For the case of the BGR18 calculation, based on a Monte Carlo set of NNPDF3.1sx~\cite{Ball:2017otu} reweighted with LHCb $D$-meson data~\cite{Bertone:2018dse}, the PDF uncertainty is defined as the standard deviation of the distribution composed by the $N_{\rm rep}=60$ (unweighted) replicas of this set---see for example~\cite{DelDebbio:2007ee}.
It is worth emphasizing that fully accounting for the PDF uncertainties both in total rates and in differential distributions has not been considered by previous studies and is presented for the first time in this work.

In addition to the predictions based on the full simulation chain summarised in the right panel of Fig.~\ref{fig:scheme}, in this section we will also present results for the scenario in which the simulation of the propagating neutrinos is stopped just after the first neutrino interaction, either NC and CC, takes place.
In other words, in such a case one assumes that the neutrinos are completely absorbed by the Earth matter after their very first interaction.
This scenario is denoted as the ``full absorption case'', and there the neutrino survival probability is exponentially suppressed.
In the full absorption limit the propagation equations reduce to the simple expression
\be
\label{eq:fullabsortion}
\frac{{\rm d}\phi(E_\nu,z)}{{\rm d}z} = -\sigma_{\rm tot}(E_\nu)\phi(E_\nu,z) \, \quad \to \quad
\phi(E_\nu)=\phi_0(E_\nu)\times \exp(-N_{A}\sigma_{\rm tot}(E_\nu)p_{L}) \, ,
\ee
where $z$ is the target column density, which depends on the zenith angle $\theta$,
and is defined as the integral along the propagation direction of the Earth density
profile $\rho_E(r)$,
\be
z(\theta) = N_A \int \rho_E(r(z,\theta)){\rm d}z = N_A p_L \, ,
\ee
with $p_{L}$ being the path length traversed by the neutrino until its first iteration and $\sigma_{\rm tot}(E_\nu)$ the total cross section (NC+CC) at a given neutrino energy $E_\nu$.
The motivation to present results for the attenuation calculation in the full absorption case is that in this way one can disentangle the effects of both NC energy degradation and CC tau regeneration from those of the full-absorption cross section.

\begin{figure}[tbp]
\centering 
\includegraphics[width=.49\textwidth]{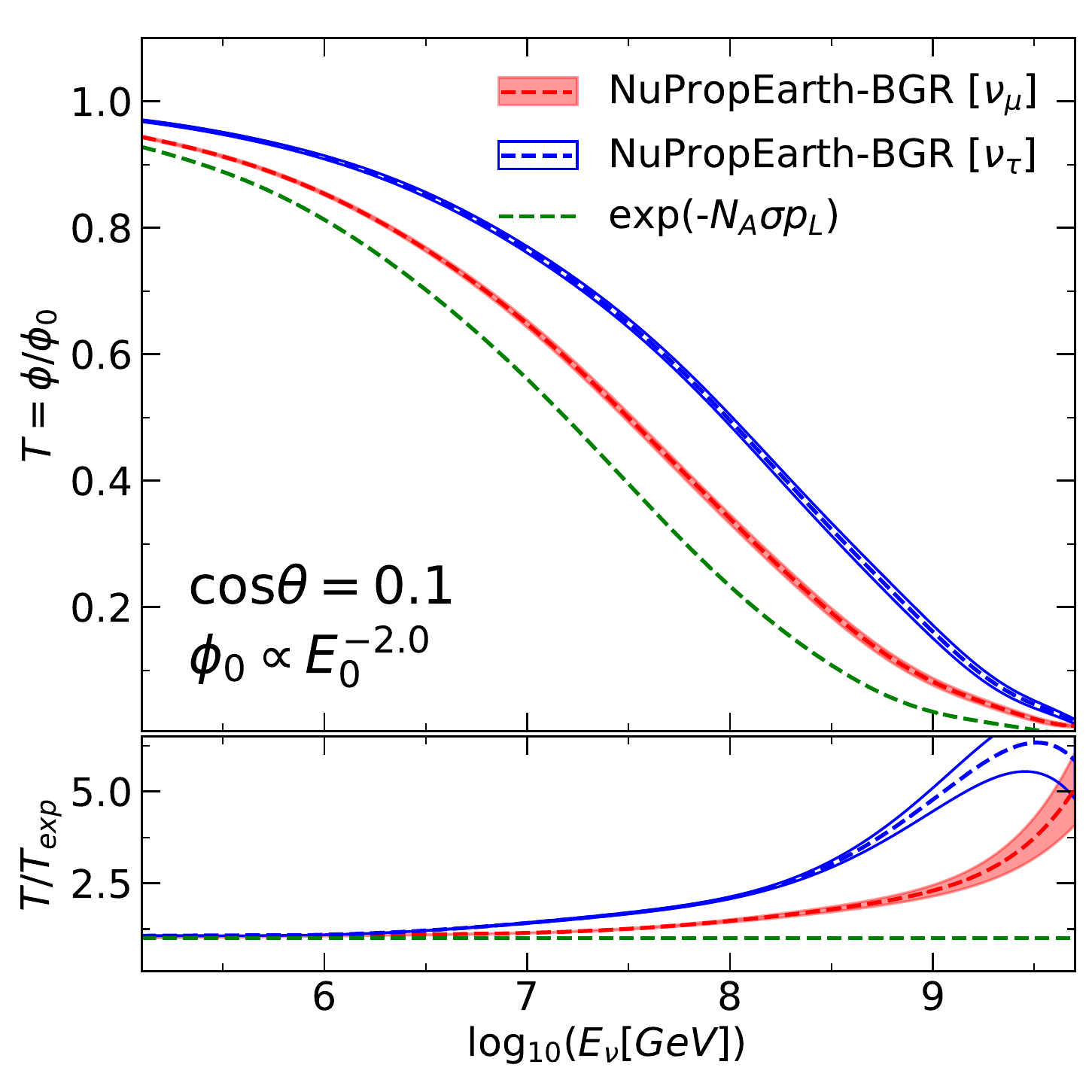}
\includegraphics[width=.49\textwidth]{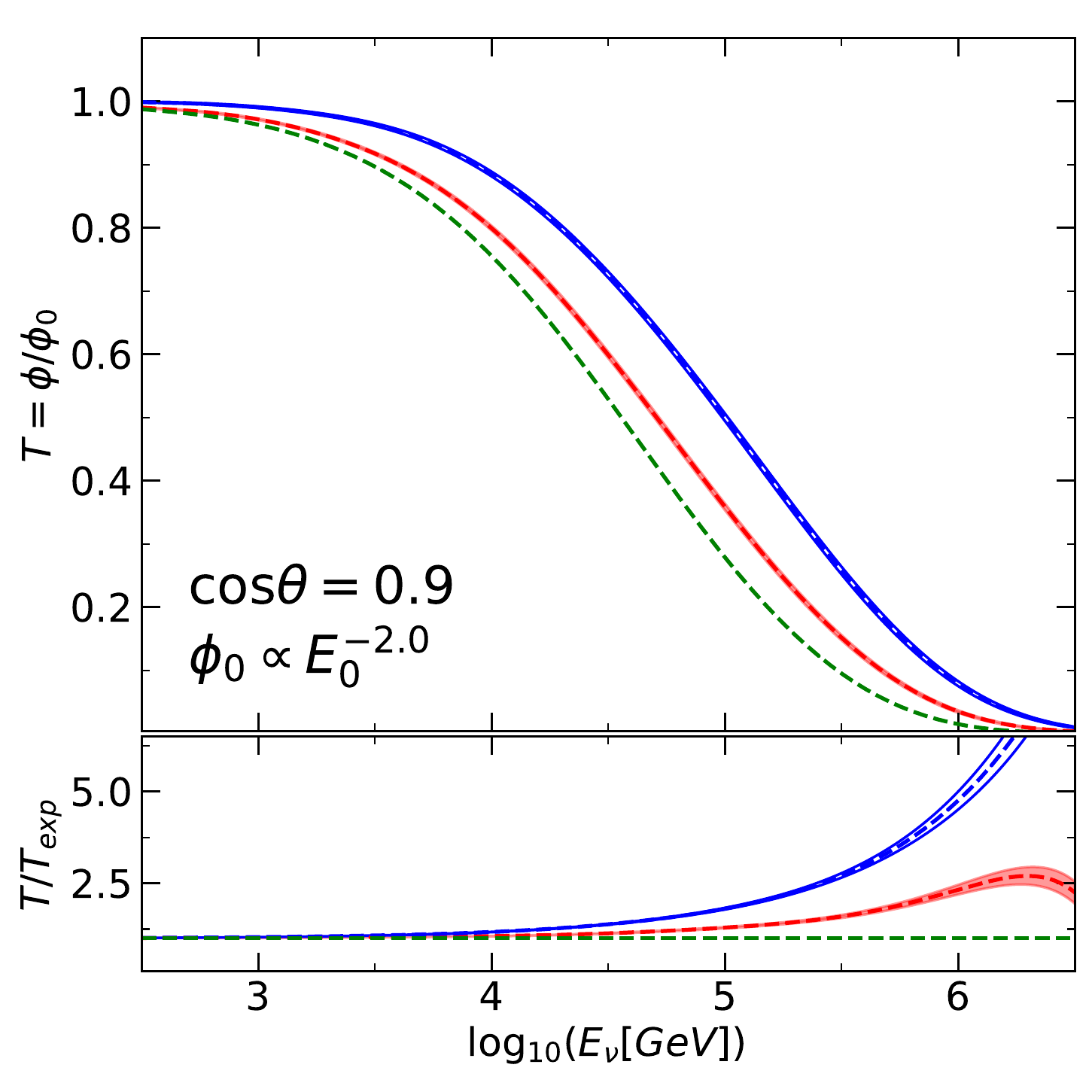}
\caption{Upper panels: the transmission coefficient,  $T=\phi/\phi_0$ as a function of the neutrino energy $E_{\nu}$ for muon $\nu_\mu$ (red filled band) and tau $\nu_\tau$ (blue empty band) neutrinos, as predicted by {\tt NuPropEarth} with the BGR18 cross section setup.
  Results are shown for two different incident angles, $\cos\theta=0.1$ (left) and $\cos\theta=0.9$ (right),
  note that the energy ranges are different in the two cases.
  The width of the $\nu_{\mu}$ and $\nu_{\tau}$ bands is dictated by the PDF uncertainties in the BGR18
  calculation.
  We also show the value of $T$ for the full absorption case
  Eq.~(\ref{eq:fullabsortion})
  for the central BGR18 cross section
  (green lines), which coincides for both $\nu_{\mu}$ and $\nu_{\tau}$.
  Bottom panels: the ratio of the transmission coefficients to the
  full absorption case, $T/T_{\rm exp}$.
}
\label{fig:nuall}
\end{figure}

In the upper panels of Fig.~\ref{fig:nuall} we display the transmission coefficient $T(E_\nu)$ computed with {\tt NuPropEarth} using the BGR18 setup for the cross sections as a function of the neutrino energy $E_{\nu}$ for both muon and tau neutrinos.
The width of the  bands in the plots indicates the free-nucleon PDF uncertainties propagated from the BGR18 calculation.
In the following, we do not show the explicit results for electron neutrinos $\nu_e$, as these coincide with those of $\nu_{\mu}$.
The only differences would arise for the scattering of electron anti-neutrinos $\bar{\nu}_e$ near the Glashow resonance, where the corresponding attenuation becomes more significant in the localized region around $E_{\nu}=6.3~\PeV$ (see also Fig.~\ref{fig:xsec_glashow}).
The role that scattering on atomic electrons plays in the predictions for the transmission coefficient $T(E_\nu)$ of electron anti-neutrinos is discussed in Sect.~\ref{sec:results_subleading}.

In the comparisons of Fig.~\ref{fig:nuall}, the predictions for $T$ are shown for two different incident angles, $\cos\theta=0.1$ and $\cos\theta=0.9$, and we assume the spectral index is fixed to $\gamma=2$.
Predictions for the attenuation rates for other values of $\gamma$ are reported in Appendix~\ref{app:spectral}.
Note that the energy ranges ($x$ axes) are different in the two cases, to reflect the fact that, depending in the incident angle $\theta$, matter effects distort the incoming flux in different ways.
For example, in the case where $\cos\theta=0.1$ one has that $T=0.9$ for muon neutrinos at $E_{\nu}\simeq 10^6~\GeV$, while for  $\cos\theta=0.9$, where neutrinos traverse a much larger amount of Earth matter, the same level of attenuation is achieved at rather lower energies, around $E_{\nu}\simeq 10^4~\GeV$.

%%%%%%%%%%%%%%%%%%%%%%%%%%%%%%%%%%%%%%%%%%%%%%%%%%
%%%%%%%%%%%%%%%%%%%%%%%%%%%%%%%%%%%%%%%%%%%%%%%%%%

In Fig.~\ref{fig:nuall} we also display the value of the transmission coefficient $T$ for the full absorption case Eq.~(\ref{eq:fullabsortion}) (green lines) using the central BGR18 cross section without PDF uncertainties. In the bottom panels of Fig.~\ref{fig:nuall} we show the ratios of the transmission coefficients to the full absorption case, $T/T_{exp}$.
Note that in this full absorption limit the predictions for the attenuation rates coincide for both $\nu_{\mu}$ and $\nu_{\tau}$, given that in such scenario the tau regeneration process is absent.
As one can see from this comparison, the attenuation is always the largest in the full absorption case.
The reason is that the two mechanisms that are excluded in this limit, namely the $\nu_\tau$ regeneration in CC scattering and the energy degradation in NC scattering, both lead to a less marked decrease or even to an increase of the value of the transmission coefficient $T$ as compared to the full absorption limit.
Therefore, evaluating separately the full absorption case allows gauging the relative impact that these two mechanisms have in the total neutrino attenuation rates.

The central value of the theory prediction $\la T \ra$ (represented with dashed lines within the bands) is obtained as the mean over the replica sample,
\be
\la T (E_\nu) \ra = \frac{1}{N_{\rm rep}}\sum_{k=1}^{N_{\rm rep}} T^{(k)}(E_\nu) \, .
\ee

The PDF uncertainties turn out to be at the few percent level for most of the relevant energy range, and become only significant at the highest energies.
For instance, for $\cos\theta = 0.9~(0.1)$ the PDF uncertainties in $T$ reach around 25\% only for $E_{\nu}\gsim 10^{10} (10^7)~\GeV$.
We therefore find that  PDF uncertainties in the {\tt NuPropEarth} calculation of the attenuation rates are reasonably small except for the highest energies.
We want to emphasize that this result is not a general feature of the attenuation calculation, but rather a direct consequence of using the BGR18 cross sections,  based on PDFs well constrained in the small-$x$ region from the LHCb charm production data.
Indeed, if a different PDF set would have been used for the cross-section calculation, such as CT18~\cite{Hou:2019efy} or MMHT14~\cite{Harland-Lang:2014zoa},  the PDF uncertainties affecting the attenuation rates would be much higher than those displayed in the bottom panels of Fig.~\ref{fig:nuall}.
   
We would also like to mention in this respect that the PDF uncertainties in the transmission coefficient $T$ arising from the DIS cross-section calculation can be estimated at first order using the full absorption limit, Eq.~(\ref{eq:fullabsortion}).
In this  case, linear error propagation indicates that the uncertainties in $T$ relate to those of the cross section via
\be
\delta T = \delta \lp \frac{\phi}{\phi_0}\rp =  \delta
\lp\exp(-N_{A}\sigma p_{L}) \rp = T \lp N_{A} p_{L} \rp \delta \sigma \,, \quad \to \quad
\frac{\delta T}{T}(E_\nu) \propto \delta \sigma(E_\nu) \, ,
\ee
and therefore the relative uncertainties on the transmission coefficient $T$ should be proportional to the uncertainties on the neutrino-nucleon DIS cross section.
This result explains why the patterns for the PDF uncertainties reported in the lower panels of Fig.~\ref{fig:nuall} follow those of the cross sections reported Fig.~\ref{fig:xsec_nucleon}.
This derivation also indicates that the role of other variations of the input settings of the calculation, such as the choice of spectral index $\gamma$ or the incident angle $\theta$, is of second order in what concerns the PDF uncertainties affecting the transmission coefficient $T$.

\subsection{Dependence on the DIS cross-section model}
\label{sec:results_xsec}

As mentioned before, several calculations of the high-energy neutrino-nucleon DIS cross sections have been reported in the literature.
These calculations have been performed either in the collinear DGLAP framework or within alternative approaches such as accounting for non-linear corrections to the QCD evolution equations.
In each case, the underlying theoretical inputs, both those of non-perturbative (such as the choice of PDF set) and perturbative (such as the treatment of heavy quark mass effects) nature will be in general different.
The origin of the resulting differences at the cross-section level between some of these calculations was investigated and discussed in~\cite{Bertone:2018dse}.

Within the {\tt NuPropEarth} framework, as explained in Sect.~\ref{sec:uheneuttheory}, it is possible to choose between different cross-section models as well as varying the input PDF set to be used in the calculation (including also nuclear PDF sets).\footnote{ 
  Some details about how to vary the input theory settings
are provided in Appendix~\ref{sec:installation}.}
Here we would like to compare the baseline {\tt NuPropEarth} predictions for the neutrino flux attenuation induced by Earth matter, based on the BGR18 calculation, with those obtained if one adopts instead the CMS11 cross-section model~\cite{CooperSarkar:2011pa} instead.
The motivation for this choice is that CMS11 is one of the default models used by the IceCube collaboration in their modelling of neutrino interaction and to compare with their cross-section measurements, see {\it e.g.}~\cite{Aartsen:2017kpd}.
Note that other cross-section models can be easily implemented in {\tt HEDIS}, either by varying the input PDF set or by providing a parametrisation of the DIS structure functions.

\begin{figure}[t]
\centering 
\includegraphics[width=.49\textwidth]{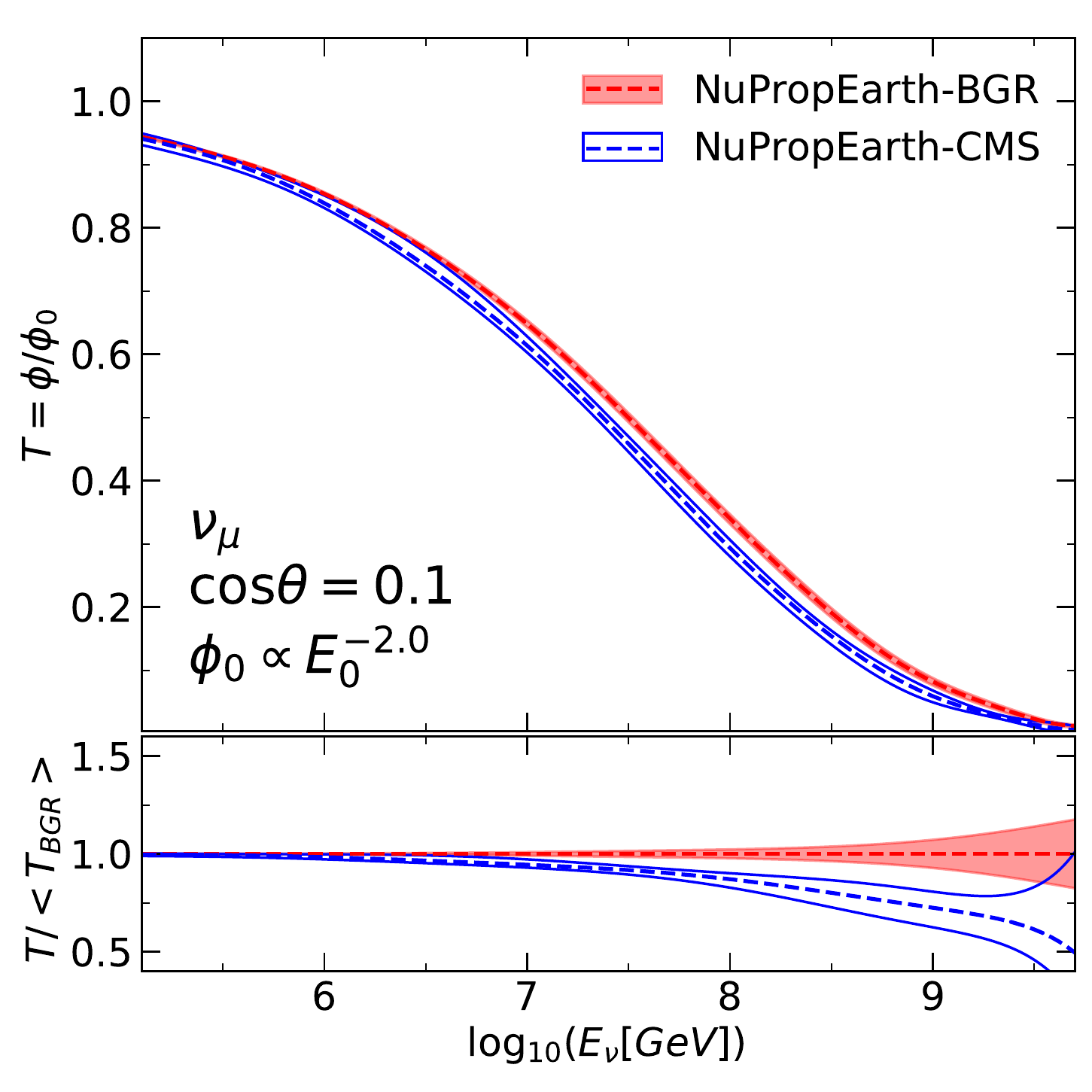}
\includegraphics[width=.49\textwidth]{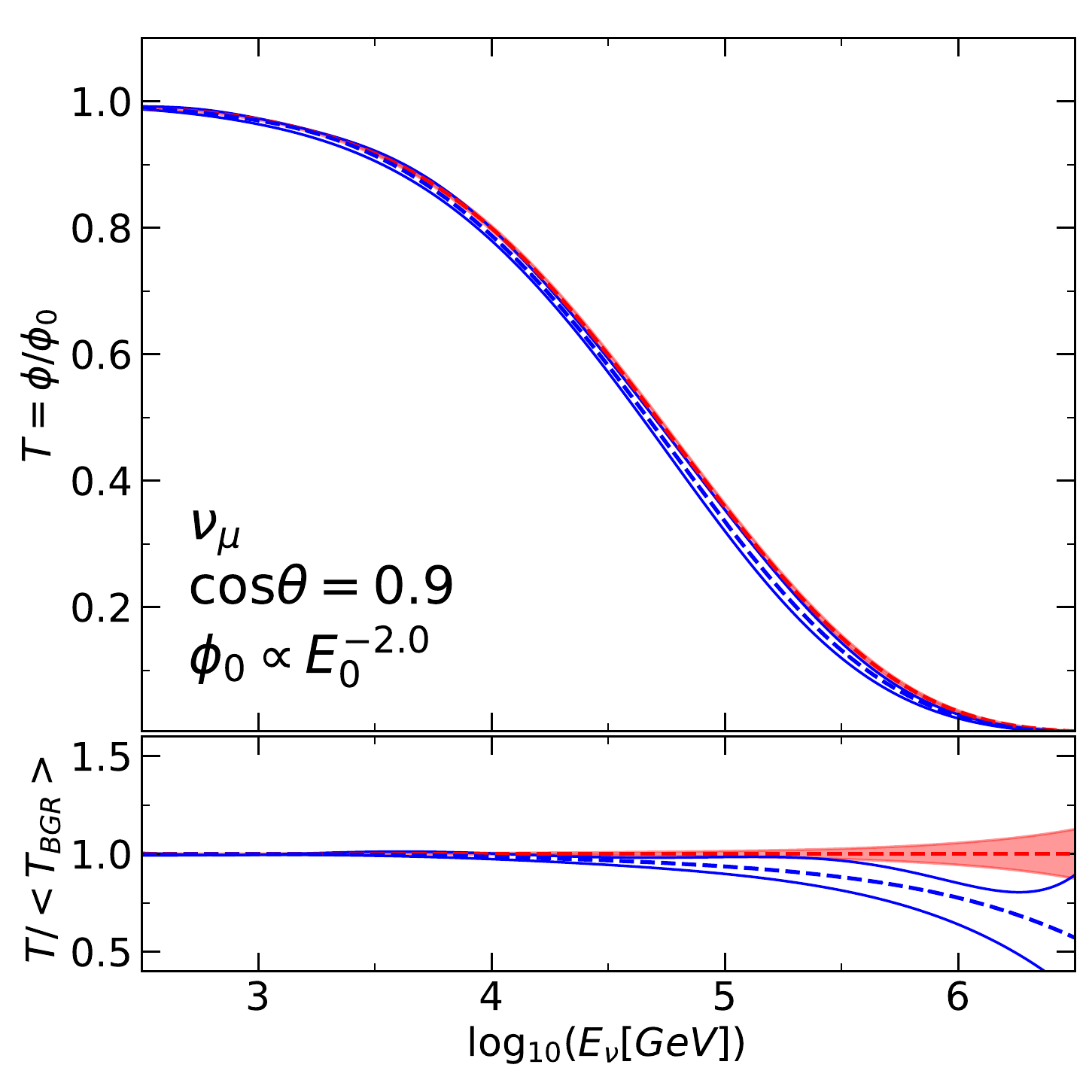}
\includegraphics[width=.49\textwidth]{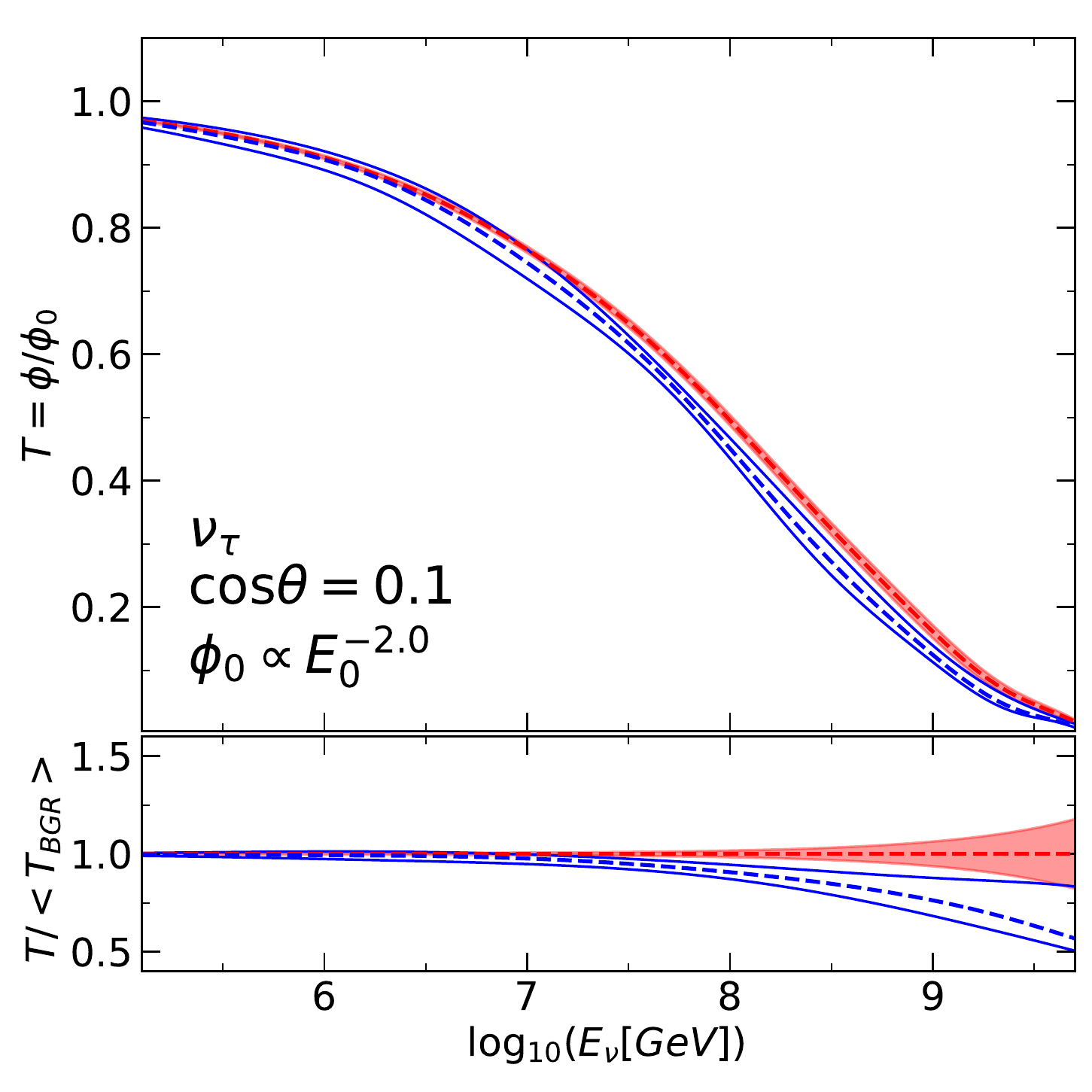}
\includegraphics[width=.49\textwidth]{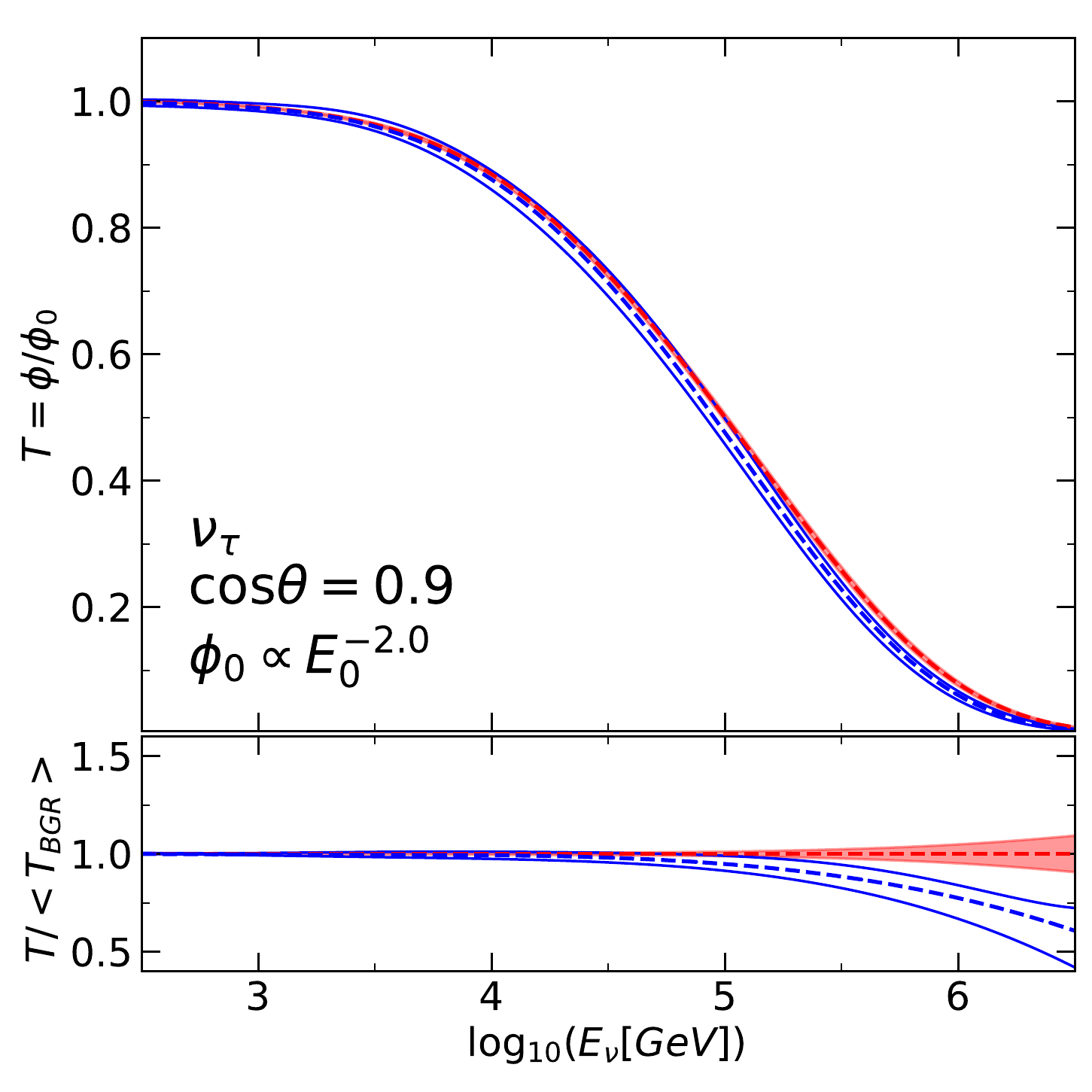}
\caption{Same as Fig.~\ref{fig:nuall}, now comparing the  {\tt NuPropEarth}
  predictions for the transmission coefficient $T$ for two different neutrino-nucleon DIS cross-section models,
  the baseline BGR18 (filled band) and CMS11 (empty band).
  The results for the $\nu_\mu$ ($\nu_\tau$)
  attenuation rates are displayed in the upper (lower) plots.
  In each of the four plots, the bottom panels display the ratio of $T$ to the
  central value of the BGR18 calculation, with the PDF uncertainties associated
  to each cross-section model.
  }
\label{fig:csms}
\end{figure}

In Fig.~\ref{fig:csms} we display a similar comparison such as the one in Fig.~\ref{fig:nuall}, now with the  {\tt NuPropEarth} predictions for the transmission coefficient $T$ corresponding to the two different neutrino-nucleon DIS cross-section models, the baseline BGR18 and CMS11.
Here we show the results for both the $\nu_\mu$  and $\nu_\tau$ attenuation rates and as before we show the outcome of the calculation for two values of the incident angle, $\cos\theta =0.1$ and $0.9$.
In each of the four plots, the bottom panels display the ratio of $T$ to the central value of the BGR18 calculation, with the corresponding relative PDF uncertainties associated to each cross-section model.

%%%%%%%%%%%%%%%%%%%%%%%%%%%%%%%%%%%%%%%%%%%%%%%%%%%%%%%%%%%%%%%%%%%%%%%%%%%%%%%%%%%%%%%%%%%%%%%%%%%%%
%%%%%%%%%%%%%%%%%%%%%%%%%%%%%%%%%%%%%%%%%%%%%%%%%%%%%%%%%%%%%%%%%%%%%%%%%%%%%%%%%%%%%%%

From the comparisons in Fig.~\ref{fig:csms} one can observe that, while at low and intermediate neutrino energies the two calculations coincide, they exhibit clear differences at higher energies.
In particular, we find that the CMS11 cross-section model predicts a more marked attenuation (reduced value of the transmission coefficient $T$) than the baseline BGR18 calculation for high neutrino energies.
For example, in the the $\nu_\mu$ case, the prediction for $T$ based on the CMS11 model is suppressed by around 25\% for $E_{\nu}=10^9~(10^6)~\GeV$ and for an incident angle of $\cos\theta=0.1~(0.9)$ as compared to the BGR18 calculation.
%To illustrate this point, we note that in the $\nu_\mu$ case, the prediction for $T$ based on the CMS11 model is suppressed by around 25\% for $E_{\nu}=10^9~(10^6)~\GeV$ and for an incident angle of $\cos\theta=0.1~(0.9)$ as compared to the BGR18 calculation.
%
Furthermore, one can also observe from this comparison that for a wide range of energies the two calculations do not agree within the corresponding PDF uncertainties.
The disagreement shown in Fig.~\ref{fig:csms} between the predictions of the two models can be traced back to the corresponding differences at the cross-section level reported for Fig.~\ref{fig:xsec_nucleon} and described in detail in Appendix~\ref{app:DIS}.

We note that in the comparisons presented in Fig.~\ref{fig:csms} the PDF uncertainties has been evaluated following the corresponding prescription from each group.
In the case of the BGR18 model we use the Monte Carlo replica prescription, while the CMS11 calculation is based on the Hessian method supplemented by additional eigenvectors to account for the model and parametrisation uncertainties.
In the region of relevance for this study, the PDF uncertainties in the two models turn out to be in general rather similar, with differences only significant at the highest energies where the BGR18 calculation becomes more precise (a result of PDF constraints from the inclusion of LHCb $D$-meson data).

\subsection{Comparison with {\tt nuFATE}, {\tt NuTauSim} and {\tt TauRunner}}

In this section we compare the predictions of the neutrino flux transmission coefficient $T(E_\nu)$ obtained from the {\tt NuPropEarth} simulations with those provided by three other software packages: {\tt nuFATE}, {\tt NuTauSim} and {\tt TauRunner}.

\paragraph{Comparison with {\tt nuFATE}.}

\begin{figure}[tbp]
\centering 
\includegraphics[width=.49\textwidth]{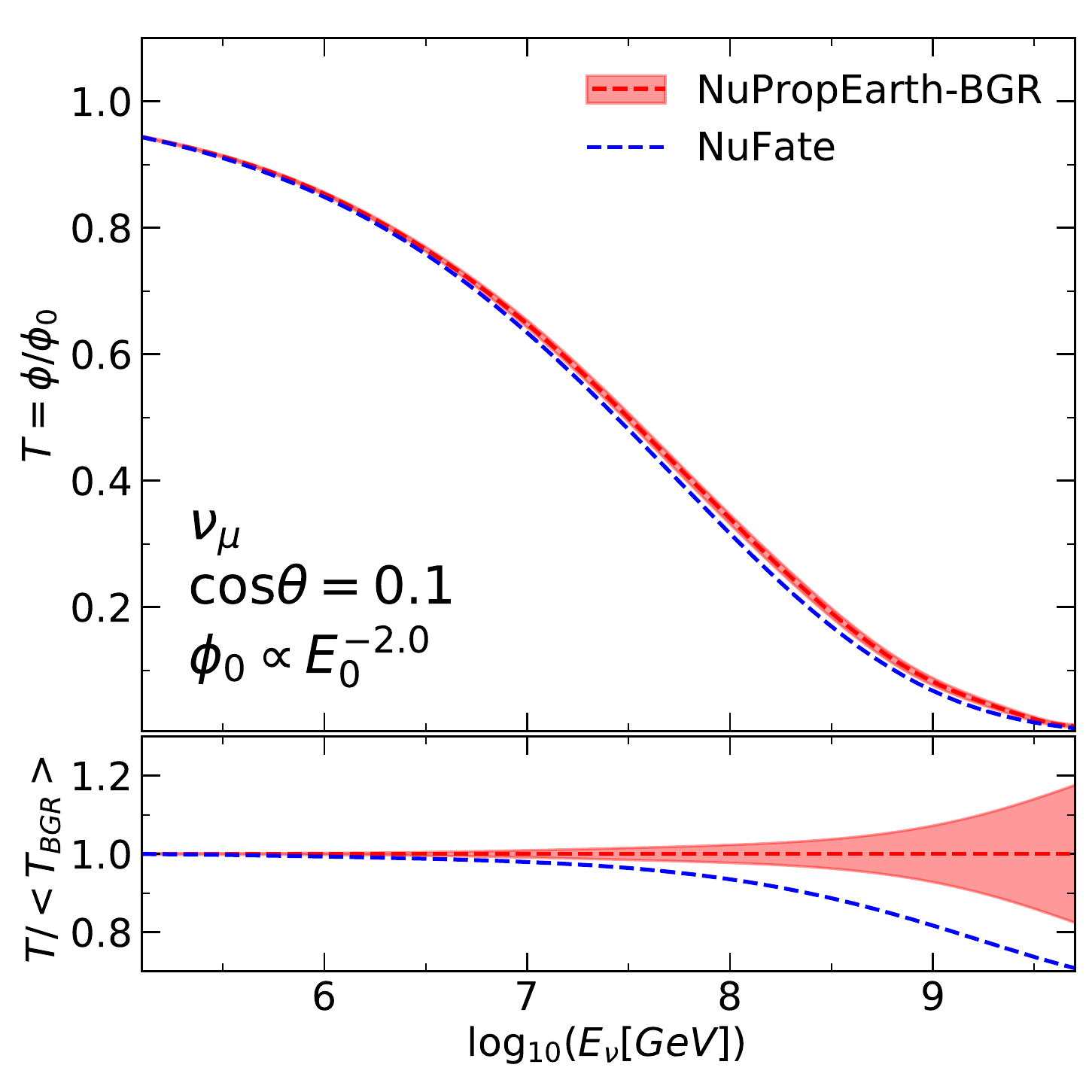}
\includegraphics[width=.483\textwidth]{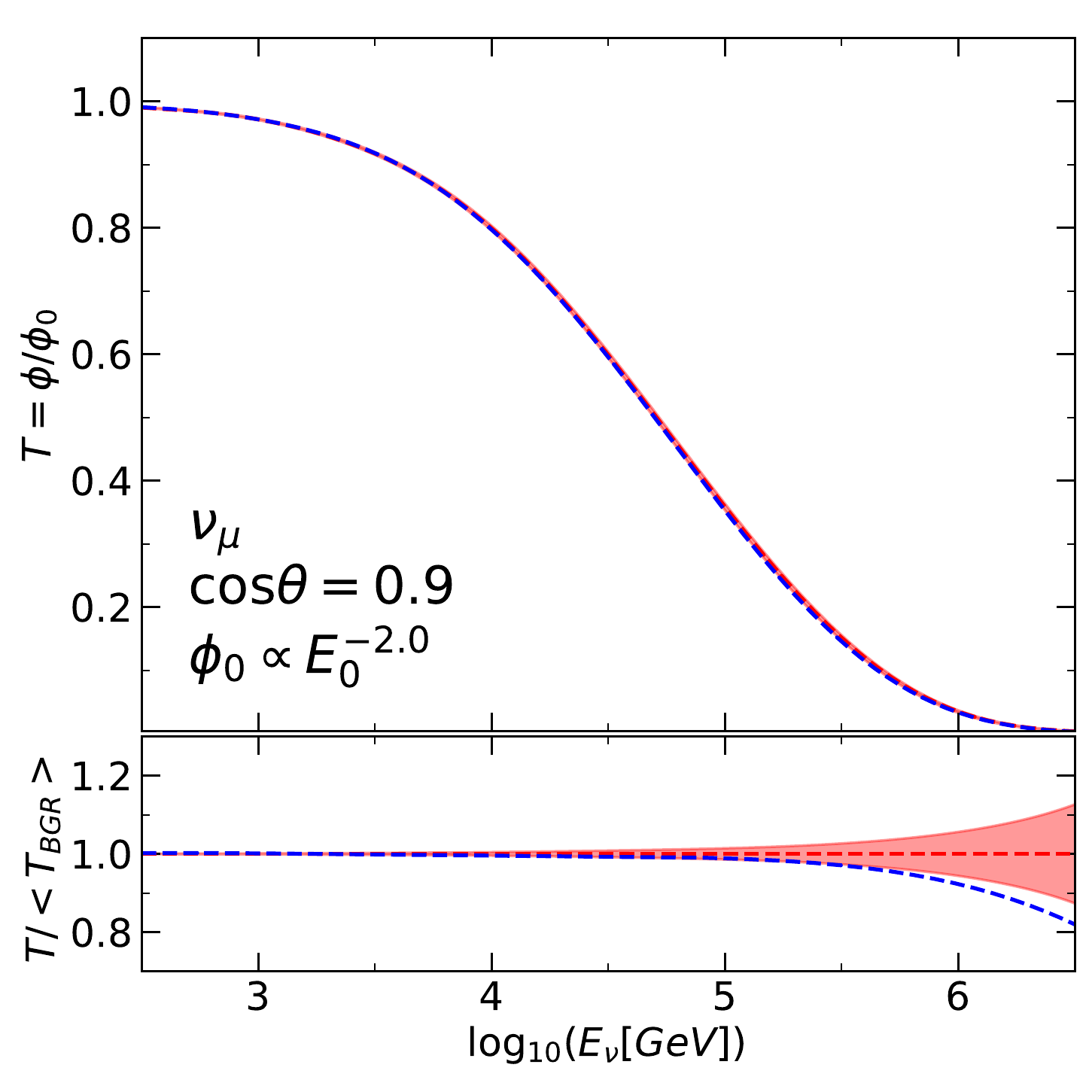}
\includegraphics[width=.49\textwidth]{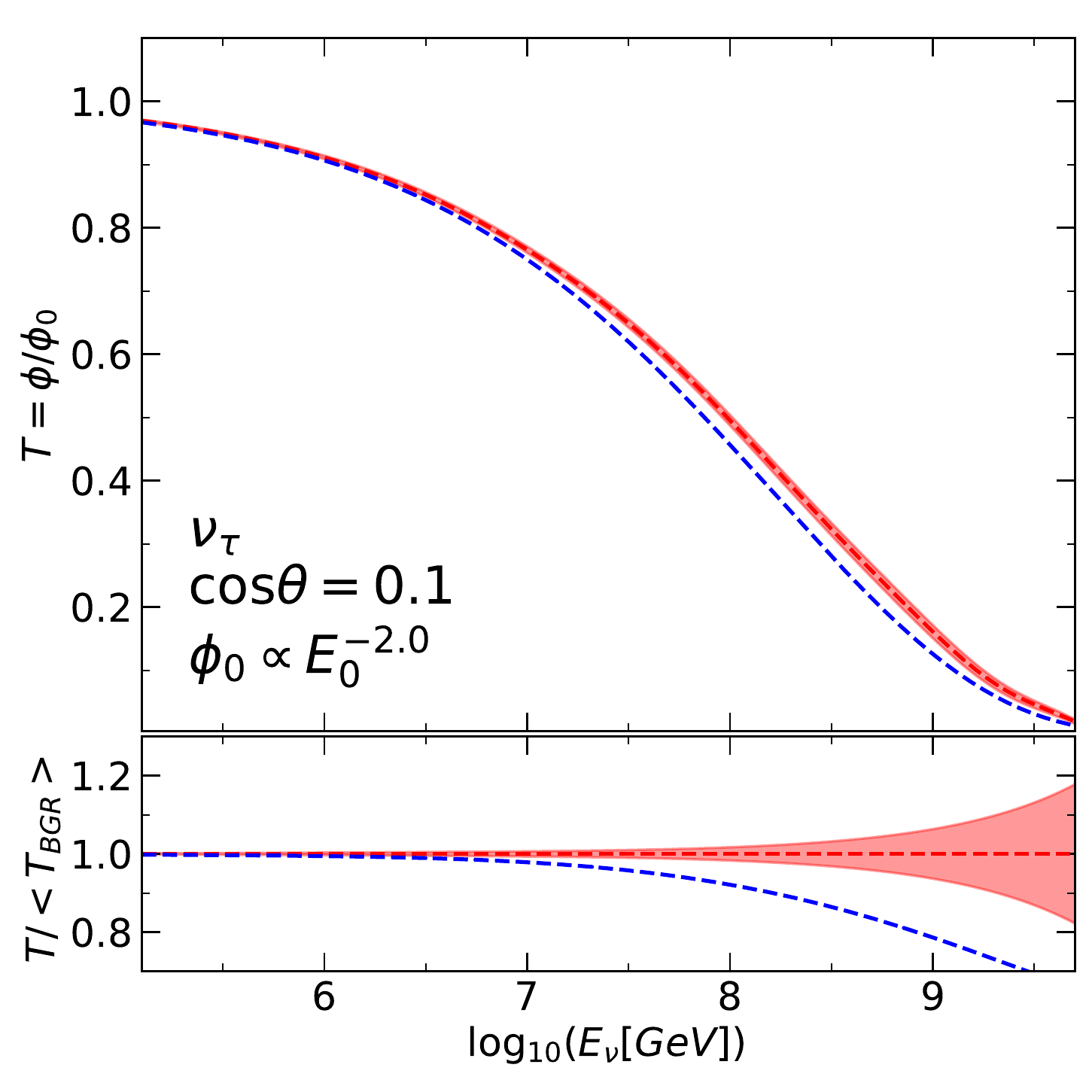}
\includegraphics[width=.483\textwidth]{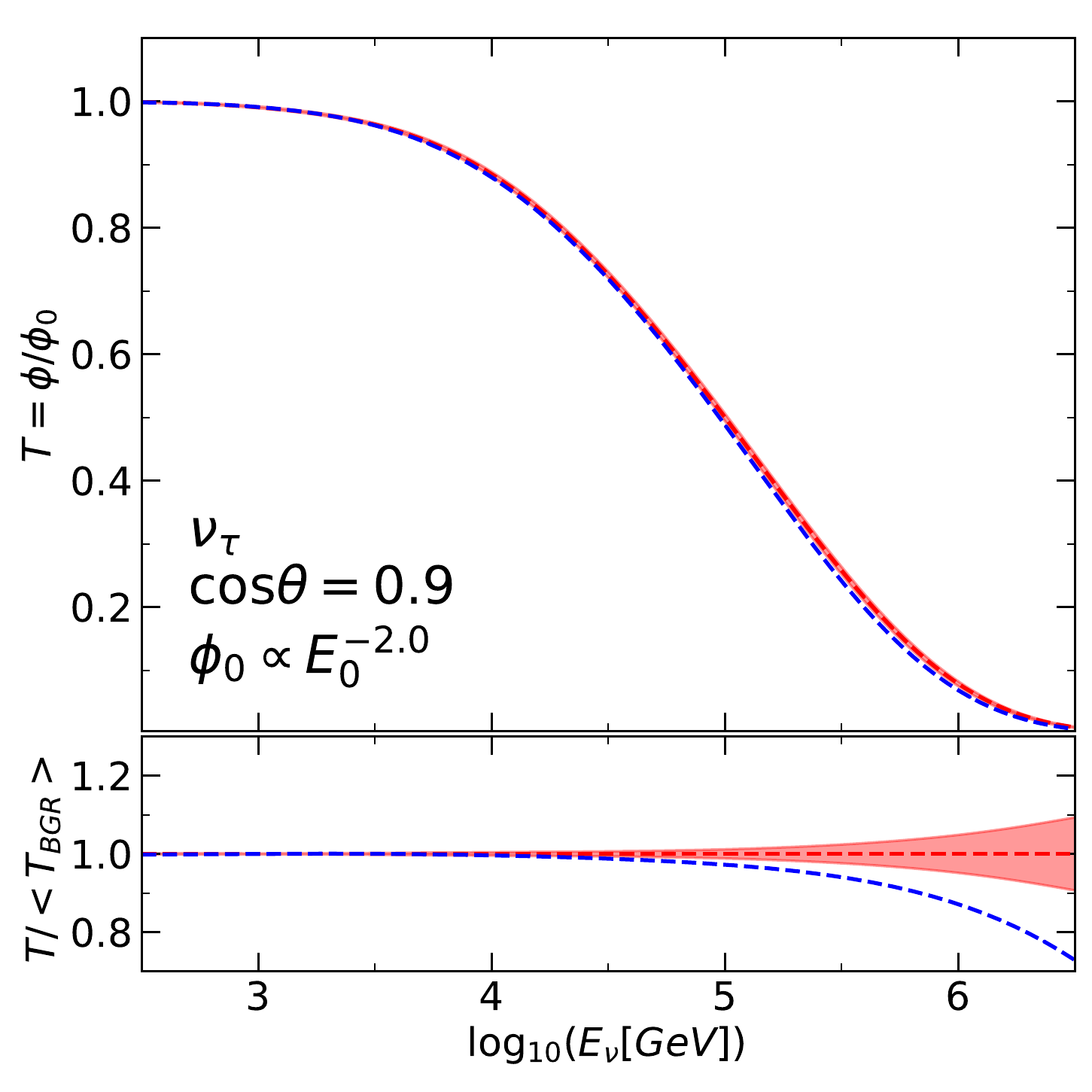}
\caption{Same as Fig.~\ref{fig:csms}, now comparing the transmission
  coefficients $T$ between the {\tt NuPropEarth} (red band) and {\tt nuFATE} (blue lines)
  calculations.
  In both cases, we use the BGR18 setup for the cross-section model, with PDF uncertainties
  only displayed for the  {\tt NuPropEarth} case.
  In the bottom panel we show these same calculation now as ratios
  to the central value of the  {\tt NuPropEarth} prediction.}
\label{fig:nufate}
\end{figure}

A drawbacks of quantifying the effects of the Earth on a particular neutrino flux using a Monte Carlo approach such as {\tt NuPropEarth} is the large computational cost.
The transmission coefficient $T$ can also be evaluated by solving the coupled cascade equations that describe the attenuation of neutrinos as they propagate through the Earth by means of linear algebra methods.
One implementation of such a strategy is provided by the \texttt{nuFATE}~\cite{Vincent:2017svp} framework.
We compare the predictions from {\tt NuPropEarth} with the baseline (BGR18) settings with those provided by the \texttt{nuFATE} program.
For a consistent comparison, the same model for the neutrino-nucleon cross section is used, namely the BGR18 calculation, so that differences are attributed to the methodologies adopted rather the DIS interaction model.
In Fig.~\ref{fig:nufate} the transmission coefficients $T$ are compared 
between the {\tt NuPropEarth} and {\tt nuFATE} calculations. The uncertainties stemming
from the PDF uncertainties are only shown for {\tt NuPropEarth}.
In the bottom panel of Fig.~\ref{fig:nufate} we display these same calculation as ratios to the central value of the  {\tt NuPropEarth} prediction.

%%%%%%%%%%%%%%%%%%%%%%%%%%%%%%%%%%%%%%%%%%%%%%%%%%
%%%%%%%%%%%%%%%%%%%%%%%%%%%%%%%%%%%%%%%%%%%%%%%%%%

We note that the comparisons presented in Fig.~\ref{fig:nufate} have been carried out by neglecting the tau energy losses in the \texttt{NuPropEarth} calculation.
The reason for this choice is to ensure a consistent comparison with the results of the  \texttt{nuFATE} framework, which currently does not account for the effects of tau energy losses.
Neglecting  tau lepton energy losses does not modify significantly the modelling of tau regeneration effects when the amount of mater traversed is large, $\cos\theta \rightarrow 1$.
However, for Earth-skimming neutrinos for which $\cos\theta \rightarrow 0$ the tau energy loss effects do change appreciably the shape of the neutrino flux that reaches the detector, as further discussed in Sect.~\ref{sec:nutausim}, and therefore it is important to take them into account for a complete calculation of the attenuation rates.

From Fig.~\ref{fig:nufate} one observes that the \texttt{nuFATE} calculation underestimates the results from {\tt NuPropEarth} by an amount that varies between  a few percent and 35\% depending on the specific value of the neutrino energy.
These discrepancies can be partly explained by the fact that the approach adopted in the \texttt{nuFATE} framework to solve the cascade equations is only exact when the neutrino energy interpolation nodes $E_{\nu, i}$ are very close among them, that is, in the limit $\lp E_{\nu, i+1} - E_{\nu, i} \rp \rightarrow0$.
In the baseline settings of~\cite{Vincent:2017svp} around 200 nodes in $E_{\nu}$ were used to solve the cascade equations.
We have verified that if we increase the number of nodes up to 2000, a factor 10 larger than in the baseline {\tt nuFATE} settings, we are able to improve the agreement with the {\tt NuPropEarth} in the full simulation case.

\paragraph{Comparison with {\tt NuTauSim} and {\tt TauRunner}.}
\label{sec:nutausim}

The growing interest in the detection of $\tau$-induced air showers with either surface or altitude detectors has lead to the development of novel  tools for the modelling of such events. Here we consider as representative examples of simulation tools for the description of  $\tau$-induced air showers the \texttt{NuTauSim}~\cite{Alvarez-Muniz:2017mpk} and \texttt{TauRunner}~\cite{Safa:2019ege} packages. Both frameworks propagate tau neutrinos in matter including regeneration effects and the energy loss of tau leptons due to electromagnetic interactions. The latter can take place via ionisation, bremsstrahlung, pair production and photo-nuclear interactions.

In this section we aim to compare the results from the {\tt NuPropEarth} baseline calculation for the transmission coefficients of tau neutrinos with those provided by \texttt{NuTauSim} and \texttt{TauRunner}. In order to disentangle the different contributions to the regeneration  effects, the simulations have been carried out with and without accounting for tau energy loss.

In general, the computational overhead required in \texttt{NuTauSim} will be smaller than that of \texttt{TauRunner} and \texttt{NuPropEarth}, being the latest the largest. The main reason for this is that \texttt{NuTauSim} and \texttt{TauRunner} adopt tabulated results for most of the physical processes that are required to model the neutrino interactions and the tau decay. In addition, the propagation of tau leptons differs between the three framework. \texttt{NuTauSim} uses a one-dimensional approach, hence it does not account for deflections but only energy losses. \texttt{NuPropEarth} and \texttt{TauRunner} use dedicated packages to propagate these particles, \texttt{TAUSIC}
~\cite{ANTONIOLI1997357} and \texttt{MMC}~\cite{Chirkin:2004hz} respectively.

%%%%%%%%%%%%%%%%%%%%%%%%%%%%%%%%%%%%%%%%%%%%%%%%%%
\begin{figure}[t]
\centering 
\includegraphics[width=.49\textwidth]{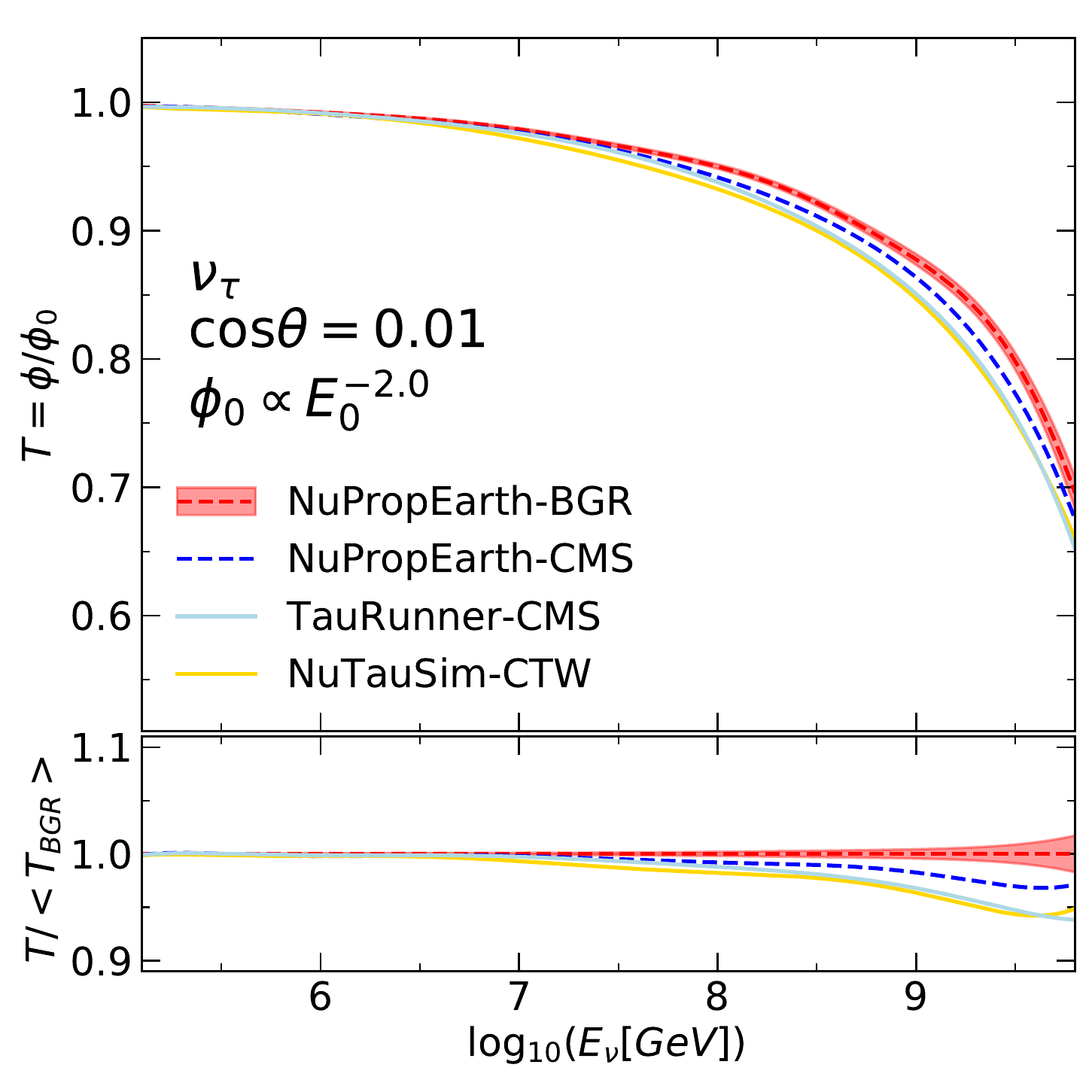}
\includegraphics[width=.49\textwidth]{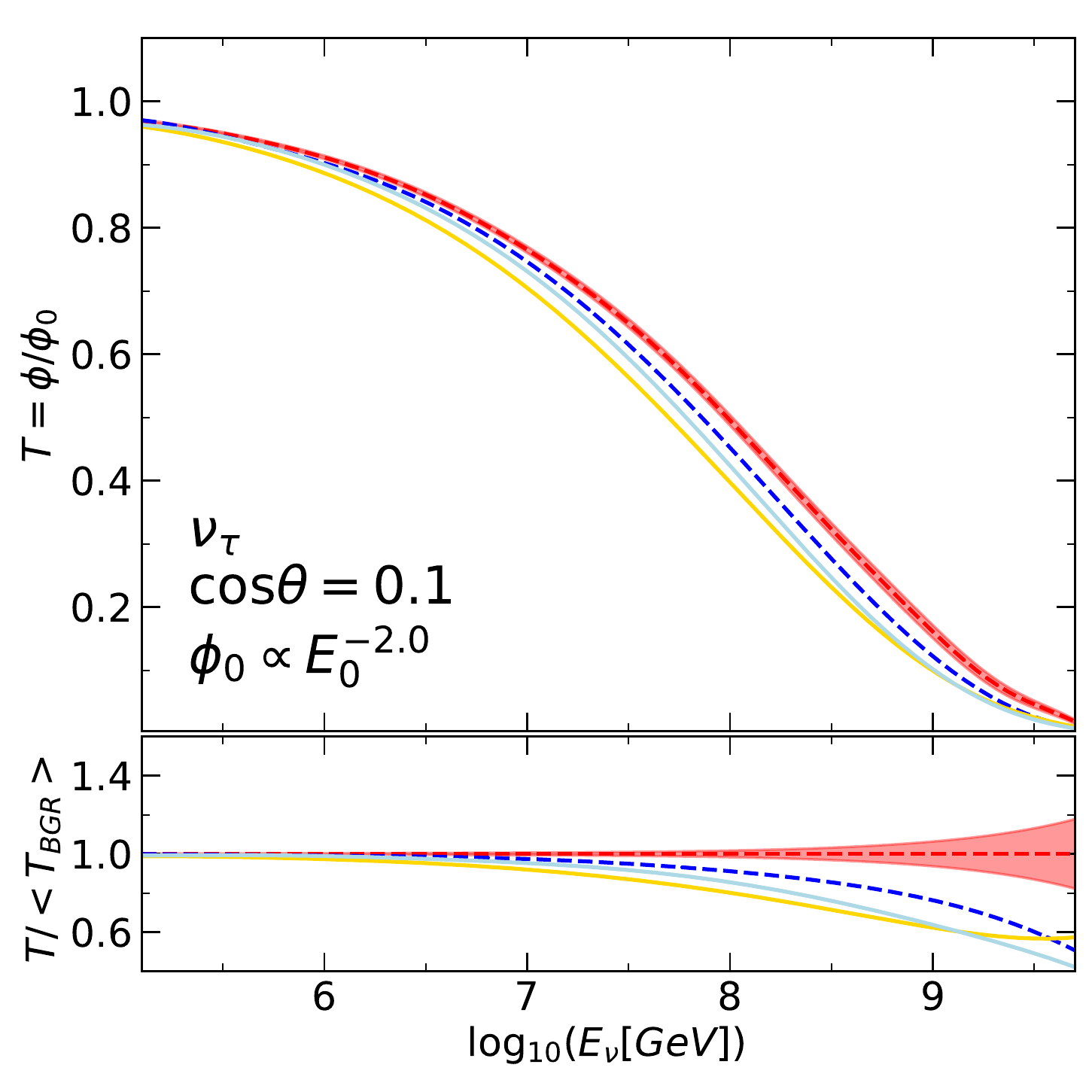}
\caption{Same as Fig.~\ref{fig:nuall}, now comparing the \texttt{NuPropEarth} predictions (using two DIS models: BGR, red bands, and CMS, dashed blue lines) with \texttt{NuTauSim} (orange lines) and \texttt{TauRunner} (light blue lines) calculations for the tau neutrino $\nu_\tau$ transmission coefficient T. The legend includes the DIS cross-section model used in each calculation. All the predictions do not account for tau energy losses. Results are shown for two different incident angles, $\cos\theta=0.01$ (left) and $\cos\theta=0.1$ (right). The bottom panels display the ratio of T to the central value of the baseline calculation, NuPropEarth-BGR.}
\label{fig:nutausim_noeloss}
\end{figure}
%%%%%%%%%%%%%%%%%%%%%%%%%%%%%%%%%%%%%%%%%%%%%%%%%%

In Fig~\ref{fig:nutausim_noeloss} we display a similar comparison to that of Fig.~\ref{fig:csms}, now restricted to the case of tau neutrinos, showing the predictions of {\tt NuPropEarth}, {\tt NuTauSim} and {\tt TauRunner} for two incident angles $\cos\theta = 0.1$ and $\cos\theta = 0.01$. Note that as opposed to previous plots, here the range of energies is the same for the two incident angles. The choice of values of the incident angles $\cos\theta$ is motivated by their relevance to $\tau$-induced shower detection, which is associated to Earth-skimming neutrinos. The results for the full simulations are provided without tau energy losses and using different DIS cross-section models, including the corresponding PDF uncertainties for our baseline model.
  
From the comparisons in Fig.~\ref{fig:nutausim_noeloss}, one can observe that for both $\cos\theta = 0.01$ and $\cos\theta = 0.1$ all the predictions are similar but they exhibit a more intense attenuation than in our baseline calculation. The reason for this behaviour can be traced back to the different cross-section models adopted in each case, given that \texttt{NuTauSim} and \texttt{TauRunner} use the CTW11~\cite{Connolly:2011} (based on the MSTW08 set of PDFs~\cite{Martin:2009iq}) and CMS11 calculations respectively. Both models predict a higher cross section for an amount between 10\% and 20\% as compared to the BGR18 calculation~\cite{Bertone:2018dse}.

The prediction of {\tt NuPropEarth} using the CSM11 model is good agreement with \texttt{NuTauSim} and \texttt{TauRunner}. The remaining discrepancy is due to the different methods used to calculate the kinematics of the outgoing leptons in the DIS interaction and the tau decay. Firstly, to model the inelasticity distribution of the outgoing lepton, \texttt{NuTauSim} uses not the CTW11 calculation but rather tabulated results of the differential cross section extracted using the obsolete PDF set CTEQ5~\cite{Lai:1999wy}. One important advantage in this respect of {\tt NuPropEarth} and \texttt{TauRunner} frameworks is that they use a consistent model for both the total and differential cross sections everywhere in the simulation chain. Secondly, {\tt NuPropEarth} uses \texttt{TAUOLA} to simulate tau decays and extract the kinematics of the decay products, whereas \texttt{TauRunner} uses a parametrization based on \cite{Dutta:2000jv} and \texttt{NuTauSim} includes tabulated results extracted using \texttt{TAUOLA}.

\begin{figure}[t]
\centering 
\includegraphics[width=.49\textwidth]{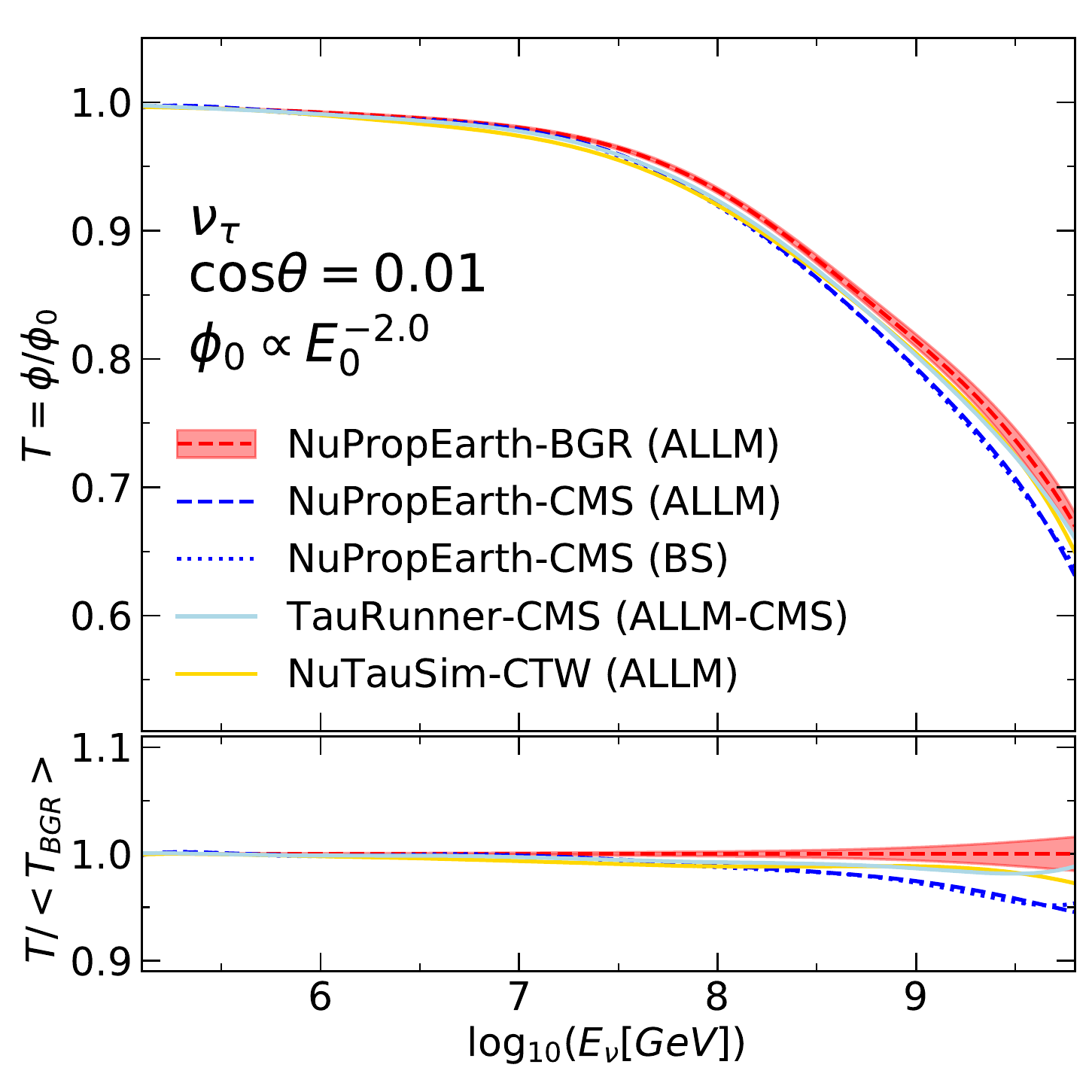}
\includegraphics[width=.49\textwidth]{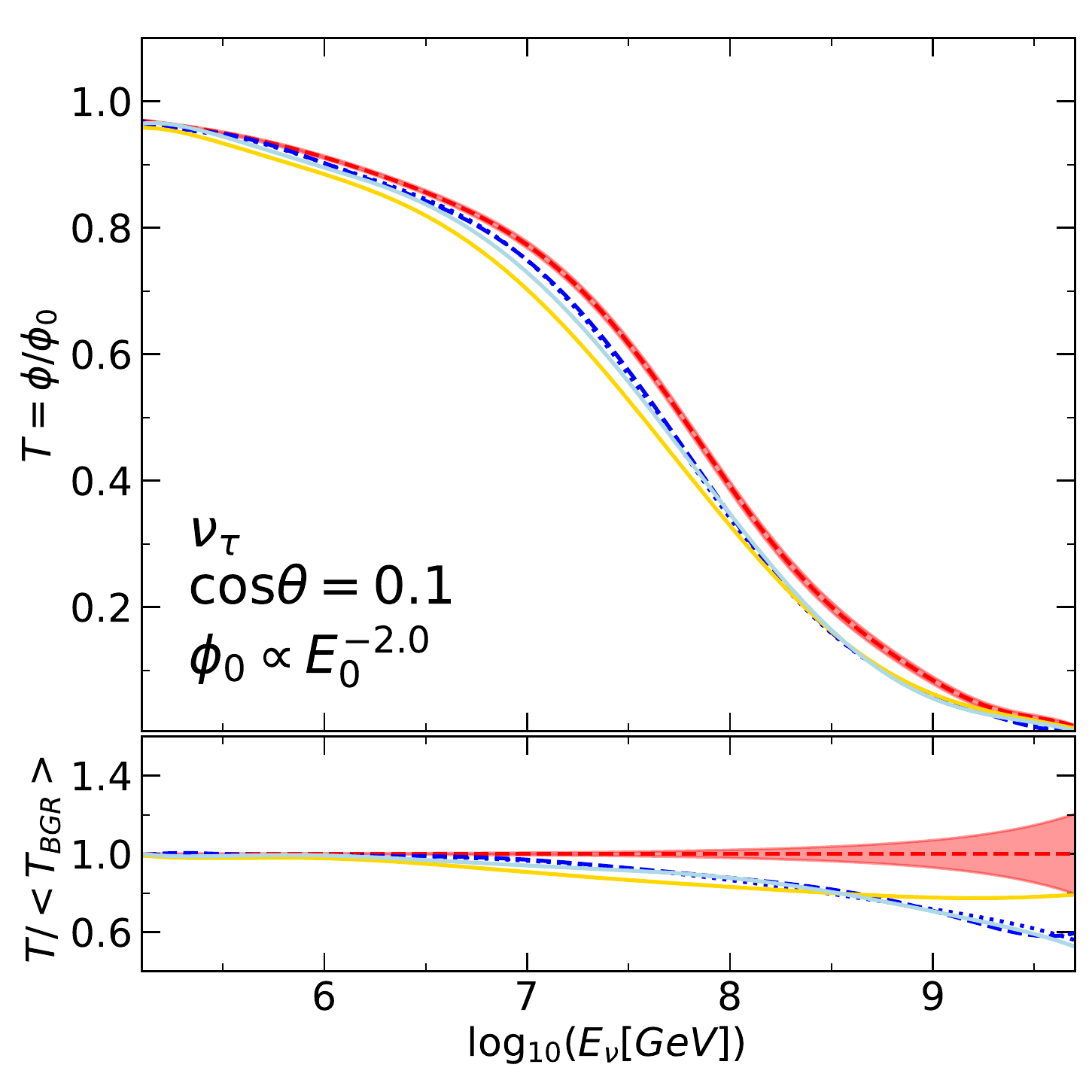}
\caption{Same as Fig.~\ref{fig:nutausim_noeloss}, now including tau energy losses in the calculations. The legend includes the DIS cross-section and the tau energy loss models used in each calculation. The bottom panels display the ratio of T to the central value of the baseline calculation, NuPropEarth-BGR (ALLM).}
\label{fig:nutausim_eloss}
\end{figure}

Finally, as we have mentioned before, the effect of the tau energy loss in the $\nu_\tau$ regeneration is important for Earth-skimming neutrinos. These losses become relevant when the tau energy is above 1 PeV, where photo-nuclear processes are dominant. In {\tt NuPropEarth}, the energy losses are computed using {\tt TAUSIC}, which includes two different models to compute the photo-nuclear cross sections (ALLM~\cite{Abramowicz:1997ms} and BS~\cite{Bugaev:2002gy}). The ALLM model can be also used in {\tt NuTauSim} and \texttt{TauRunner}, but the latest includes a modification to use the CSM11 structure functions. The comparisons in Fig.~\ref{fig:nutausim_eloss} include the $\nu_\tau$ transmission coefficient taking into account the tau energy loss.

In the three frameworks, one observes that the transmission coefficient $T$ becomes smaller at energies above $10^7~\GeV$ when the energy loss is included, enhancing the transmission coefficient for smaller energies. Similarly to Fig.~\ref{fig:nutausim_noeloss}, the differences in the cross section models remains as the main source of discrepancy between the baseline and other predictions. Also, the predictions of {\tt NuPropEarth} using the CSM11 model, \texttt{NuTauSim} and \texttt{TauRunner} show good agreement, although the reduction of the transmission coefficient is slightly higher in {\tt NuPropEarth}. This is observed with both ALLM and BS models, which show very similar predictions.

We can conclude that the agreement of the three frameworks is acceptable considering the different approaches that they use. The main source of discrepancy between them at energies below 10 EeV is the DIS cross section model, while effects such as tau energy losses and decay are subdominant. At energies above 10 EeV, which are relevant for proposed experiments like ARIANNA or GRAND, tau energy losses will become more relevant. Hence, in the future, it will be crucial to quantify with these frameworks the impact of different tau propagation models in the transmission coefficient.

%%%%%%%%%%%%%%%%%%%%%%%%%%%%%%%%%%%%%%%%%%%%%%%%%%
%%%%%%%%%%%%%%%%%%%%%%%%%%%%%%%%%%%%%%%%%%%%%%%%%%

\subsection{Impact of the sub-dominant interaction processes}
\label{sec:results_subleading}

Up to this point, our predictions for the transmission coefficients $T$ evaluated with {\tt NuPropEarth} have included only the dominant neutrino-matter interaction mechanisms: charged- and neutral-current deep inelastic scattering.
Now we assess the impact on the attenuation rates of the subleading interaction processes discussed in Sect.~\ref{sec:formalism}.
For the dominant DIS contributions, we will adopt our baseline settings, based on the BGR18 calculation displayed in Fig.~\ref{fig:nuall}.
We will begin by quantifying the impact of these subleading interaction processes for muon and electron neutrinos, and then move to discuss the case of electron anti-neutrinos, where the impact of the Glashow resonance will be present.

\begin{figure}[t]
\centering 
\includegraphics[width=.49\textwidth]{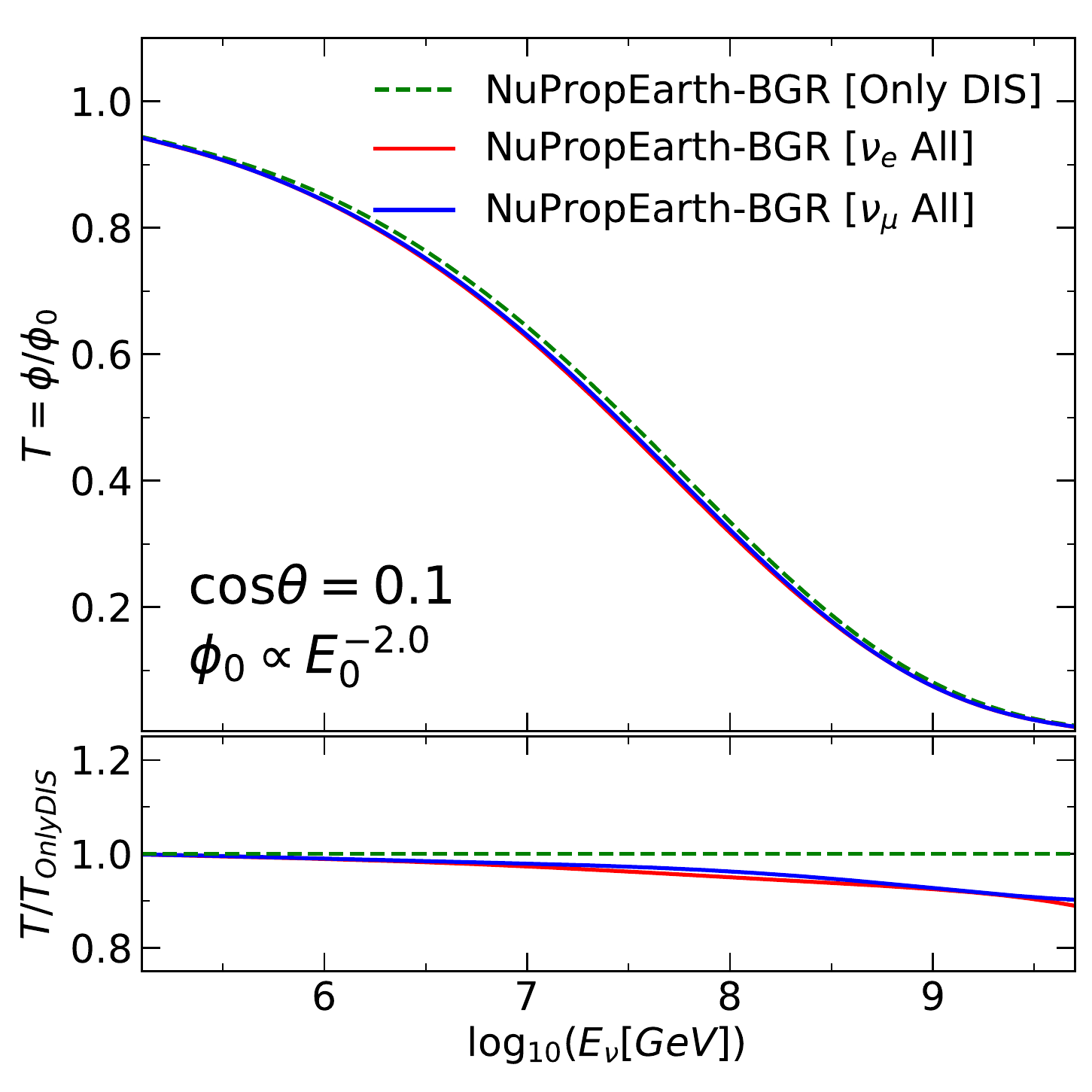}
\includegraphics[width=.483\textwidth]{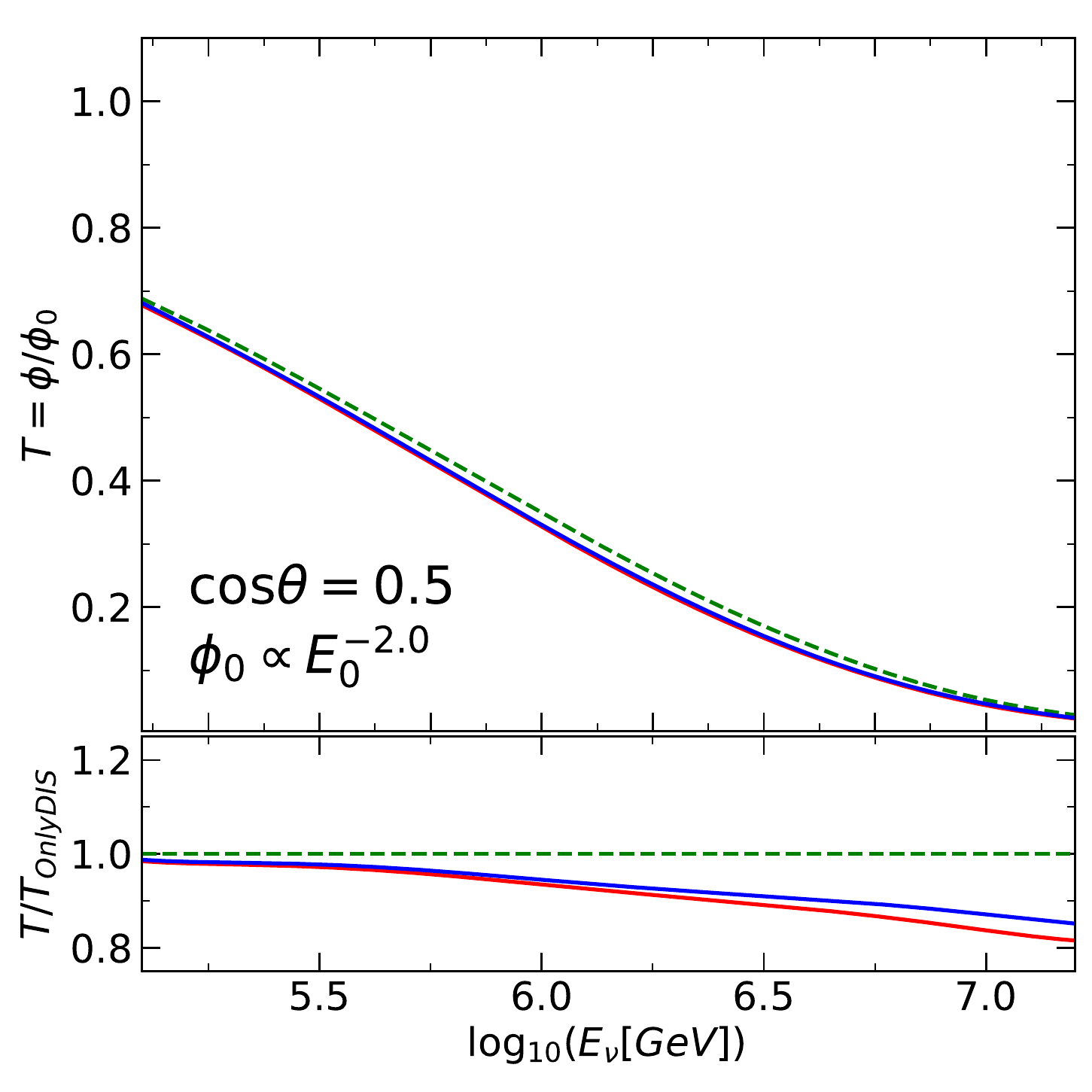}
\caption{The transmission coefficient $T(E_\nu)$ 
  for $\nu_\mu$ (blue lines) and $\nu_e$ (red lines)
  and for $\cos\theta=0.1$ (left) and $\cos\theta=0.5$ (right plots)
  computed with {\tt NuPropEarth} when different contributions to the total the neutrino-matter
  interaction cross section are taken into account.
  We compare the results obtained when only the DIS NC and CC cross sections
  are accounted for
  (``only DIS'') with those where DIS is supplemented
 with  the subleading
 interaction mechanisms (``all'').
 The bottom panels display the ratio $T/T_{\rm DIS}$ between the transmission
 coefficient evaluated with
 all interaction processes and with only the DIS cross sections.}
\label{fig:attsubleading}
\end{figure}

In Fig.~\ref{fig:attsubleading} we display the transmission coefficient $T(E_\nu)$  for muon  and electron neutrinos for incident angles of $\cos\theta=0.1$  and $\cos\theta=0.5$ computed with {\tt NuPropEarth} for different assumptions on the neutrino-matter interaction cross sections.
Note that the energy range is different for each value of $\cos\theta$.
We compare the results obtained with the dominant DIS cross sections (same as those presented in Fig.~\ref{fig:nuall}) with the corresponding predictions where DIS is supplemented with the subleading interaction mechanisms.
The bottom panels then display the ratio $T/T_{\rm DIS}$ between the transmission coefficient evaluated with all interaction processes and with only the DIS cross sections.

%%%%%%%%%%%%%%%%%%%%%%%%%%%%%%%%%%%%%%%%%%%%%%%%%%
%%%%%%%%%%%%%%%%%%%%%%%%%%%%%%%%%%%%%%%%%%%%%%%%%%

From the results of Fig.~\ref{fig:attsubleading}, one can observe that the impact of the subleading interaction mechanisms is more marked for high neutrino energies.
Their net effect is a suppression of the transmission coefficient $T$, given that the subleading channels lead to an enhancement of the total interaction cross section as indicated in Fig.~\ref{fig:xsec_subleading}.
This suppression can be as large as 10\%~(15\%) in the case of $\cos\theta=0.1~(0.5)$ for neutrino energies of $E_{\nu}\simeq 10^{10}~(10^7)~\GeV$.
We also find that the effects of the subleading channels are very similar for electron and muon neutrinos, as expected since the interaction mechanisms in Sect.~\ref{sec:formalism} depend only mildly on the neutrino flavor.
We can conclude that accounting for the contribution of the subleading interaction mechanisms is important for a precision calculation of the neutrino attenuation rates, in particular to study the high-energy tail of the astrophysical neutrino flux.

\begin{figure}[t]
\centering 
\includegraphics[width=.49\textwidth]{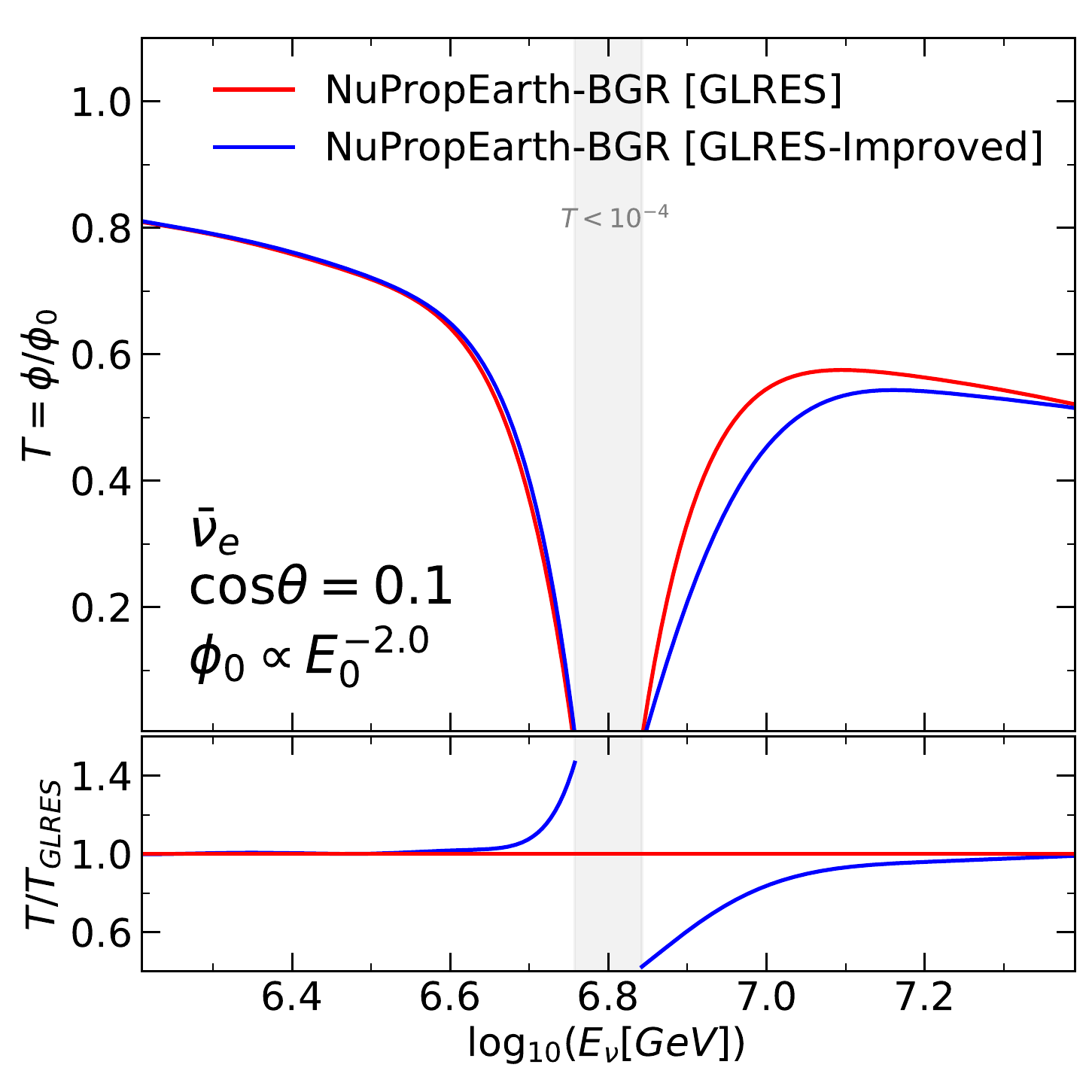}
\caption{The transmission coefficient $T(E_\nu)$ 
  for electron anti-neutrinos $\bar{\nu}_e$ 
  for $\cos\theta=0.1$ 
  computed with {\tt NuPropEarth} with all subleading
  interaction channels taken into account.
  The range of neutrino energy has been restricted to the vicinity
  of the Glashow resonance at $E_{\nu}\simeq 6.3~\PeV$.
  Results are shown for the cases in which the Glashow resonance cross section
  is computed at the Born level (``GLRES'') and when including
  higher-order electroweak corrections  (``GLRES-Improved'').
  The grey region indicates the range of energies for which the transmission
  coefficient is smaller than $T<10^{-4}$.
  The bottom panel displays the ratio of the {\tt NuPropEarth}
calculations to the central value of the ``GLRES'' curve.}
\label{fig:attglashow}
\end{figure}

Then in Fig.~\ref{fig:attglashow} we display the transmission coefficient $T(E_\nu)$  for electron anti-neutrinos $\bar{\nu}_e$ for $\cos\theta=0.1$ computed with {\tt NuPropEarth} with all subleading interaction channels taken into account.
The range of neutrino energies has been restricted to the vicinity of the Glashow resonance at $E_{\nu}\simeq 6.3~\PeV$.
We recall from Fig.~\ref{fig:xsec_glashow} that in the vicinity of the Glashow resonance the cross section becomes very large, with neutrino scattering on atomic electrons dominating over the DIS component by two orders of magnitude.
Results are shown for the cases in which the Glashow resonance cross section is computed at LO (``GLRES'') and when higher-order corrections have been included (``GLRES-Improved'').
The bottom panel displays the ratio of the {\tt NuPropEarth} calculations to the central value of the ``GLRES'' curve.

%%%%%%%%%%%%%%%%%%%%%%%%%%%%%%%%%%%%%%%%%%%%%%%%%%
%%%%%%%%%%%%%%%%%%%%%%%%%%%%%%%%%%%%%%%%%%%%%%%%%%

As expected, the impact of the Glashow resonance is negligible except in the vicinity of $E_{\nu}\simeq 6.3~\PeV$, where it dominates over all other the components of the cross section leading to a vanishing transmission coefficient.
We also observe how the higher-order corrections to the Glashow resonance scattering are visible in the predictions for $T(E_\nu)$ around the resonance, region.
However, we also note that this is a rather localised effect, and for neutrino energies outside the resonance region this interaction mechanism can be neglected.

\section{Nuclear PDF effects in the attenuation rates}
\label{sec:results_nucl}

The cross sections for neutrino-matter interactions, as discussed
in Sect.~\ref{sec:uheneuttheory}, are dominated by
charged- and neutral-current deep-inelastic scattering off nucleons bound within
the nuclei of the atoms that compose Earth matter.
The results for the attenuation rates presented in
Sect.~\ref{sec:results_p} are based on DIS cross sections that assume
that the PDFs of these bound nucleons can be identified
with those of their free-nucleon counterparts.
However, it is well known that the quark and gluon PDFs of bound
nucleons in heavy nuclei are modified as compared to the free-proton case --- see~\cite{Ethier:2020way} for a recent review.
In this section, we study how the results for the high-energy neutrino attenuation in matter
are modified in the presence of nuclear modifications of the partonic structure of nucleons.
As we will demonstrate, the limited  experimental constraints available on small-$x$ nuclear PDFs
implies that these represent now the dominant source of theoretical uncertainties
in the calculation of the attenuation rates.

\subsection{Nuclear parton distributions at small-$x$}
\label{sec:pdfrev}

The DIS interactions between high-energy neutrinos and Earth matter probe the proton PDFs for momentum fractions $x$ and virtualities $Q^2$ in regions outside those constrained by the lepton-proton scattering data from HERA~\cite{Abramowicz:2015mha}.
For instance, a neutrino with energy $E_\nu = 10^9~\GeV$ is sensitive to partonic momentum fractions in the region of $x\simeq 10^{-7}$ and $Q^2 \simeq m_{\PW}^2$. 
While the low-$x$ HERA data ($x\simeq 3\times 10^{-5}$ at small $Q^2$-values) provide some relevant information through correlations introduced as part of the DGLAP evolution of these PDFs, these constraints are mild.
The situation is far worse for lepton-nucleus scattering, where available constraints are restricted to $x\gsim 10^{-3}$.

\begin{figure}[tbp]
\centering 
\includegraphics[width=1.\textwidth]{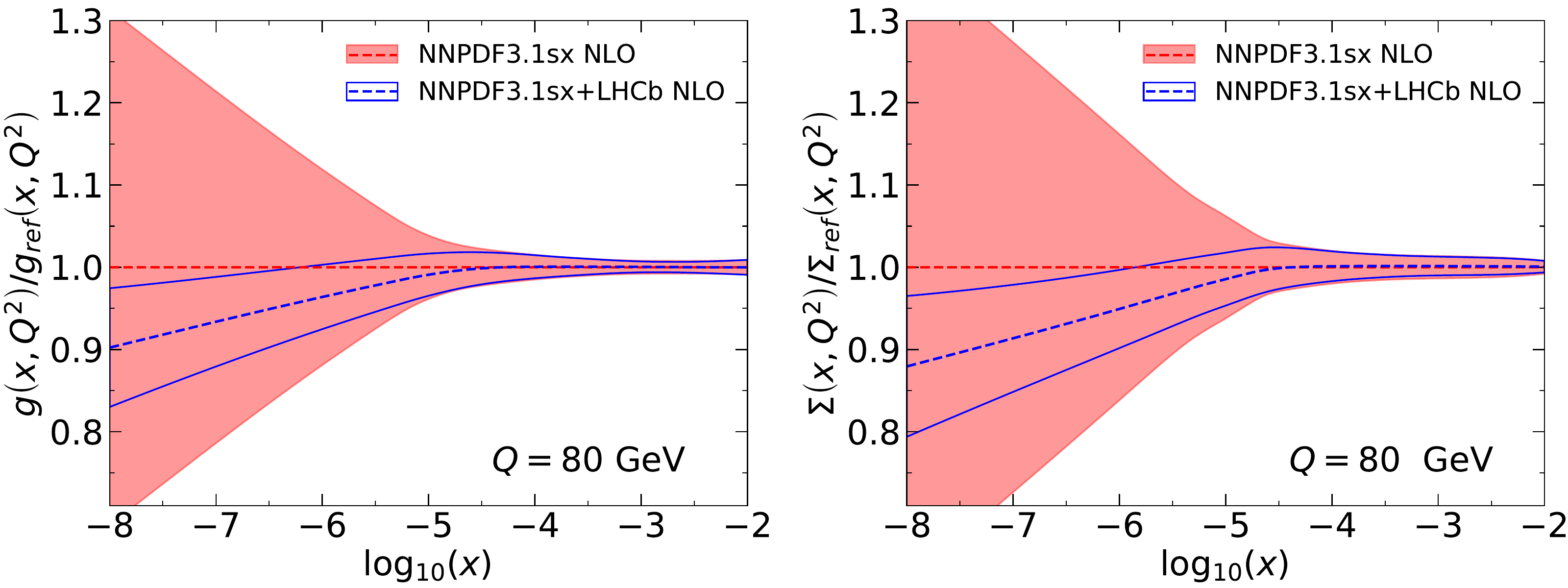}
\caption{The gluon (left) and quark singlet (right) in two variants
  of the NNPDF3.1sx NLO fit with and without the LHCb $D$ meson
  production measurements included.
  Results are shown at $Q=80$ GeV and normalised to the central value
  of the NNPDF3.1sx baseline.
}
\label{fig:NNPDF31sx}
\end{figure}

In order to constrain the proton PDFs in this small-$x$ region, one can exploit the information provided by $D$-meson production from LHCb~\cite{Zenaiev:2015rfa,Gauld:2015yia,Cacciari:2015fta,Gauld:2016kpd,Zenaiev:2019ktw}.
To illustrate the relevance of the LHCb charm data to  constrain the small-$x$ proton PDFs,
Fig.~\ref{fig:NNPDF31sx} compares the gluon and the total quark singlet ($\Sigma=\sum_q \lp f_q+f_{\bar{q}}\rp$) in two variants of the NNPDF3.1sx NLO fit with and without including the LHCb $D$-meson measurements.
As can be seen, for $x\simeq 10^{-7}$ the PDF uncertainties are reduced by up to a factor five thanks to the  constraints from the LHCb data.
In this plot, the similarities between $\Sigma$ and $g$ arise since the latter drives the evolution of the former via the DGLAP equations in the small-$x$ limit~\cite{Ball:2016spl}.

%%%%%%%%%%%%%%%%%%%%%%%%%%%%%%%%%%%%%%%%%%%%%%%%%%%%%%%%%%%%%%%%%%%%%%%%%%%%
%%%%%%%%%%%%%%%%%%%%%%%%%%%%%%%%%%%%%%%%%%%%%%%%%%%%%%%%%%%%%%%%%%%%%%%%%%%%%%

Given the limited amount of experimental information available, the PDFs of bound nucleons,
known as nuclear PDFs (nPDFs), exhibit rather larger uncertainties
as compared to their free-proton counterparts.
Nuclear PDFs can be parametrised as
\be
\label{eq:isoscalarnPDFs2}
f_a^{(N)}(x,Q^2,A) = R_a(x,Q^2,A) \times \lp \frac{Z}{A} f_a^{(p)}(x,Q^2) + \lp 1-\frac{Z}{A}\rp
f_a^{(n)}(x,Q^2)  \rp \, , 
\ee
where $\{ f_a^{(p)} \}$ and $\{ f_a^{(n)} \}$ represent the free proton and neutron PDFs,
$Z$ and $A$ are the atomic and mass  number of the target nucleus, $a$ is a flavour
index that runs over all active quarks and the gluon, and nuclear effects are
encoded in the modification factors $ R_a(x,Q^2,A)$.
Up to now, most studies of neutrino-matter interactions have assumed that nuclear effects
are negligible, that is, that $R_a\simeq 1$ represents
a good  approximation.\footnote{Nuclear effects in high-energy
  neutrino interactions have been recently studied in~\cite{Klein:2020nuk}.}

\begin{figure}[tbp]
\centering 
\includegraphics[width=1.\textwidth]{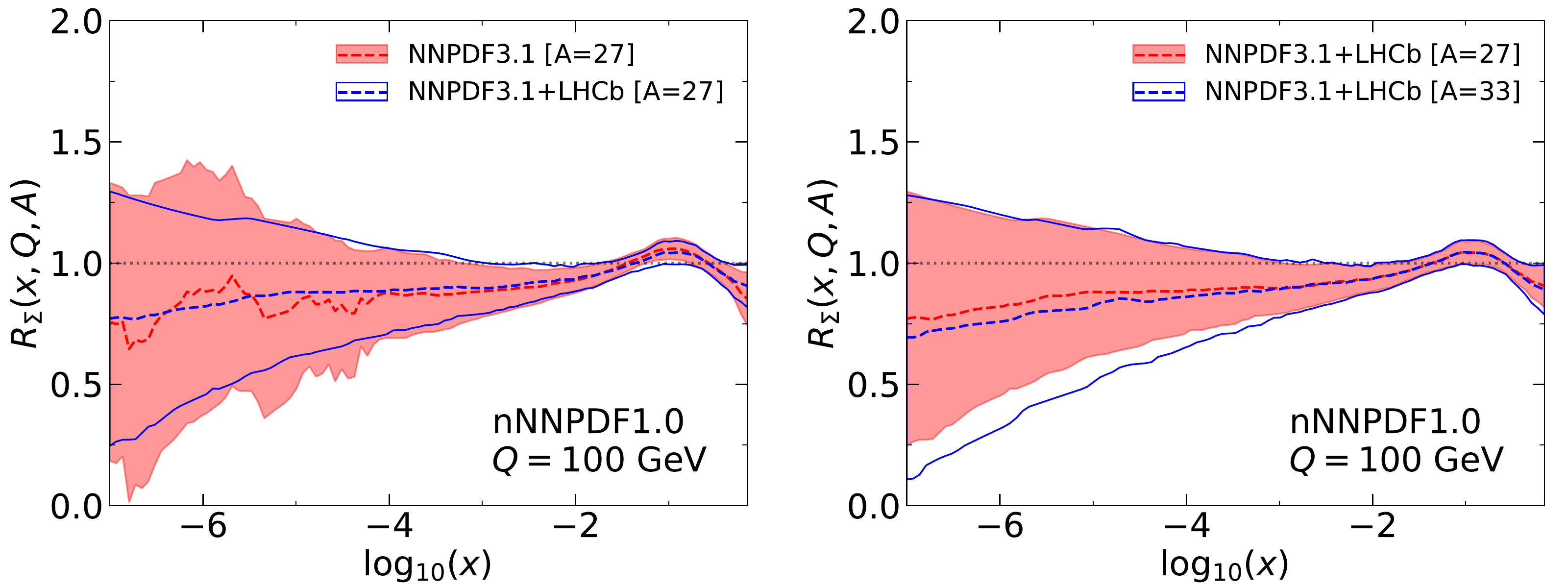}
\caption{The nuclear ratio $R_{\Sigma}$, Eq.~(\ref{eq:isoscalarnPDFs2}),
  evaluated for the singlet
  quark nPDF at $Q=100~\GeV$.
  Left: comparison of two variants of the nNNPDF1.0 fit,
  one using NNPDF3.1 and the other NNPDF3.1+LHCb as the proton boundary
  conditions.
  Right: comparison of $R_{\Sigma}$ between
  $A=27$ and $33$ for the fits based on
  NNPDF3.1+LHCb as boundary condition.
  The bands indicate the 68\% confidence level uncertainties.
}
\label{fig:nNNPDF10-singlet}
\end{figure}

Nuclear PDFs can be either extracted from experimental data
by means of global fits~\cite{AbdulKhalek:2019mzd,deFlorian:2011fp,Eskola:2016oht,Kovarik:2015cma}
in the same manner as in the proton case, or described
with phenomenological models as in~\cite{Kulagin:2004ie}.
To illustrate the typical nuclear PDF uncertainties,
in the left plot of Fig.~\ref{fig:nNNPDF10-singlet} we compare $R_{\Sigma}$ for $A=27$
between two variants of the
nNNPDF1.0 fit~\cite{AbdulKhalek:2019mzd},
one using NNPDF3.1 and the other NNPDF3.1+LHCb as the proton boundary conditions.
In both cases, uncertainties reach up to 50\% in the small-$x$ region.
In the right plot, we compare $R_{\Sigma}$ between
$A=27$ and $A=33$,  which bracket
the average mass number of Earth matter provided
by the composition model used here.
This comparison
demonstrates that the
dependence of $R_\Sigma$ with $A$ is mild for this range of atomic numbers.

%%%%%%%%%%%%%%%%%%%%%%%%%%%%%%%%%%%%%%%%%%%%%%%%%%
%%%%%%%%%%%%%%%%%%%%%%%%%%%%%%%%%%%%%%%%%%%%%%%%%%

\begin{figure}[tbp]
\centering 
\includegraphics[width=1.\textwidth]{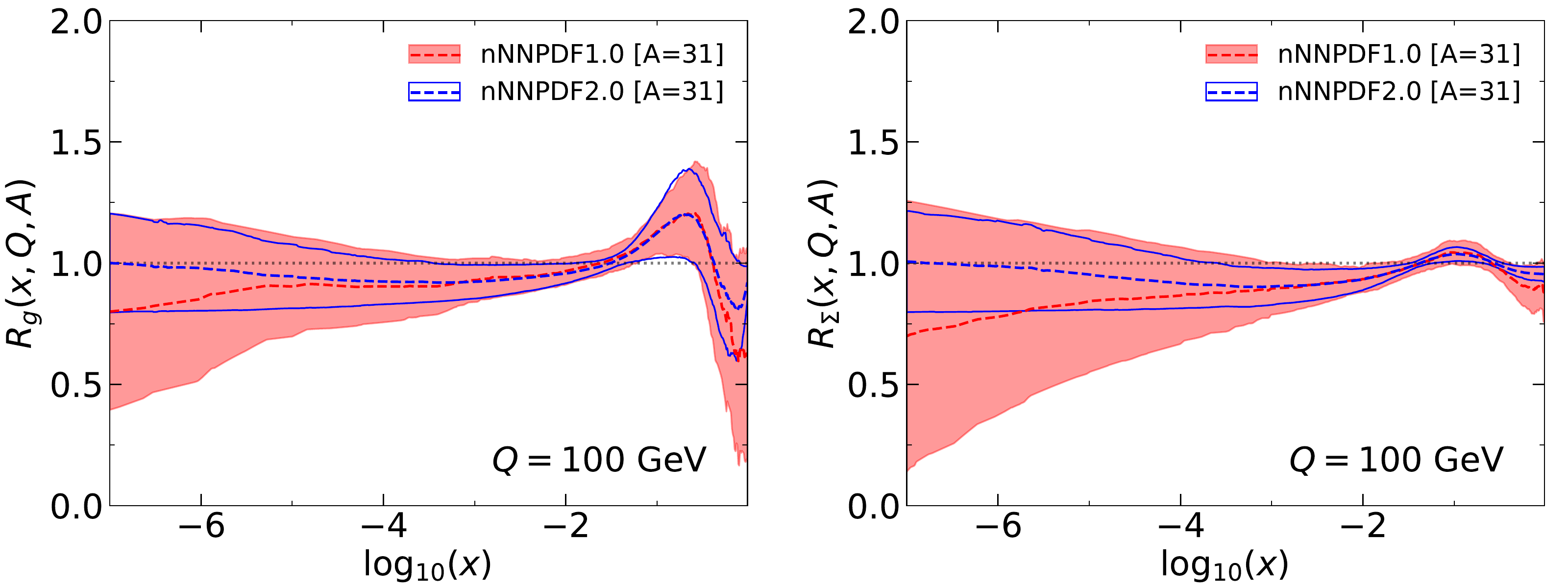}
\caption{Comparison between the
  nNNPDF1.0 and nNNPDF2.0 predictions for the nuclear modification ratios $R_f$
  of the gluon (left) and quark singlet (right) nPDFs for $Q=100~\GeV$ and $A=31$.
  The bands indicate the 68\% confidence level uncertainties.
}
\label{fig:pdfplot-nnpdf10_vs_nnpdf20_68cl}
\end{figure}

It was already noted in~\cite{Bertone:2018dse,Gauld:2019pgt} that nPDF uncertainties
represent one of the dominant theory uncertainties
for high-energy neutrino-nucleus cross sections.
It is thus important to quantify how nuclear PDFs
modify the {\tt NuPropEarth} predictions for the neutrino flux attenuation
presented in Sect.~\ref{sec:results_p}.
In this work, we will estimate these effects by means of the recent nNNPDF2.0 determination~\cite{AbdulKhalek:2020yuc}, which combines NC and CC lepton-nucleus DIS structure functions with the constraints from W and Z production data in proton-lead collisions at the LHC.
Further, in nNNPDF2.0 the positivity of cross-sections in lepton-nucleus and nucleus-nucleus scattering is enforced.
Fig.~\ref{fig:pdfplot-nnpdf10_vs_nnpdf20_68cl} displays the comparison between the nNNPDF1.0 and nNNPDF2.0 predictions for the nuclear modification ratios $R_f$ of the gluon  and quark singlet for $Q=100~\GeV$ and $A=31$.
One observes a marked reduction of PDF uncertainties in nNNPDF2.0, justifying our choice of baseline nPDF for the {\tt NuPropEarth} calculations.

%%%%%%%%%%%%%%%%%%%%%%%%%%%%%%%%%%%%%%%%%%%%%%%%%%
%%%%%%%%%%%%%%%%%%%%%%%%%%%%%%%%%%%%%%%%%%%%%%%%%%

In the same way as in the proton case, it has been shown that
$D$-meson production in proton-lead
collisions from LHCb~\cite{Aaij:2017gcy} can be used
to constrain nPDFs down to $x \simeq 10^{-6}$~\cite{Gauld:2015lxa,Kusina:2017gkz,Eskola:2019bgf}.
However, these studies are based on proton PDF baseline sets that do not
include the corresponding LHCb charm data from proton-proton collisions, and
thus do not treat consistently the small-$x$ PDFs in the $A=1$
and $A=208$ cases.
Ideally, one would need a combined proton and nuclear PDF analysis
with a consistent interpretation of $D$-meson
production from both $\Pp\Pp$ and $\Pp\Pb$ collisions.
More towards the future, several proposed experiments (or extensions to current experiments)
could provide constraints on nuclear PDFs in the small-$x$ region.
One example is the Forward Calorimeter (FoCal) extension
of the ALICE detector, where small-$x$ PDFs can be accessed
via direct photon production at low $p_T$ and forward
rapidities~\cite{Campbell:2018wfu,vanLeeuwen:2019zpz}.
Such an extension would be highly valuable for data collection in future $\Pp A$ runs, as well as in Oxygen collisions which is foreseen~\cite{Citron:2018lsq}.
On that note, recent studies of jet production in this environment~\cite{Huss:2020dwe} have demonstrated such data will also be relevant for understanding energy-loss and nuclear modification effects.
Structure function measurements at proposed lepton-nucleus colliders, such as the Electron
Ion Collider (EIC)~\cite{Boer:2011fh} and the Large Hadron electron Collider (LHeC)~\cite{AbelleiraFernandez:2012cc}, would provide
powerful constraints on nPDFs in the small-$x$ regime.

\subsection{Impact on the attenuation rates}
\label{sec:nPDFattenuation}

Following this discussion on nuclear PDFs at small-$x$, we now move to assess their
impact on the neutrino attenuation rates as compared to the results based on the free-proton PDFs
presented in Sect.~\ref{sec:results_p}.
In this work, the double-differential DIS neutrino cross sections for bound nucleons within
a nuclei $N$ with mass number $A$ are related to the free-proton BGR18 cross sections by means of
the following prescription
\be
\label{eq:nuclearprescription}
\frac{{\rm d}^2\sigma^{\nu N}(x,Q,A)}{{\rm d}x\,{\rm d}y} = \frac{{\rm d}^2\sigma^{\nu p}(x,Q)}{{\rm d}x\,{\rm d}y}\Bigg|_{\rm BGR18} \times
R_\sigma(x,Q,A) \, ,
\ee
in terms of a cross-section nuclear modification factor defined as
\be
R_\sigma(x,Q,A)\equiv \lp \frac{{\rm d}^2\sigma^{\nu N}(x,Q,A)}{{\rm d}x\,{\rm d}y} \Bigg/ \frac{{\rm d}^2\sigma^{\nu N}(x,Q,A=1)}{{\rm d}x\,{\rm d}y}  \rp\Bigg|_{\rm nNNPDF2.0} \, ,
\ee
that is, as the ratio of the double-differential CC or NC DIS cross sections (per nucleon) for
a target with atomic number $A$ to those of an isoscalar $A=1$ target.
This nuclear ratio $R_\sigma$ is evaluated using the nNNPDF2.0 set and
represents the cross section equivalent to the nPDF ratios displayed in
Fig.~\ref{fig:pdfplot-nnpdf10_vs_nnpdf20_68cl}.
When calculating Eq.~(\ref{eq:nuclearprescription}), we compute
the nPDF uncertainties  from the standard deviation over the $\widetilde{N}_{\rm rep}=400$ Monte Carlo
replicas of nNNPDF2.0.
For the rest of the section, when presenting the calculations
with nuclear effects, we neglect the sub-dominant proton PDF errors.

\begin{figure}[t]
\centering 
\includegraphics[width=1.\textwidth]{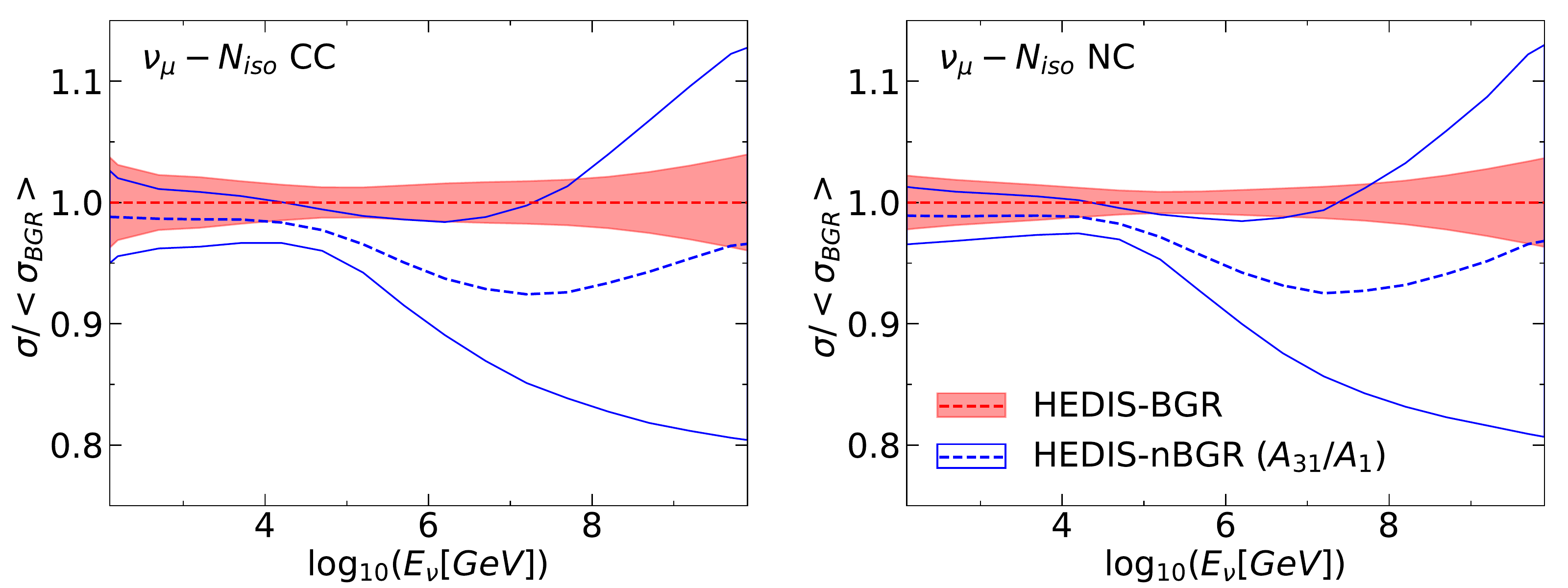}
\caption{The CC (left) and NC (right)
  neutrino-nucleus DIS cross sections with their corresponding
  (n)PDF uncertainties
as a function of $E_{\nu}$, computed with {\tt HEDIS}.
We compare the BGR18 predictions without and with
nPDF effects, the latter evaluated
according to Eq.~(\ref{eq:nuclearprescription}) with nNNPDF2.0,
as ratios to the central value of the free-proton calculation.
}
\label{fig:xsec_nuclear}
\end{figure}

Fig.~\ref{fig:xsec_nuclear} displays the {\tt HEDIS}
calculation of the charged- and neutral-current 
neutrino-nucleon DIS cross sections  with the
corresponding one-sigma (n)PDF uncertainties.
We adopt $A=31$ as a representative atomic mass number for the elements that compose Earth matter.
We compare the results of the BGR18 free-nucleon cross sections with
those with the same calculation extended by the nNNPDF2.0 nuclear effects
using the prescription of Eq.~(\ref{eq:nuclearprescription}), normalised in both cases
to the central prediction of the free-proton case.
The uncertainty band of the ``BGR'' (``nBGR'') calculation has been
obtained from the standard deviation
over the $N_{\rm rep}=60$ ($\widetilde{N}_{\rm rep}=400$) replicas of
the NNPDF3.1+LHCb (nNNPDF2.0) set.

%%%%%%%%%%%%%%%%%%%%%%%%%%%%%%%%%%%%%%%%%%%%%%%%%%
%%%%%%%%%%%%%%%%%%%%%%%%%%%%%%%%%%%%%%%%%%%%%%%%%%

From this comparison, one can observe that, on the one hand,
for neutrino energies $E_{\nu}\lsim 10^4~\GeV$ the two calculations
turn out to be rather similar, implying that nuclear effects can be neglected in this kinematic region.
On the other hand, for $E_{\nu}\gsim 10^4~\GeV$,  we observe two main differences
common to the CC and NC cross sections.
First of all, a suppression of the central value, which follows from the nuclear shadowing
(suppression of the small-$x$ quark and gluon PDFs) displayed by nPDFs.
This cross-section suppression can be as large as 8\% for $E_{\nu}=10^{8}~\GeV$.
Secondly, an increase in the uncertainty band, from
a few percent in BGR18  to up to 15\% once nuclear effects are taken into account.
Such a increase is a direct consequence of the large nPDF uncertainties that affect the small-$x$ region due to the limited constraints, as illustrated in Fig.~\ref{fig:pdfplot-nnpdf10_vs_nnpdf20_68cl}.
We note that results for the nuclear modifications in neutrino-nucleon cross-section computed with the EPPS16 nPDF set~\cite{Eskola:2016oht} have been provided and tabulated in~\cite{Bertone:2018dse,Gauld:2019pgt}. These latter results have smaller uncertainties, and are consistent with the nNNPDF2.0 results presented here.

%%%%%%%%%%%%%%%%%%%%%%%%%%%%%%%%%%%%%%%%%%%%%%%%%%
\begin{figure}[t]
\centering 
\includegraphics[width=.49\textwidth]{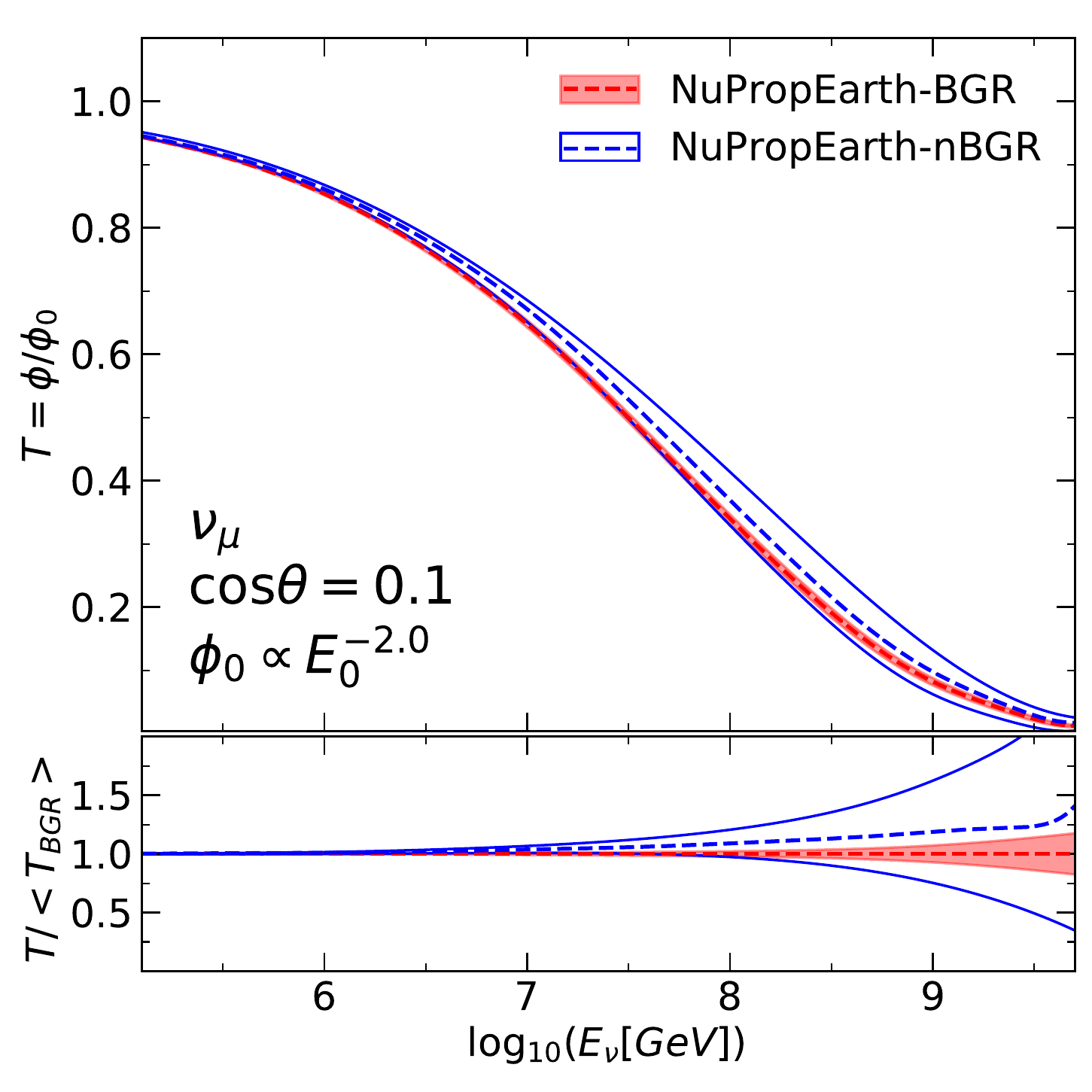}
\includegraphics[width=.483\textwidth]{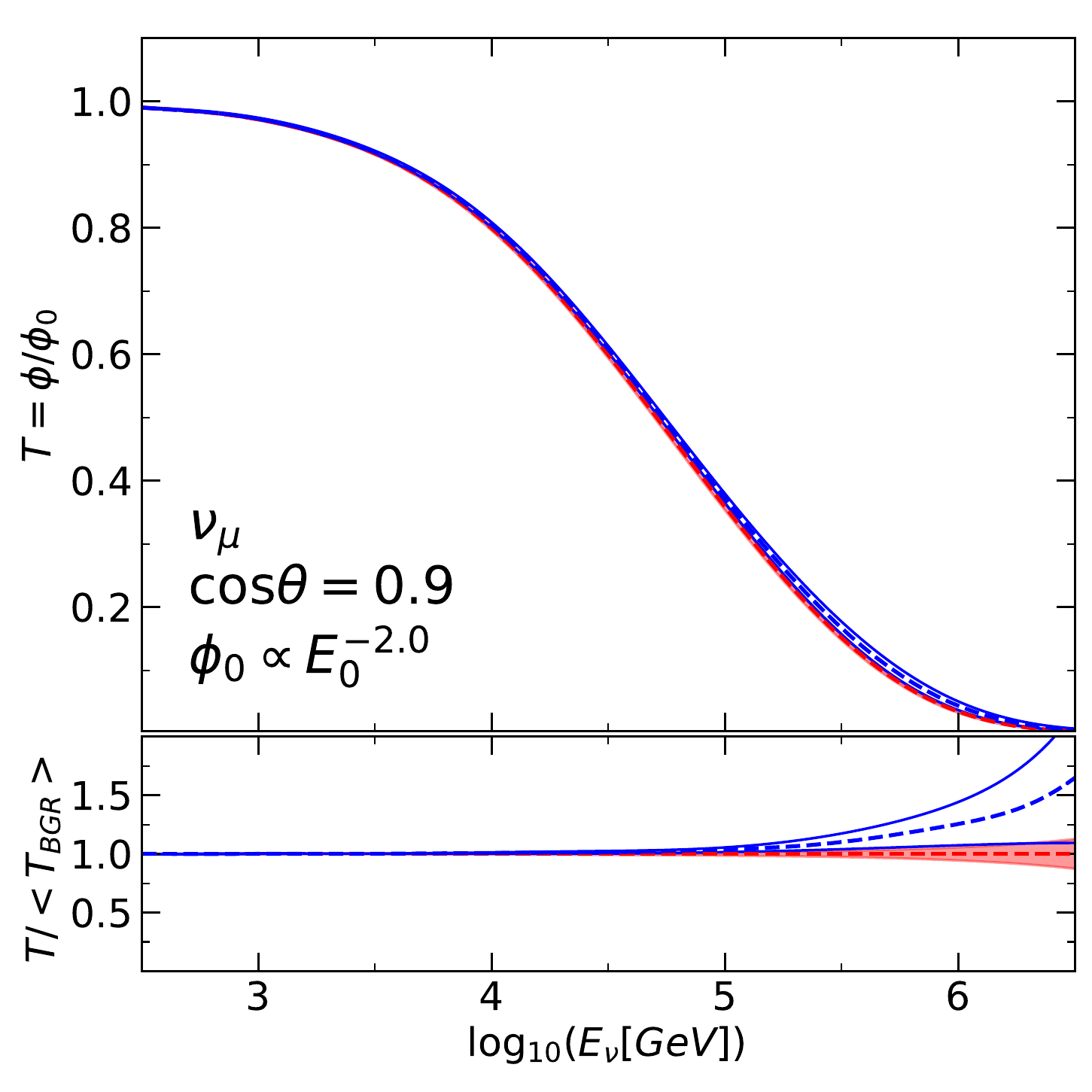}
\includegraphics[width=.49\textwidth]{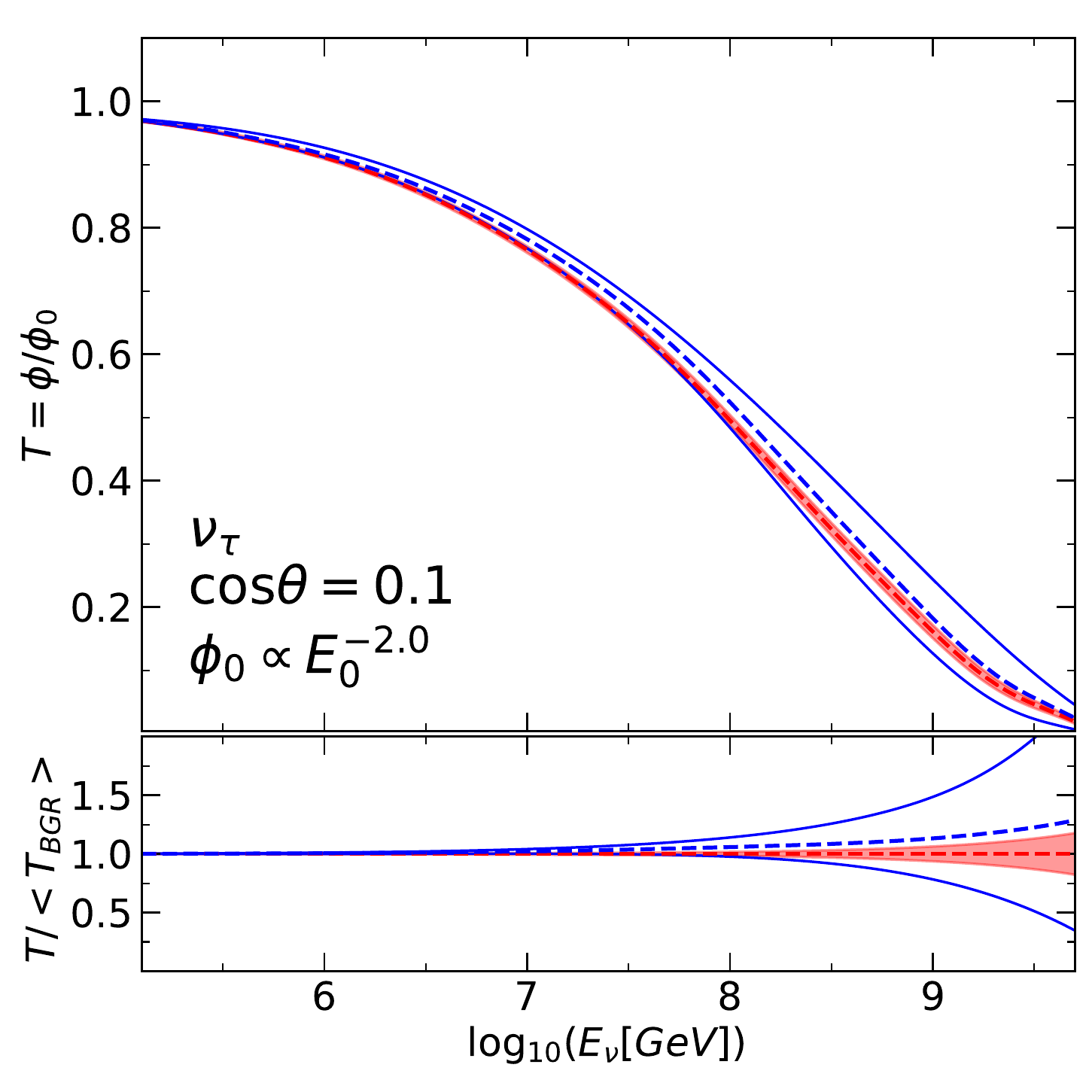}
\includegraphics[width=.483\textwidth]{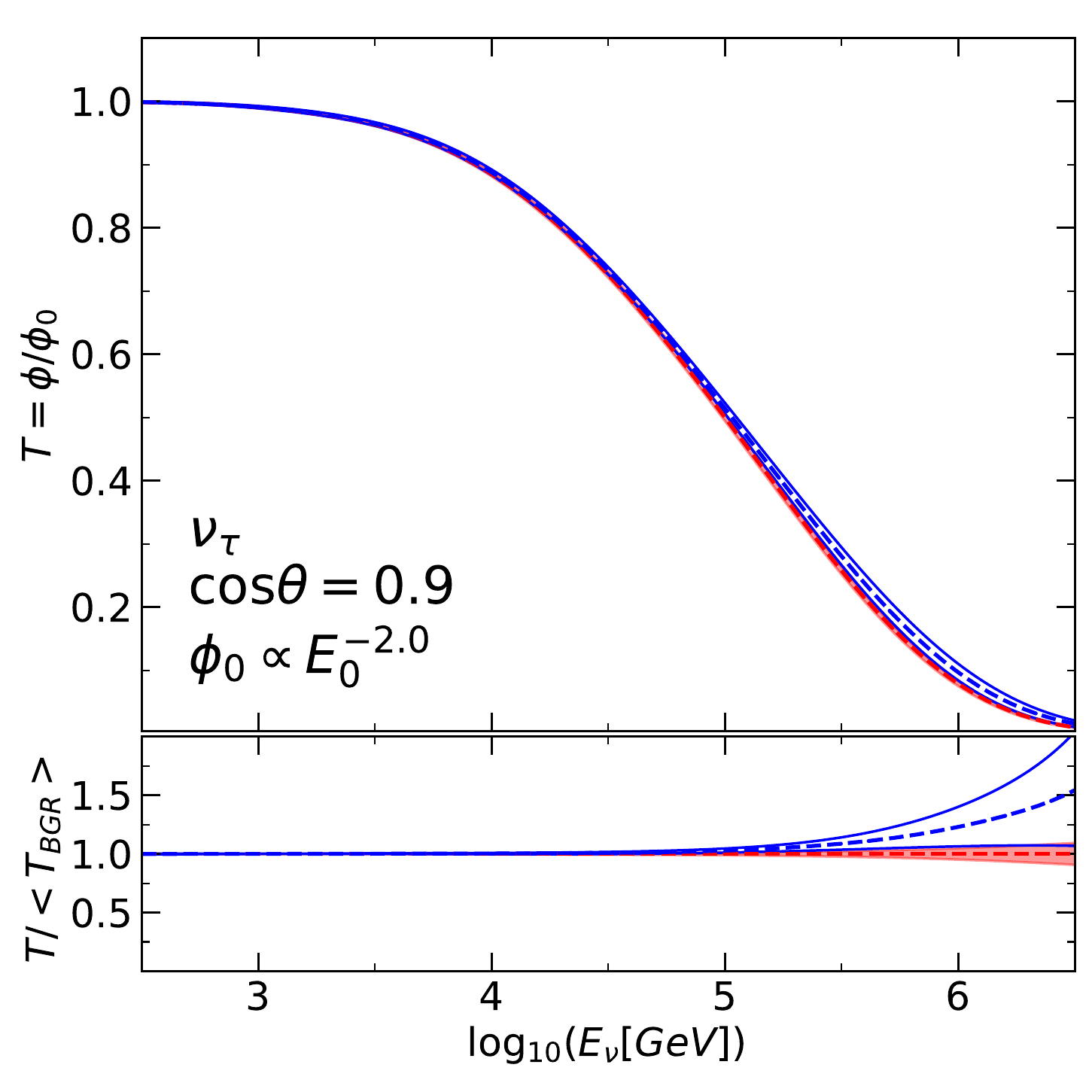}
\caption{The transmission
  coefficients $T(E_\nu)$ for $\nu_\mu$ (upper)
  and $\nu_\tau$ (lower plots) obtained with {\tt NuPropEarth} using the
  BGR18 DIS cross sections without and with
nPDF effects, the latter evaluated
according to Eq.~(\ref{eq:nuclearprescription}), together with the
corresponding PDF uncertainties in each case.
The bottom panel displays the same results normalised to
  the central value of the BGR18 free-nucleon calculation.
}
\label{fig:attnuclear}
\end{figure}
%%%%%%%%%%%%%%%%%%%%%%%%%%%%%%%%%%%%%%%%%%%%%%%%%%

Thanks to the prescription in Eq.~(\ref{eq:nuclearprescription}), we can quantify
how the {\tt NuPropEarth} predictions for the neutrino attenuation rates are
modified once these nuclear PDF effects are taken into account.
In Fig.~\ref{fig:attnuclear} we display the transmission
  coefficients $T$ for muon and tau neutrinos
  obtained with {\tt NuPropEarth} for two values of the incidence angle
  $\cos\theta$ (note that the $E_\nu$ range is different in the left and right plots).
  We compare the predictions obtained by 
using the  BGR18 DIS cross sections without and with
nuclear effects, the latter evaluated according to Eq.~(\ref{eq:nuclearprescription}).
We also display  the corresponding (n)PDF uncertainties in each case, evaluated
in the same way as in Fig.~\ref{fig:xsec_nuclear}.
The bottom panel displays the same results normalised to
the central value of the BGR18 free-nucleon calculation.

From the comparisons in Fig.~\ref{fig:attnuclear}, one finds that accounting for nuclear 
effects in the attenuation calculation has two main implications.
First of all, at large neutrino energies the attenuation rates decrease (the transmission
coefficient $T$ increases) as a consequence
of the shadowing-induced suppression of the neutrino-nucleus cross section displayed
in Eq.~(\ref{fig:xsec_nuclear}).
Secondly, there is a marked increase in the associated PDF uncertainties at large $E_\nu$.
One also finds that nuclear effects are somewhat less important for
larger values of  $\cos\theta$, that is, for neutrinos that propagate
though a sizable amount of Earth matter.
This latter result suggests that the measurement of the interaction
cross sections of energetic Earth-skimming neutrinos would represent
a promising probe of small-$x$ nuclear structure.
For instance, a rececent study estimates that next-generation
UHE neutrino experiments such as GRAND and POEMMA~\cite{Denton:2020jft}
could measure this cross-section for neutrinos of $E_{\nu}\simeq 10^9$ GeV
with 20\% precision.
The comparison with Fig.~\ref{fig:xsec_nuclear} then
demonstrates that such measurements would bring
in important information on models of nuclear PDFs.

\section{Summary and outlook}
\label{sec:summary}

A central ingredient for the interpretation of the results of neutrino telescopes are the theoretical predictions for the attenuation rates that affect the incoming neutrinos as they traverse the Earth towards the detector.
In this work, we have presented predictions for high-energy neutrino propagation through Earth matter.
Our results are based on state-of-the-art calculations for the high-energy neutrino-matter interaction cross-sections, which we have implemented in an updated version of the {\tt HEDIS}
module of the {\tt GENIE} generator.
In addition to the dominant interaction process, DIS off quarks and gluons, we have included all other relevant subdominant channels: (in)elastic scattering off the photon field of nucleons, coherent scattering off the photon field of nuclei, as well as the scattering on atomic electrons via the Glashow resonance.

Our predictions for the neutrino attenuation rates, based on this calculation of the interaction cross-sections, have been obtained by means of a newly developed Monte Carlo simulation framework, {\tt NuPropEarth}.
The flexibility of {\tt NuPropEarth} has allowed us to study in detail the sensitivity of
the predicted attenuation rates with respect to the DIS cross-section model, including nuclear
PDF effects,
the value of the spectral index $\gamma$, the incidence angle, and
the impact of the sub-leading interaction processes.
We have also compared the predictions of {\tt NuPropEarth} with those from related tools, in particular with {\tt nuFATE},
{\tt NuTauSim} and {\tt TauRunner}, and traced back the origin of the observed differences.

We have demonstrated that, given the sensitivity of the attenuation rates
on the high-energy neutrino-nucleus cross sections, adopting precise
calculations for the latter is essential.
The baseline results of {\tt NuPropEarth} are based on  the BGR18 DIS cross sections,
constructed from PDFs constrained at small-$x$ by the LHCb $D$-meson data.
This choice ensures that PDF uncertainties in the attenuation rates
are limited to the few-percent level except for the highest neutrino energies, where
the flux is strongly attenuated.
We also found that nuclear PDF effects have a significant effect
in the predictions for the attenuation rates.
The large uncertainties that affect small-$x$ nPDFs translate into large
PDF uncertainties in the transmission coefficient $T$ for high values of $E_{\nu}$.
Improving the precision of nuclear PDF determinations is therefore important
in order to enhance the robustness of calculations
of the neutrino attenuation rates.
In the longer term, measurements of high-energy
neutrino event rates could be used to test QCD calculations and
models of nucleon and nuclear structure, for instance by means of
Earth-skimming neutrinos with the proposed GRAND experiment.

As neutrino astronomy evolves from discovery to precision science,
improving our modelling of neutrino-matter interactions
is becoming increasingly crucial.
The availability of the {\tt NuPropEarth} package should thus represent a powerful
resource for theorists and experimentalists interested in the modelling 
of high-energy neutrino interactions, propagation, and detection.

\vspace{0.7cm}
\hrule
\vspace{0.7cm}

\noindent
The {\tt NuPropEarth} code can be downloaded from its
public {\tt GitHub} repository:

\begin{center}
\url{https://github.com/pochoarus/NuPropEarth}
\end{center}
Some information about the installation and usage of
the code is provided in  Appendix~\ref{sec:installation}.
{\tt NuPropEarth} will be also integrated in a novel framework called {\tt gSeaGen}~\cite{gseagen} for the modelling of neutrino and lepton event rates for custom detector geometries.
The installation of {\tt NuPropEarth} requires the {\tt HEDIS} module
of {\tt GENIE}.
A modified non-official version of {\tt GENIE} with the {\tt HEDIS} module
integrated can be obtained from:
\begin{center}
  \url{https://github.com/pochoarus/GENIE-HEDIS/tree/nupropearth}
\end{center}
    {\tt HEDIS} will become part of the official {\tt GENIE} distribution
    with its next major release.

    In addition to the code itself, in the {\tt HEDIS} repository
    we also make available pre-computed cross-section tables for a variety
    of theory settings as a function of the neutrino energy.
    We provide these look-up tables available both in XML format
    and as {\tt ROOT} n-tuples.
    In particular, our calculation of the neutrino-matter
    cross sections including all subleading processes
    and presented in Sect.~\ref{sec:results_subleading}
    can be obtained from:
    \begin{center}
  \url{https://github.com/pochoarus/GENIE-HEDIS/tree/nupropearth/genie_xsec}
\end{center}

\acknowledgments

We are grateful to the GENIE collaboration for their help implementing {\tt HEDIS} in {\tt GENIEv3}. We also acknowledge C. Di Stefano for facilitating the integration of {\tt NuPropEarth} in {\tt gSeaGen}. 
We thank Rabah Abdul Khalek and Jake Ethier for providing tailored versions of the nNNPDF1.0 set for this study, and to Valerio Bertone for previous collaboration relevant to this work.
The research of R.~G. is supported by the Dutch Organisation for Scientific Research (NWO) through the VENI grant 680-47-461.
J.~R. is supported by an European Research Council Starting
Grant ``PDF4BSM'' and by the NWO.

\appendix
\section{Dependence of the attenuation rates on the spectral index}
\label{app:spectral}

The results presented in Sect.~\ref{sec:results_p}
assumed that the spectral index $\gamma$ that characterizes
the energy scaling of the incoming neutrino flux, $\phi_0(E_\nu)\propto E_{\nu}^{-\gamma}$,
was fixed to be $\gamma=2$.
The motivation for this choice was that such a value for the spectral index
is consistent with the latest measurements
of the energy dependence of astrophysical
neutrinos as measured by IceCube.
In this Appendix we study how the {\tt NuPropEarth} predictions
vary as the value
of the spectral index $\gamma$ is changed, that is, how the attenuation rates
are modified if the incoming neutrino flux is harder or softer
as compared the assumed baseline scenario.

\begin{figure}[t]
\centering 
\includegraphics[width=.49\textwidth]{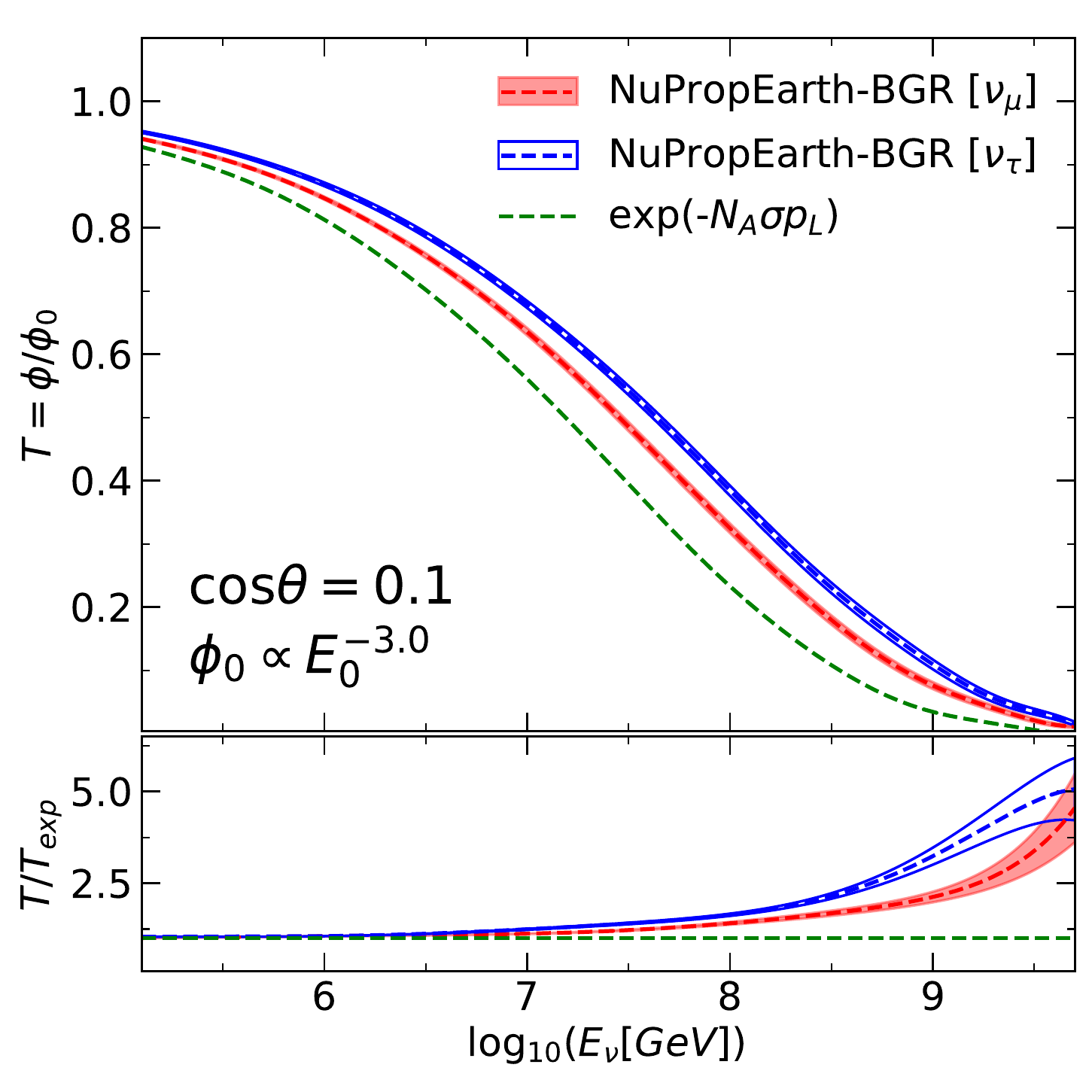}
\includegraphics[width=.49\textwidth]{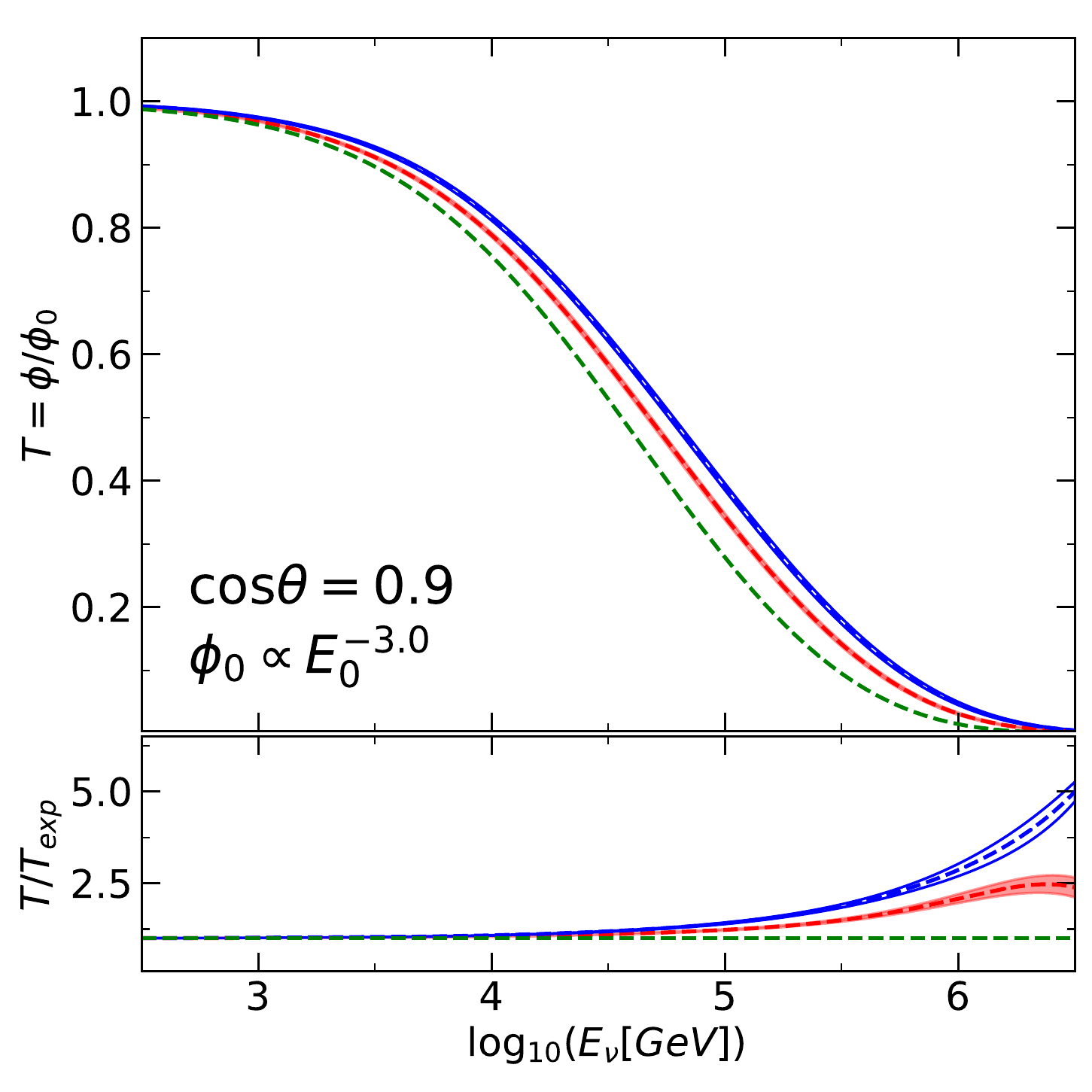}
\includegraphics[width=.49\textwidth]{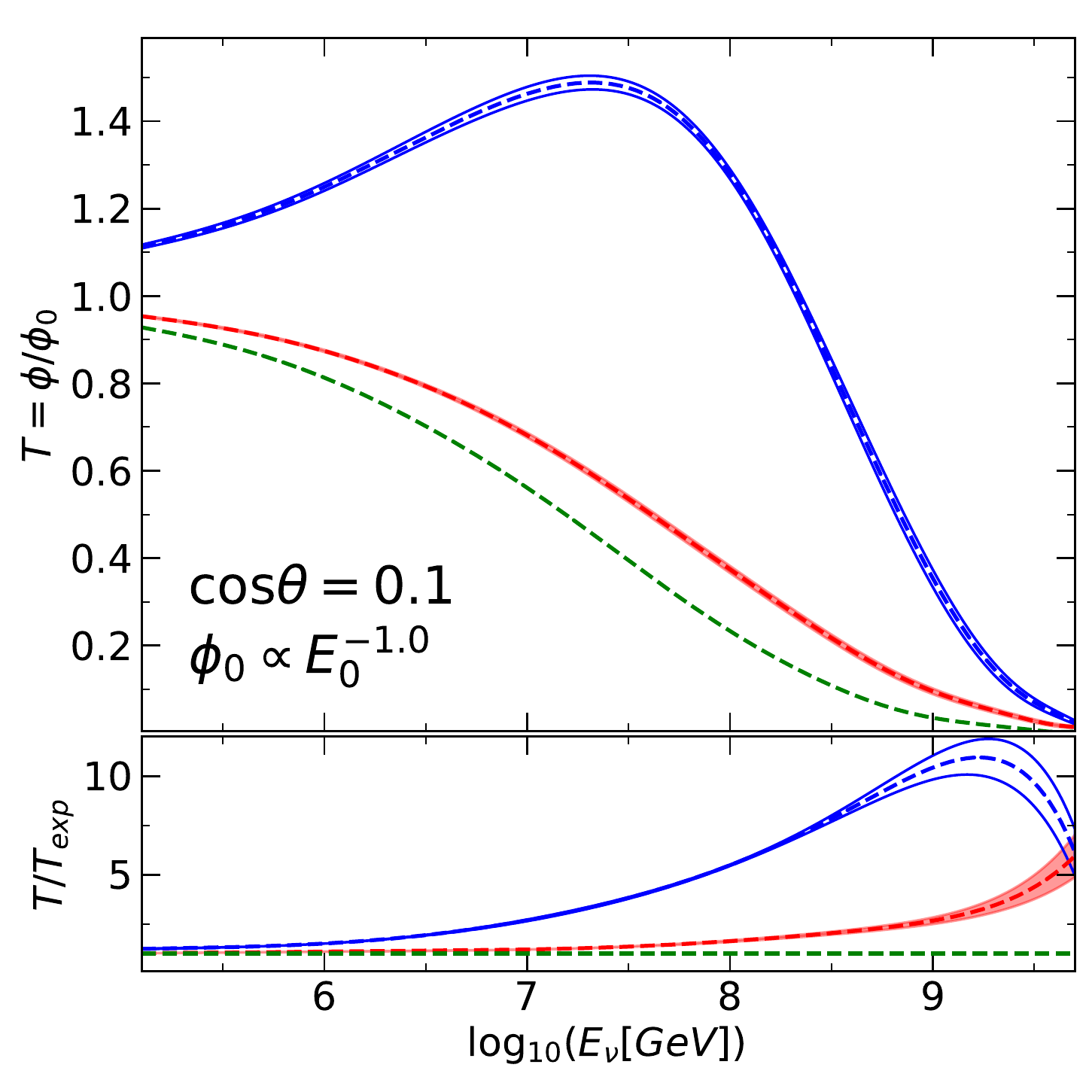}
\includegraphics[width=.49\textwidth]{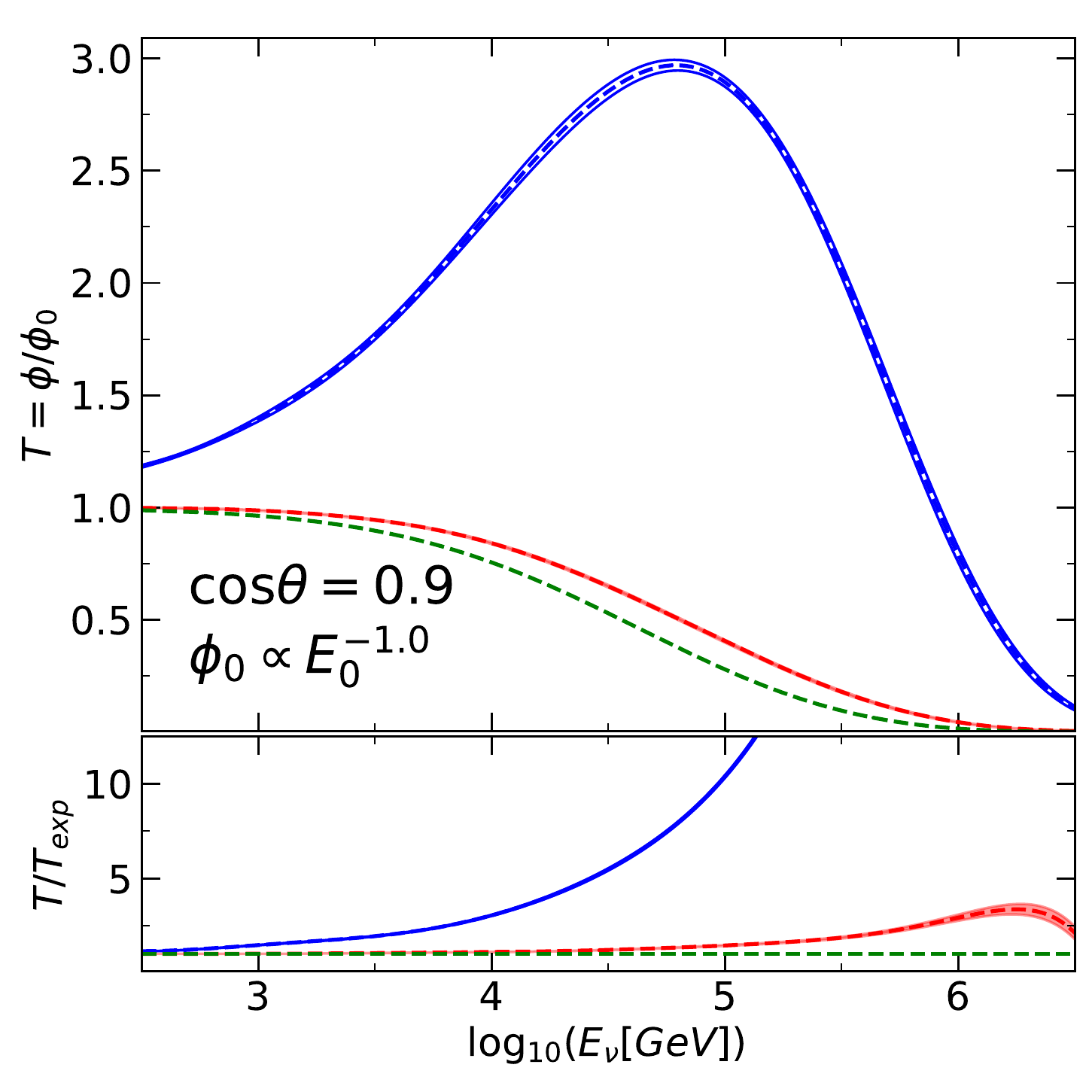}
\caption{The transmission coefficient $T$ for muon and tau
  neutrinos for two values of the incident angle $\cos\theta$.
  The upper (lower) plots display the {\tt NuPropEarth} predictions
  for a spectral index of $\gamma=3$~(1).
  The bottom panels display the ratio to the full absorption case.
}
\label{fig:spectral}
\end{figure}

Fig.~\ref{fig:spectral} displays a similar comparison as that 
in Fig.~\ref{fig:nuall} now for two
different values of the spectral index,  $\gamma=3$
and  $\gamma=1$, rather than the $\gamma=2$ value
using in the baseline.
All other input theoretical  settings of the calculation, such as the DIS cross-section model,
are kept the same as those used in Fig.~\ref{fig:nuall}.
Results are provided for two values of the incident angle
$\cos\theta$, note the different range of $E_\nu$ in each case.
The bottom panels display the corresponding relative PDF uncertainties.
 
%%%%%%%%%%%%%%%%%%%%%%%%%%%%%%%%%%%%%%%%%%%%%%%%%
%%%%%%%%%%%%%%%%%%%%%%%%%%%%%%%%%%%%%%%%%%%%%%%%%

The comparison between Figs.~\ref{fig:spectral} and~\ref{fig:nuall}
reveals that in the case where the incoming astrophysical neutrino flux
is softer ($\gamma=3$), the results
for the attenuation rates are reasonably similar than those
obtained in the baseline scenario ($\gamma=2$).
The situation is different
in the case of a harder incoming neutrino flux ($\gamma=1$),
where the predictions
for $T$ change significantly for tau neutrinos.
Indeed, one finds that in this case there is a range
of values of $E_\nu$ for which the transmission
coefficient becomes larger than unity.
The explanation of this behaviour is that  $\nu_\tau$ regeneration effects
are  markedly enhanced for hard spectra, where high-energy
tau neutrinos are effectively converted into lower-energy
ones, in such a way that the overall neutrino flux for a given range
of energies is larger than the incoming
flux.
As one can see from  Fig.~\ref{fig:spectral}, in the case
of $\cos\theta=0.1$~(0.9) the transmission coefficient for
$\nu_\tau$ is larger than unity for $E_{\nu} \lsim 10^{8.5}$~($10^6$) GeV.
Interestingly, for such a hard spectral index ($\gamma=1$), the interactions with
Earth matter actually enhance, rather than attenuate,
the incoming astrophysical neutrino flux for a wide range of energies.

It is also worth mentioning that the PDF uncertainties
associated to the transmission coefficients are essentially
independent on the value assumed for the spectral index $\gamma$.
The same considerations apply to the case where nuclear corrections
are taken into account.

\section{Recommended settings for the neutrino-nucleon cross-section}
\label{app:DIS}

Throughout this paper, we have displayed results for the neutrino-nucleon cross sections obtained with {\tt HEDIS}, which use the theoretical model for the computation of the DIS structure functions (input PDF sets and theoretical settings) as presented in the CMS11~\cite{CooperSarkar:2011pa} and BGR18~\cite{Bertone:2018dse} calculations.
The comparison of these two models is given in Fig.~\ref{fig:xsec_nucleon}, where the implementation of CMS11 and BGR18 calculations within {\tt HEDIS} is validated (the central values and PDF uncertainties of the original NLO calculations are reproduced).

There are two important points which require comment.
Firstly, in Fig.~\ref{fig:xsec_nucleon} the CMS11 calculation is compared to the {\tt HEDIS} implementation of the BGR18 calculation in the region $E_{\nu} \lesssim 5\times10^3~\GeV$ (below the recommended use of BGR18). In this region both CC and NC computations differ by up to $\approx15\%$.
The second is that large differences are observed for the CC process across all shown energies, this ranges from $\approx 6\%$ at $E_{\nu} = 5\times10^3~\GeV$, and exceeds 10\% for $E_{\nu} \gtrsim 10^7~\GeV$. 
We clarify the origin of the observed differences in what follows. A recommendation  (where appropriate) for future neutrino-nucleon DIS predictions is also provided---this recommendation has been made default within {\tt HEDIS}.

% Low Q2 integration
\subsection{The role of the low-$Q^2$ region}
As noted above, there are large differences observed between the CMS11 and {\tt HEDIS} implementation of the BGR18 calculation in the energy range of $E_{\nu} \lesssim 5\times10^3~\GeV$.
In this energy range, the predicted neutrino-nucleon cross-section is small, and the impact on the attenuation rate is negligibly small (see Fig.~\ref{fig:csms}).
However, as {\tt HEDIS} allows the user to compute the neutrino-nucleon cross-section in this region, we comment on the validity of making predictions in this regime.

\begin{figure}[tbp]
\centering 
\includegraphics[width=1.\textwidth]{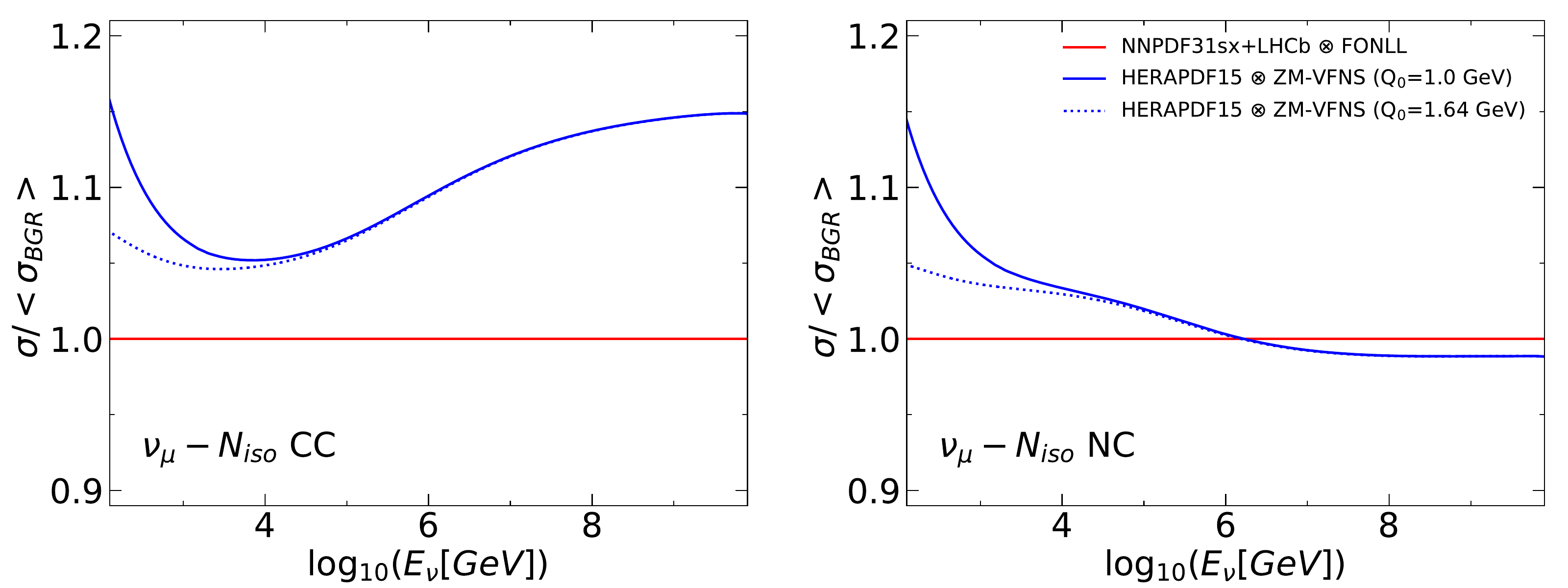}
\caption{Same as Fig~\ref{fig:xsec_nucleon}, now comparing the central values of the CMS11 and BGR18
  calculations with the default theory settings.
  In the former case, we use either
$Q_{\rm min} = 1.64~\GeV$ or $1.0~\GeV$ as the lower limit of the $Q^2$ integration.}
\label{fig:HERA_lowQ2}
\end{figure}

In the energy range of $E_{\nu} \lesssim 5\times10^3~\GeV$, the main difference between the predictions shown in Fig.~\ref{fig:xsec_nucleon} is the choice of $Q_0$ of the associated input PDF set, which determines the lower bound of the $Q^2$ integration entering the inclusive cross-section prediction obtained from integrating Eq.~\eqref{eq:CCxsec}. The relevant settings of the input PDF sets for these two models are summarised in Table~\ref{tab:DIS}.
\begin{table}[ht]
  \caption{Summary of relevant input settings for the CMS11 and BGR18 DIS models.}
  \renewcommand{\arraystretch}{1.45}
\begin{center}
  \begin{tabular}{c|c|c|c}
\toprule
Model & PDF set & $Q_0~[\GeV]$ & $x_{\rm min}$ \\ \midrule
CMS11 & HERAPDFNLO1.5 & $1.0$ & $10^{-8}$  \\
BGR18 & NNPDF3.1sx+LHCb & $1.64$ & $10^{-9}$  \\
\bottomrule
\end{tabular}
\end{center}
\label{tab:DIS}
\end{table}%
As noted in Sect.~2.2 of Ref.~\cite{Bertone:2018dse}, the contribution to the inclusive cross section for $E_{\nu} \lesssim 5\times10^3~\GeV$ from the low-momentum transfer (low-$Q^2$) region is non-negligible.
To better quantify this point, in Fig.~\ref{fig:HERA_lowQ2} we compare the cross-section predictions obtained with CMS11 and BGR18 models, where we have included the predictions from CMS11 with either $Q_{\rm min} = 1.64~\GeV$ or $1.0~\GeV$ as the lower limit of the $Q^2$ integration.
This demonstrates the relevance of the low-momentum region to the inclusive cross section for $E_{\nu} \lesssim 5\times10^3~\GeV$.
For neutrino energies above $E_{\nu} \simeq 5\times10^3~\GeV$, the contribution of this low-$Q^2$ region to the inclusive cross section is small ($\approx 1\%$ or below).
It is for this reason that the BGR18 model can only be considered reliable (as explicitly discussed in~\cite{Bertone:2018dse}) for $E_{\nu} \gtrsim 5\times10^3~\GeV$, as indicated by the red solid filled region in Fig.~\ref{fig:xsec_nucleon}.

%%%%%%%%%%%%%%%%%%%%%%%%%%%%%%%%%%%%%%%%%%%%%%%%%%%%%%%%%%%%%%%%%%%%%%%%%%%%%%
%%%%%%%%%%%%%%%%%%%%%%%%%%%%%%%%%%%%%%%%%%%%%%%%%%%%%%%%%%%%%%%%%%%%%%%%%%%%%%
%%%%%%%%%%%%%%%%%%%%%%%%%%%%%%%%%%%%%%%%%%%%%%%%%%%%%%%%%%%%%%%%%%%%%%%%%%%%%%%%%%
%%%%%%%%%%%%%%%%%%%%%%%%%%%%%%%%%%%%%%%%%%%%%%%%%%%%%%%%%%%%%%%%%%%%%%%%%%%%%%

Clearly, a description of this low-$Q^2$ is region is required to reliably predict the inclusive cross section (with percent level accuracy) for neutrino energies in the range $E_{\nu} \lesssim 5\times10^3~\GeV$. 
This is however a kinematic region where the framework of collinear factorisation starts to break down, and hence also the computation of structure functions according to Eq.~\eqref{eq:CCxsec}.
Practically, this breakdown is observed as a deterioration in the value of $\chi^2/n_{\rm dat}$ of a collinear PDF fit when data in the low-$Q^2$ region is included. This effect has been observed by several PDF fitting groups---see for example Fig.~2 and 3 of Ref.~\cite{Abt:2016vjh}.
As a result, it is difficult to determine how reliable the predictions of the structure functions $F_i^{\nu N}(x,Q^2)$ within the range of $Q^2\in[1.0,(1.64)^2]~\GeV^2$ are.
Also, the contribution from integration region of $Q^2\leq 1~\GeV^2$ is still absent in this approach.

As {\tt HEDIS} accommodates the use of any input PDF set, the user is free to choose which inputs they wish to use for the calculation in this low-energy region.
However, care must be taken when interpreting the results for the reasons discussed above.
In the future, improved predictions for this low-$Q^2$ region may be assessed through a combined fit of collinear PDFs which smoothly transitions onto a fit of structure functions for $Q \lsim 2~\GeV$, or a model-dependent extrapolation as in~\cite{Bodek:2002ps}.

%In the meantime, we would recommend to use the internal {\tt GENIE} model for $E_{\nu} \lesssim 5\times10^3~\GeV$, based on the  Bodek-Yang calculation~\cite{Bodek:2002ps} of low-$Q^{2}$ structure functions.
%
%In the next iteration of the {\tt HEDIS} framework we plan to provide the structure functions combining both the Bodek-Yang prescription (or a suitable variant) for $Q \le 2~\GeV$ and an input PDF set such as NNPDF3.1sx+LHCb.
%
%This way, a single  single set of DIS structure functions will be able to describe the neutrino-nucleon cross sections from $E_\nu \simeq 10~\GeV$ up to $E_{\nu}\simeq 10^{10}~\GeV$.

%  Q2 = 10^2.5 = 320 GeV2 and x > 0.1
\subsection{Dependence on the input PDF set}

\begin{figure}[tbp]
\centering 
\includegraphics[width=1.\textwidth]{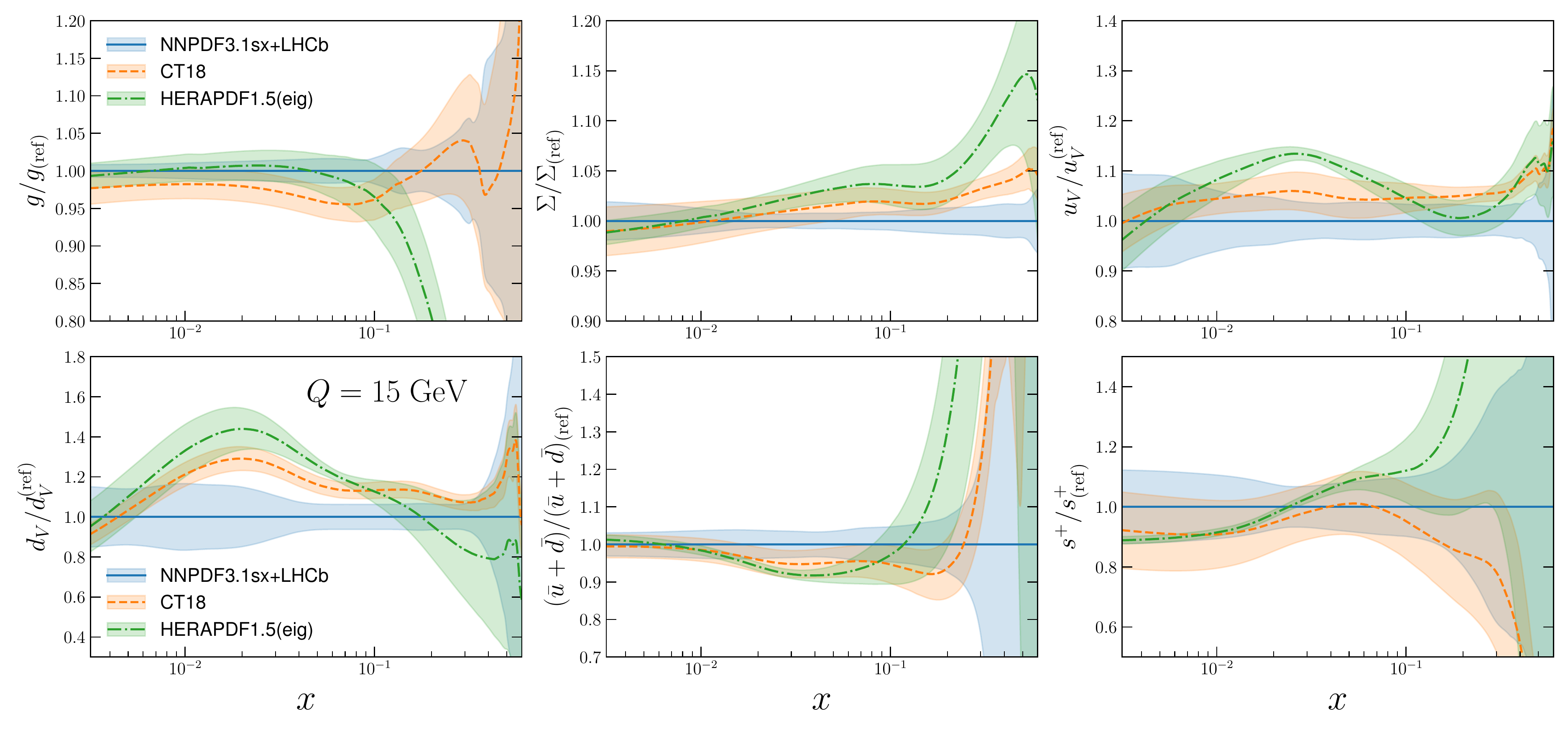}
\caption{Comparison between NNPDF3.1sx+LHCb (used as reference for the normalisation), CT18, and HERAPDF1.5,
  with the corresponding one-sigma uncertainty bands.
  We display the gluon, the total quark singlet, the up and down quark valence distributions,
  the up and down quark sea, and the total strangeness at $Q=15~\GeV$ in the $x$ region
  relevant for neutrino-nucleus scattering with $E_\nu \simeq 5 \times 10^3~\GeV$.
  The HERAPDF1.5 error band includes only the contribution from the experimental
  uncertainties but not from the model and parametrisation errors.
}
\label{fig:pdfcomp}
\end{figure}

As indicated in Table~\ref{tab:DIS}, another difference between the CMS11 and BGR18 calculations
is the input PDF set, HERAPDF1.5 and NNPDF3.1+LHCb respectively.
While the former is restricted to HERA structure functions, the latter is a global analysis which also includes the constraints from fixed-target DIS and weak gauge-boson production at colliders, among other processes.
For neutrinos of energy $E_\nu \simeq 5 \times 10^3~\GeV$, it can be shown that the inclusive cross-section is dominated by the region around $Q\sim15~\GeV$ and $0.04 \lsim x \lsim 0.6$.
The comparison between these two PDF sets in this kinematic region is displayed in Fig.~\ref{fig:pdfcomp}, where the corresponding results for the recent CT18 global analysis~\cite{Hou:2019efy} are also shown.
Specifically, we show the following (combinations of) PDFs: the gluon; the total quark singlet; the up and down quark valence distributions, the up and down quark sea, and the total strangeness. 

%%%%%%%%%%%%%%%%%%%%%%%%%%%%%%%%%%%%%%%%%%%%%%%%%%
%%%%%%%%%%%%%%%%%%%%%%%%%%%%%%%%%%%%%%%%%%%%%%%%%%

Before studying the differences in the PDF sets, it is useful to note the PDF combinations which enter the leading-order structure function predictions for an isoscalar target---see Eq.~(4) and Eq.~(5) of~\cite{CooperSarkar:2011pa}.
For example, the prediction for $F_2$ is proportional to the quark singlet combination, while the leading contribution to $F_3$ is from the valence content.
From the comparisons in Fig.~\ref{fig:pdfcomp} one can observe that in general the quark PDFs are higher in HERAPDF1.5 as compared to NNPDF3.1+LHCb.
For instance, the quark singlet is 5\% larger at $x=0.2$, and the up valence
distribution is 10\% larger at $x=0.04$.
Combined with the above information, the $\simeq 5\%$ difference found between the CMS11 and BGR18 calculations in the region $E_{\nu}\sim 5\times 10^{3}~\GeV$ can be traced back to the use of different input PDFs.

Given that in the $x\gsim 0.04$ region the HERAPDF1.5 determination has limited information on quark flavour separation (coming from the CC structure functions at high $Q^2$), and that NNPDF3.1+LHCb includes the $D$-hadron data which constrains the small-$x$ gluon PDF (which is relevant at large neutrino energies), we consider the global determination NNPDF3.1+LHCb to be a more reliable input for the neutrino-nucleon cross-section for $E_{\nu} \gtrsim 5\times10^{3}~\GeV$.
%
%The relevance of a global dataset to pin down the quark and antiquark flavour separation in the valence region is emphasized by the comparison with the CT18 analysis, which displays an improved agreement with NNPDF3.1sx+LHCb within uncertainties.

\subsection{Mass effects in top quark production in CC scattering}
The other major difference between the CMS11 and BGR18 calculations is the treatment of heavy quark mass effects, which is most relevant for the description of top quark production in charged-current scattering.

In the CMS11 model, the top quark contribution is included as part of a
zero-mass variable flavour number (ZM-VFN) scheme, where all quarks are treated as massless. 
As part of this calculation, a threshold constraint is also included with the requirement $W^2=Q^2(1/x-1) \geq m_t^2$, which that ensures the mass of the out-going hadronic system exceeds the top quark production threshold.

In the BGR18 calculation top quark mass effects are instead included as part of the FONLL general-mass VFN scheme as implemented within {\tt APFEL} and described in~\cite{Forte:2010ta,Ball:2011mu}.
This is the same mass scheme which is used for the extraction of the NNPDF3.1 set~\cite{Ball:2017otu}.
In the FONLL formalism, the prediction obtained with the massive charm quark structure function ($n_f = 3$) is improved through the inclusion higher-order terms which are resummed as part of the massless computation---see Eq.~(28) onwards of Ref.~\cite{Ball:2011mu} for details. 
The impact of bottom and top quark mass effects are included via single-mass contributions to the structure function in the $n_f = 3$ scheme.
As a consequence, the impact of the leading logarithmic $b$-quark initiated contribution to the sub-process $\nu b \to \ell t$ is absent within this implementation.

%%%%%%%%%%%%%%%%%%%%%%%%%%%%%%%%%%%%%%%%%%%%%%%%%%
\begin{figure}[tbp]
\centering 
\includegraphics[width=1.\textwidth]{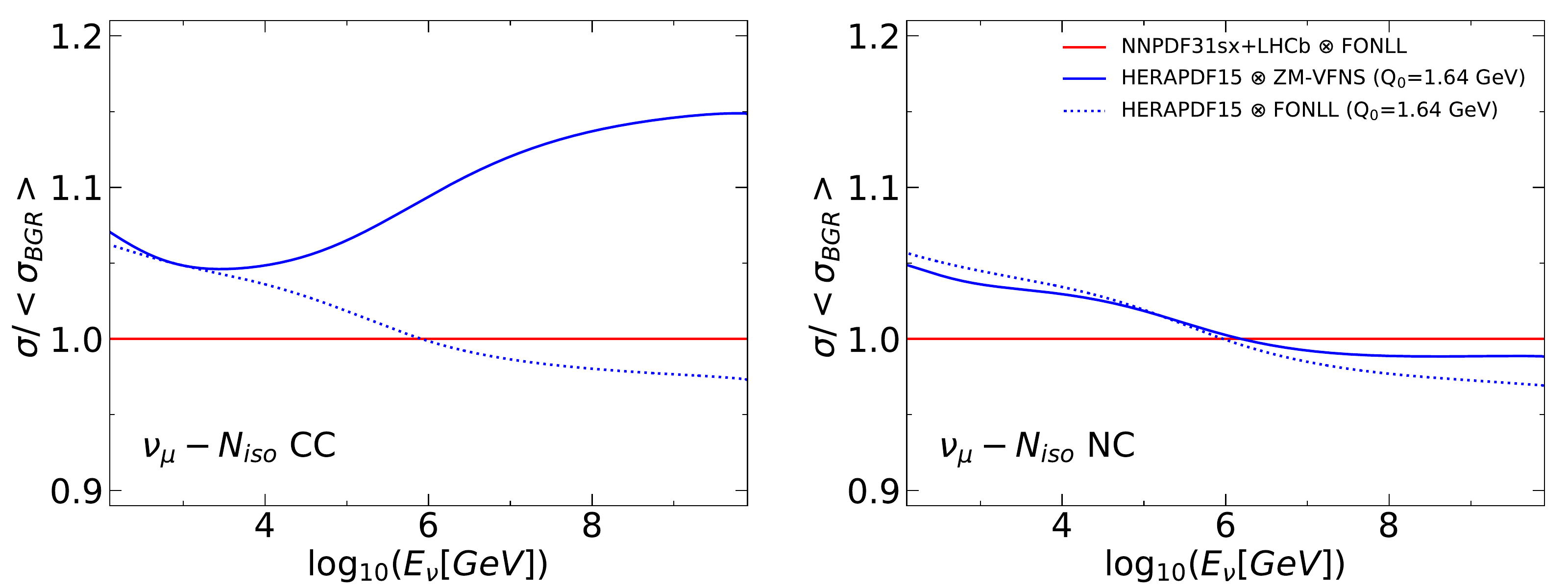}
\caption{Same as Fig.~\ref{fig:HERA_lowQ2} now comparing
  the BGR18 calculation with the CMS11 default (based on the ZM-VFN scheme)
  and with a calculation which also uses HERAPDF1.5 as input PDF
  but now with structure functions evaluated in the FONLL
  general-mass scheme.
}
\label{fig:HERA_pdf}
\end{figure}
%%%%%%%%%%%%%%%%%%%%%%%%%%%%%%%%%%%%%%%%%%%%%%%%%%

In Fig.~\ref{fig:HERA_pdf} we compare the BGR18 calculation with the CMS11 default (based on the ZM-VFN scheme) and with a calculation which uses HERAPDF1.5 as input PDF but with structure functions evaluated in the FONLL general-mass scheme.
One finds that, once a common treatment of heavy quark mass effects is adopted, the predictions for the CC cross-sections are in much better agreement for $E_{\nu}\gsim 10^4~\GeV$ (and as discussed before, the differences for $E_{\nu}\lsim 10^4~\GeV$ are explained by the value of $Q_0$ and the input PDFs).
%
%Indeed, the FONLL-based calculations of the CC cross-section agree within a few percent for up to $E_\nu\simeq 10^{10}~\GeV$, with the remaining differences traced back also to the input PDFs.
%
In the case of NC scattering, mass effects are much less important and the CMS11 calculation is only mildly modified when structure functions are evaluated with the FONLL scheme.

\begin{figure}[tbp]
\centering 
\includegraphics[width=1.\textwidth]{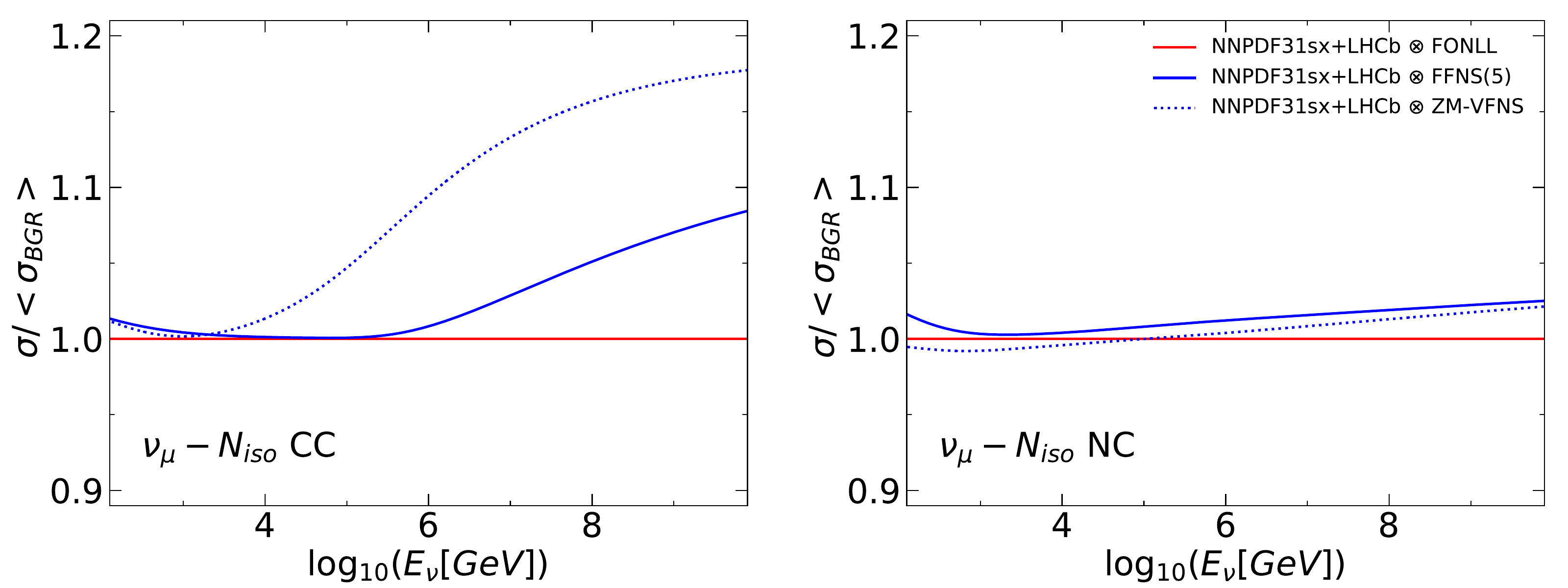}
\caption{Same as Fig.~\ref{fig:HERA_lowQ2} now comparing
  the predictions for the NC and CC inclusive DIS cross-sections
  with NNPDF3.1sx+LHCb as input PDF and three different heavy quark mass schemes:
  the massless calculation (ZM-VFN), the FONLL calculation
  as implemented in~\cite{Bertone:2018dse}, and a fully massive calculation
  of top quark production in the $n_f=5$ scheme.
  See text for more details.
}
\label{fig:top}
\end{figure}

Fig.~\ref{fig:top} then displays a comparison of the predictions for NC and CC inclusive DIS cross-sections using NNPDF3.1sx+LHCb as input PDF for three different heavy quark mass schemes: the massless scheme, the FONLL general-mass scheme as implemented in~\cite{Bertone:2018dse}, and a fully massive calculation of top quark production in the $n_f=5$ FFN scheme.
In the latter case, top quark mass effects are accounted for exactly, all other quarks are treated as massless, and all $b$-quark initiated contributions are included.
One can observes how the ZM-VFNS prediction greatly overestimates the rate of top-quark production. This is a consequence of the approximate treatment of the heavy quark threshold at small $Q^2$  values.
At leading order, the relation between $x$ (the kinematic DIS variable) and  $\chi$
(the argument of the PDF in a massive computation) is $\chi = x\left(1 + m_t^2/Q^2\right)$.
This means that requiring a production threshold of $Q^2(1/x-1) \geq m_t^2$ leads to an over-estimation of the kinematically allowed region of the top quark.
In contrast, the FONLL prescription underestimates the cross-section by a few percent for $E_{\nu}\sim 10^7~\GeV$ and up to 8\% at $E_{\nu}\sim 10^{10}~\GeV$, as a result of the missing leading-logarithmic correction mentioned above. 

While the latter issue can in principle be addressed, for example by including two-quark mass contributions to the FONLL structure functions, our current recommendation is that the best predictions of the neutrino-nucleon cross section in CC scattering are those obtained with the FFNS ($n_f = 5$) prediction of the structure functions.
An inconsistency is introduced in this approach if VFNS inputs PDF sets are used. However, as top quark mass effects are much more relevant for the inclusive cross-section at high energies, this is a preferable option.
This has been made the default for the CC process within {\tt HEDIS}.
We refer the reader to~\cite{Barge:2016uzn} for a discussion on the feasibility of measuring exclusive top-quark production with IceCube.

%%%%%%%%%%%%%%%%%%%%%%%%%%%%%%%%%%%%%%%%%%%%%%%%%%
%%%%%%%%%%%%%%%%%%%%%%%%%%%%%%%%%%%%%%%%%%%%%%%%%%
\section{Installation and usage of {\tt NuPropEarth}}
\label{sec:installation}

In this appendix we provide some brief instructions concerning
the installation and usage of {\tt NuPropEarth}.

\paragraph{Installation.}
As mentioned above,
the code can be downloaded from its {\tt GitHub} repository:
\begin{center}
\url{https://github.com/pochoarus/NuPropEarth}
\end{center}
\noindent
which also contains the corresponding installation instructions.
The main dependencies of {\tt NuPropEarth}
are {\tt GENIE}v3 with the {\tt HEDIS} module included, {\tt TAUOLA++}~\cite{DAVIDSON2012821} and {\tt TAUSIC}~\cite{ANTONIOLI1997357}.
A modified non-official version of {\tt GENIE}v3 with  {\tt HEDIS}
integrated can be obtained from:
\begin{center}
  \url{https://github.com/pochoarus/GENIE-HEDIS/tree/nupropearth}
\end{center}
We note that {\tt HEDIS} will become part of the official {\tt GENIE} distribution
    with its next major release.
The installation of {\tt GENIE}/{\tt HEDIS} 
relies on a number of external packages:
{\tt ROOT6}, {\tt Pythia6}~\cite{Sjostrand:2007gs}, {\tt LHAPDF6}~\cite{Buckley:2014ana},
and the PDF evolution library {\tt APFEL}~\cite{Bertone:2013vaa,Carrazza:2014gfa}.

Once these external packages have been installed, one can proceed with the
installation of {\tt GENIE3}/{\tt HEDIS} and {\tt NuPropEarth} by using the following commands:\\

\noindent
{\tt
cd \$GENIE \\
./configure --enable-lhapdf6 --enable-apfel --with-lhapdf6-inc=\$LHAPDF/include \\ --with-lhapdf6-lib=\$LHAPDF/lib --with-apfel-inc=\$APFEL/include \\--with-apfel-lib=\$APFEL/lib\\
make \\
cd \$NUPROPEARTH \\
make \\
}

\paragraph{Usage.}
In order to run {\tt NuPropEarth}, one should enter the following commands:\\

\vspace{-0.3cm}
\noindent
{\tt 
NUMBEROFEVENTS="1e5" \\
NUPDG="14" \# 14=numu, 12=nue, 16=nutau, -14=anumu, -12=anue, -16=anutau, \\
COSTHETA="0.1" \\
TUNE="GHE19\_00a\_00\_000"  \\
MODULE="HEDIS"  \\
ComputeAttenuation -n \$NUMBEROFEVENTS -t \$COSTHETA -p \$NUPDG ---cross-sections \\
\${GENIE}/genie\_xsec/\${TUNE}.xml \\
---event-generator-list \$MODULE ---tune \$TUNE \\
}

\noindent
where the various inputs that the code requires for its execution are the described
in the following.
First of all, {\tt NUMBEROFEVENTS} indicates the number of Monte Carlo events
to be generated.
To ensure that the code produces sufficient statistics for the applications discussed in
this work, we find that one
should generate at least $10^6$ events per simulation.
{\tt NUPDG} indicates which of the six possible (anti-)neutrino flavours
compose the incoming flux for which its attenuation is being computed
and {\tt COSTHETA} stands for the cosine of the nadir (incident) angle $\cos\theta$.
The next input that needs to be selected is
the parameter {\tt TUNE} indicates the model for the neutrino-nucleon DIS
cross section adopted.
For example, {\tt GHE19\_00a\_00\_000} stands for the member 0 of
the BGR18 calculation while {\tt GHE19\_00b\_00\_000} for the member
0 of the CMS11 calculation.
Additional cross-section models, for example for new PDF sets, can be obtained
by adjusting the corresponding source code.

The parameter {\tt MODULE} allows selecting the specific neutrino-matter interaction
processes to be included in the  calculation of the attenuation rates.
This parameter can be used to reproduce some of the comparison studies that
we presented in Sect.~\ref{sec:results_p}. If {\tt MODULE="HEDIS"} is used,
then only neutrino-nucleon DIS interactions will be take into account. For $\bar{\nu}_e$, the
scattering of atomic electrons is relevant, so {\tt "HEDISGLRES"} is more convenient.
To incorporate all the sub-leading processes
described in Sect.~\ref{sec:formalism}, then {\tt "HEDISGLRESAll"} should be used.
Varying other settings of the simulation, such as the value of the incoming neutrino flux
spectral index, can also be modified using command line arguments. 

\paragraph{Precomputed  cross-section tables.}
The most CPU-intensive task when simulating the neutrino interaction events required
for the attenuation calculation
is the evaluation of the interaction cross sections.
Therefore, pre-computed cross section data are used to reduce the computational time. The cross section for each process is tabulated as function of the incoming neutrino energy and written out in XML format. Those files are converted to ROOT format to facilitate its usage. Both formats can be found the non-official repository of {\tt GENIE},
\begin{center}
  \url{https://github.com/pochoarus/GENIE-HEDIS/tree/nupropearth/genie_xsec}
\end{center}
Specifically, the pre-computed neutrino-matter interaction cross-section files
for all the subprocesses that we discuss in this work
can be found in the that link.
They include the neutrino-nucleon DIS model (for the member 0 of both the BGR18 and CSM11 calculations) and all the sub-leading contributions for different neutrino flavors and nuclei.
In the pre-computed tables that we provide, the neutrino energy $E_\nu$ ranges from 100 GeV to $10^{10}$ GeV. 
In addition, Table~\ref{tab:xsec} provides
the results for the muon neutrino-nucleon total cross sections for CC and NC interactions as function of the neutrino energy, assuming an isoscalar target and using the optimal {\tt HEDIS} settings described in Appendix~\ref{app:DIS}.
  For reference, we also display the corresponding predictions (and their uncertainties)
  for the {\tt HEDIS} calculations based on the original BGR18 and CMS11 settings.

\begin{table}[ht]
\renewcommand{\arraystretch}{1.3}
\setlength{\tabcolsep}{3pt}
\small
\begin{center}
\begin{tabular}{c|ccc|ccc}
\toprule
$E_{\nu}~[\GeV]$ &  & $\sigma_{\rm CC}/A~[10^{-38}~ {\rm m}^2]$ & & & $\sigma_{\rm NC}/A~[10^{-38}~{\rm m}^2]$ &  \\ \midrule
                &  {\tt HEDIS}~(opt)&    BGR18 &    CMS11 & {\tt HEDIS}~(opt) & BGR18 & CMS11 \\
$5\times10^{3}$  &   0.252 &  $0.252\pm0.004$ & $0.265^{+0.006}_{-0.003}$ &  0.084 &  $0.084\pm0.001$ & $0.087^{+0.001}_{-0.001}$ \\
$1\times10^{4}$  &   0.440 &  $0.439\pm0.007$ & $0.462^{+0.009}_{-0.006}$ &  0.150 &  $0.150\pm0.002$ & $0.155^{+0.002}_{-0.001}$ \\
$5\times10^{4}$  &   1.317 &  $1.316\pm0.017$ & $1.395^{+0.020}_{-0.016}$ &  0.480 &  $0.480\pm0.005$ & $0.491^{+0.005}_{-0.004}$ \\
$1\times10^{5}$  &   1.97  &  $1.96 \pm0.02$  & $2.09^{+0.03}_{-0.02}$    &  0.738 &  $0.738\pm0.007$ & $0.752^{+0.006}_{-0.006}$ \\
$5\times10^{5}$  &   4.51  &  $4.49 \pm0.06$  & $4.88^{+0.08}_{-0.06}$    &  1.79  &  $1.79 \pm0.02$  & $1.81^{+0.011}_{-0.014}$ \\
$1\times10^{6}$  &   6.28  &  $6.23 \pm0.09$  & $6.82^{+0.12}_{-0.08}$    &  2.55  &  $2.55 \pm0.03$  & $2.56^{+0.019}_{-0.020}$ \\
$5\times10^{6}$  &  13.0   &  $12.7 \pm0.2$   & $14.15^{+0.30}_{-0.16}$   &  5.44  &  $5.44 \pm0.06$  & $5.41^{+0.06}_{-0.05}$ \\
$1\times10^{7}$  &  17.5   &  $17.0 \pm0.3$   & $19.0^{+0.4}_{-0.2}$      &  7.39  &  $7.39 \pm0.09$  & $7.34^{+0.09}_{-0.07}$ \\
$5\times10^{7}$  &  33.7   &  $32.2 \pm0.6$   & $36.5^{+0.9}_{-0.5}$      & 14.5   & $14.5  \pm0.2$   & $14.31^{+0.20}_{-0.16}$ \\
$1\times10^{8}$  &  44.1   &  $41.9 \pm0.8$   & $47.7^{+1.3}_{-0.8}$      & 19.0   & $19.0  \pm0.3$   & $18.8^{+0.3}_{-0.2}$ \\
$5\times10^{8}$  &  80     &  $75   \pm2$     & $86^{+3}_{-3}$            & 34.7   & $34.7  \pm0.8$   & $34.3^{+0.6}_{-0.6}$ \\
$1\times10^{9}$  & 102     &  $96   \pm3$     & $110^{+3}_{-4}$           & 44     & $44    \pm1$     & $44.0^{+0.8}_{-0.9}$ \\
$5\times10^{9}$  & 177     & $164   \pm6$     & $188^{+6}_{-13}$          & 77     & $77    \pm3$     & $76^{+1}_{-2}$ \\
$1\times10^{10}$ & 222     & $205   \pm8$     & $235^{+8}_{-19}$          & 97     & $97    \pm4$     & $96^{+2}_{-4}$ \\
\bottomrule
\end{tabular}
\end{center}
\caption{The muon neutrino-nucleon total cross sections for CC and NC interactions as function of the neutrino energy, assuming an isoscalar target and using the optimal {\tt HEDIS} settings described in Appendix~\ref{app:DIS}.
  For reference, we also display the corresponding predictions (and their uncertainties)
  for the {\tt HEDIS} calculations based on the BGR18 and CMS11 settings.
}
\label{tab:xsec}
\end{table}

%\bibliographystyle{bib/JHEP}
%\bibliography{bib/UHEneut_attenuation}

\providecommand{\href}[2]{#2}\begingroup\raggedright\endgroup

\end{document}